\documentclass[11pt]{iopart}

\usepackage[top=2.8cm, bottom=2.8cm, left=2.4cm, right=2.4cm]{geometry}
\usepackage[utf8]{inputenc}  
\usepackage[T1]{fontenc}     

\expandafter\let\csname equation*\endcsname\relax
\expandafter\let\csname endequation*\endcsname\relax

\usepackage{amsmath,amssymb,lmodern} 
\usepackage{cite}
\usepackage{sidecap}
\usepackage[caption=false]{subfig}
\usepackage{graphicx}
\usepackage{tabularx}
\usepackage{enumerate} 
\usepackage{makecell}
\usepackage{hyperref}\hypersetup{
    colorlinks,%
    citecolor=blue,%
    filecolor=blue,%
    linkcolor=blue,%
    urlcolor=blue
}
\usepackage{color}
\usepackage{cleveref}

\newcommand{\vc}[1]{\boldsymbol{#1}}

\begin{document}
%\preprint{PI/UAN-2018-640FT}
\title[Arbitrarily coupled $p-$forms in cosmological backgrounds]{Arbitrarily coupled $p-$forms in cosmological backgrounds}

\author{Juan P. Beltr\'an Almeida}
\address{Departamento de F\'isica, Universidad Antonio Nari\~no, \\ Cra 3 Este \# 47A-15, Bogot\'a DC, Colombia}
\ead{juanpbeltran@uan.edu.co}
%\author{Author 2}
%\email{author2@uan.edu.co}

\author{Alejandro Guarnizo}
\address{Departamento de F\'isica, Universidad Antonio Nari\~no, \\ Cra 3 Este \# 47A-15, Bogot\'a DC, Colombia}
\address{Departamento de F\'isica, Universidad del Valle,\\
Ciudad Universitaria Mel\'endez, Santiago de Cali 760032, Colombia}
\ead{alejandro.guarnizo@correounivalle.edu.co}

\author{C\'esar A. Valenzuela-Toledo}
\address{Departamento de F\'isica, Universidad del Valle,\\
Ciudad Universitaria Mel\'endez, Santiago de Cali 760032, Colombia}
\ead{cesar.valenzuela@correounivalle.edu.co}

\date{\today}

\begin{abstract}
In this paper we consider a model based on interacting $p-$forms and explore some cosmological applications. Restricting to gauge invariant actions, we build a general Lagrangian allowing for arbitrary interactions between the $p-$forms (including interactions with a $0-$form, scalar field) in a given background in $D$ dimensions. For simplicity, we restrict the construction to up to first order derivatives of the fields in the Lagrangian.      
We discuss with detail the four dimensional case and devote some attention to the mechanism of topological mass generation originated by couplings  of the form $B\wedge F$ between a $p-$form and a $(3-p)-$form. As a result, we  show the system of the interacting $p-$forms $(p=1,2,3)$ is equivalent to a parity violating, massive, Proca vector field model.  
Finally, we discuss some cosmological applications. In a first case we study a very minimalistic system composed by a $3-$form coupled to a $0-$form. The $3-$form induces an effective potential which acts as a cosmological constant term suitable to drive the late time accelerated expansion of the universe dominated by dark energy. We study the dynamics of the system and determine its critical points and stability. Additionally, we study a system composed by a scalar field and a $1$-form. This case is interesting because the presence of a coupled $1-$form can generate non vanishing anisotropic signatures during the late time accelerated expansion. We discuss the evolution of cosmological parameters such as the equation of state in this model. %  Among the results, we show that this system offers an interesting arena to cosmological applications, such as dark energy, due to . }
\end{abstract}

\submitto{\CQG}

\maketitle
%%%%%%%%%%%%%%%%%%%%%%%%%%%%%%%%%%%%%%%%%%%%%%%%%%%%%%%%%%%%%%%%%
%%%%%%%%%%%%%%%%%%%%%%%%%%%%%%%%%%%%%%%%%%%%%%%%%%%%%%%%%%%%%%%%%
\section{Introduction}
%\cite{Ade:2015ava,Ade:2015hxq,Akrami:2018odb}

The inflationary paradigm \cite{Guth:1982,Starobinsky1982,Hawking1982}  successfully predicts the statistical properties of the fluctuations in temperature of the cosmic microwave background (CMB) and the formation and distribution of large scale structures (LSS), whose properties have been measured with a significantly increase of precision during the last decades \cite{Aghanim:2018eyx}.  In its simple form, based on a single scalar field %(herefater the \textit{inflaton}) 
with a slow-roll potential, inflation predicts a nearly scale invariant, nearly Gaussian and statistically isotropic distribution of the primordial perturbations. However, some anomalies in the data, suggest that  models beyond the standard slow-roll description are needed in order to fully account for its presence. These anomalies are in principle related to non-gaussianity,  statistical anisotropy, parity violation, the power deficit ta low multipoles, the hemispherical power asymmetry, the alignment of low multipoles in the CMB angular power spectrum, among others \cite{Aghanim:2018eyx,Akrami:2019izv,Akrami:2019bkn,Schwarz:2015cma,Perivolaropoulos:2014lua} (see Refs \cite{Schwarz:2015cma} and \cite{Perivolaropoulos:2014lua}  for a  review of the cosmological anomalies).  Models based on vector fields \cite{Dimastrogiovanni:2010sm,Soda:2012zm, Maleknejad:2012fw}, $p-$forms \cite{Mulryne:2012ax, Ohashi:2013qba, Ohashi:2013mka, Kumar:2016tdn,Farakos:2017jme, Obata:2018ilf, Almeida:2019xzt}, higher spin fields \cite{Arkani-Hamed:2015bza,Kehagias:2017cym,Bartolo:2017sbu,Franciolini:2017ktv,Baumann:2017jvh,Franciolini:2018eno,Bordin:2018pca,Anninos:2019nib} or axion monodromy inflationary models \cite{Cai:2014vua,Cai:2015xla,Ibanez:2015fcv,Valenzuela:2016yny} have been considered as a plausible explanation of some of those anomalies.  The use of vectors, or general gauge fields is strictly constrained by the cosmic no-hair theorem \cite{Wald:1983ky}, which states that these fields dilute rapidly in the presence of a cosmological constant, which render them,  in principle, irrelevant during the inflationary expansion. Nevertheless, one can evade the conditions behind this theorem by introducing couplings of the form  $f(\phi) F^{\mu\nu} F_{\mu\nu}$ and $f(\phi) F^{\mu\nu} \tilde{F}_{\mu\nu}$ where $f(\phi)$ is an arbitrary function of the scalar inflaton field, $F_{\mu\nu}=\partial_{\mu} A_{\nu}- \partial_{\nu} A_{\mu}$ is the field strength of the $1-$form $A_{\mu}$ and $\tilde{F}_{\mu\nu}$ is its dual. %\cite{Watanabe:2009ct}. 
Those couplings allow for the introduction of anisotropic non-diluting signals in the correlation functions of the primordial curvature perturbation during the inflationary era.  %The idea behind it, is that these fields can naturally induce anisotropic signals, and therefore possible signatures in statistical correlators describing observables in inflationary physics. A very well-known example of these models is based in a Maxwell term coupled to a  
The model  $f(\phi) F^{\mu\nu} F_{\mu\nu}$ has been extensively studied in the literature  \cite{Yokoyama:2008xw, Watanabe:2009ct, Dimopoulos:2009vu, Dimopoulos:2009am, Watanabe:2010bu, Bartolo:2012sd, Biagetti:2013qqa, Shiraishi:2013vja, Abolhasani:2013zya, Lyth:2013kah, Rodriguez:2013cj, Lyth:2013vha, Shiraishi:2013oqa, Chen:2014eua, Almeida:2014ava, Fujita:2017lfu,Thorsrud:2012mu}. Parity violating signals in the correlators and potential applications to primordial magnetogenesis can be achieved with a prototypical term of  form $f(\phi) F^{\mu\nu} \tilde{F}_{\mu\nu}$. This term has also been studied with great interest in the recent literature \cite{Sorbo:2011rz, Dimopoulos:2012av, Anber:2012du, Bartolo:2014hwa, Caprini:2014mja, Bartolo:2015dga, Namba:2015gja, Shiraishi:2016yun, Caprini:2017vnn, Almeida:2017lrq, Almeida:2018pir,Almeida:2019hhx}. \\

In the specific context of inflation, a general study of cosmological perturbations and stabilities at background level has been done in Refs. \cite{Ohashi:2013qba, Ohashi:2013mka,  Obata:2018ilf,Almeida:2019xzt}, in which, a coupling of a function of  a scalar field and a $1-$ and $2-$form, is considered. At perturbative level, the statistical anisotropy induced by these models can be parameterized through the power spectrum of curvature perturbations  \cite{Ackerman:2007nb}
 \begin{equation}\label{ps}
 P_{\zeta}(\bi{k}) = P^{(0)}_{\zeta}(k) \left[ 1 + g_{*} \cos^2 \theta_{\bi{k},\bi{V}} \right],
 \end{equation}
being $P^{(0)}_{\zeta}(k) $ the isotropic power spectrum, $g_{*}$ the anisotropy parameter, $\bi{V}$ is the preferred direction, $\bi{k}$ the wave vector and $\theta_{\bi{k,}\bi{V}}$ the angle between $\bi{k}$ and $\bi{V}$. % \cite{Ohashi:2013qba,Ohashi:2013mka}. 
It has been shown that the  different types of anisotropies (coming from the vector and $2-$form field) affect the sign of $g_{*}$, being positive for the $2-$form field, and negative for the $1-$form case \cite{Ohashi:2013qba}.  These features allow to constraint different inflationary models  
with current CMB bounds on $g_{*}$.   {Interestingly enough, some oscillatory features in the spectrum can be derived from certain axion monodromy inflationary models \cite{Flauger:2009ab, Flauger:2014ana} yielding in some cases to a step-like profile on the inflaton's potential  \cite{Cai:2014vua}. These type of models lead to observational predictions for the power spectrum which exhibits great similarities with the predictions from the models discussed here. Specifically, the power spectrum of the curvature perturbation can be parametrized in a similar form as the one in  Eq. \eqref{ps}  and their predictions are highly degenerated with the predictions of inflation in presence of $p-$forms \cite{Cai:2015xla}.  Aside of  offering a nice explanation of the power deficit anomaly at low multipoles, these kind of models  predict specific patterns in the temperature and polarization spectra which can be tested and constrained by forthcoming CMB observations \cite{Cai:2015xla}.} \\

A different approach to the $p-$forms that we consider here, that induce statistically anisotropic signatures in the power spectrum and the bispectrum, is the use of higher spin fields (HSF) 
 \cite{Arkani-Hamed:2015bza,Kehagias:2017cym,Bartolo:2017sbu,Franciolini:2017ktv,Baumann:2017jvh,Franciolini:2018eno,Bordin:2018pca}. As in the model with vector fields, HSF does not generate long-lived perturbations with spin $s$ unless they are coupled to the inflaton field through terms of the form $f(\phi) \sigma_{\mu_1\cdots \mu_s}\sigma^{\mu_1\cdots \mu_s}$ where $\sigma_{\mu_1\cdots \mu_s}$ is a symmetric tensor representing a spin $s$ field. Possible observables and signatures from HSF in LSS probes and future galaxy surveys %and yielding a perfect arena  to test those models
have been discussed in \cite{Schmidt:2015xka,MoradinezhadDizgah:2017szk,MoradinezhadDizgah:2018pfo,MoradinezhadDizgah:2018ssw}. A peculiar difference between the $p-$forms and the HSF are the symmetries  involved: while $p-$forms are built with antisymmetric objects, HSF are built out of symmetric tensors. This implies that, for instance, it is not possible to introduce non vanishing parity violating signals by using HSF since contractions of the antisymmetric tensor $\epsilon^{\mu_1 \mu_2 \mu_3 \mu_4}$ with any symmetric tensor is trivially zero. $p-$forms on the other hand, due to their antisymmetric structure, are better suited to study parity breaking signatures.    \\

Going beyond the $1-$ and $2-$form cases in four dimensions mentioned before, studies of general $p-$forms in Bianchi cosmologies has  been carried out recently in \cite{Normann:2017aav}, and  studies specifically related with $3-$forms had been carried out in  \cite{Brown:1987dd, Brown:1988kg, Duncan:1989ug, Duff:1989ah, Bousso:2000xa, Dvali:2005an, Kaloper:2008qs, Kaloper:2008fb, Koivisto:2009sd, Koivisto:2009ew, Koivisto:2009fb, Bielleman:2015ina, Ibanez:2015fcv, Valenzuela:2016yny,Farakos:2017jme}.  A property exploited in those references relies on the fact that the field strength of the $3-$form in four dimensions is proportional  to the volume element and can be seen as a cosmological constant term. This makes three form relevant for the discussion of the cosmological constant problem and also viable dark energy candidates \cite{Amendola2010}. Vector fields, or $1-$forms, had also been used as dark energy candidates  \cite{Clifton:2011jh,Koivisto:2008xf,ArmendarizPicon:2004pm,Boehmer:2007qa,Thorsrud:2012mu}.  On more theoretical grounds, a generalization of the scalar $0-$form \textit{Galileons} \cite{Fairlie:1991qe,Fairlie:1992nb,Nicolis:2008in,Deffayet:2009wt,Deffayet:2009mn} to arbitrary $p-$forms in $D$-dimensions has been carried out in \cite{Deffayet:2010zh,Deffayet:2017eqq}.
%In the scalar field sector (%or more properly
%the $0-$form), 
%many attempts has been made to generalize the inclusion of scalars with certain symmetries as broad as possible. A well known example are the so called \textit{Galileons}, which are scalar fields satisfying a shift symmetry, and most important, their action entail to equations of motion of at most of second order, a must have property to avoid instabilities \cite{Deffayet:2009mn,Deffayet:2011gz}.   In this context, a generalization to $p-$form Galileons in $D$-dimensions as $p-$forms with second order equations of motion, has been carried out in \cite{,Deffayet:2010zh,Deffayet:2017eqq}. 
In the previous construction, it is allowed to have an action with second derivatives of the $p$-form, in close resemblance with Gauss-Bonnet-Lovelock actions. This procedure leads to counterterms and non-minimal couplings with the gravity sector when writing the action in curved space. Differently from this $p-$form Galileon generalization which includes second order derivatives in the Lagrangian, our aim here is to study general $p-$forms models with up to first order derivatives.  \\

It is desirable to fully describe a theory which couples different types of $p-$forms demanding only first derivatives of the field strengths, unlike the approach of the $p-$form Galileons, and classifies their possible imprints in statistical correlators.  The main purpose of this paper is to set up a general Lagrangian based on coupled $p-$forms, restricted by gauge invariance. Additionally, we aim to highlight the role of $p-$form models in cosmological backgrounds. In \cref{sec:pform} we start with the basic definitions of $p-$forms, their field strengths and duals. After discussing the gauge invariance of the $p-$forms the construction of the Lagrangian starts using as building blocks the field strengths coupled with a function of a scalar field (a $0-$form), as well as couplings between different $p-$forms in $D$ dimensions. We briefly discuss the existence of topological terms and their natural appearance in the Lagrangian. Section \ref{sec:4d} is devoted to exploring the Lagrangian in 4 dimensions, with a detailed description of the field equations and the energy-momentum tensor. We devote a subsection to describe the mechanism of topological mass generation that arises due to the interplay of the $1-$form and the $2-$form. Some cosmological applications are shown in  \cref{sec:cosmology}, with special attention to the effect of the coupled $3-$form-scalar field term in the dynamics;  we also discuss a late time evolution of the universe in presence of an anisotropic source such as a $1-$form; lastly, we also  comment very shortly on the signatures of $p-$forms in statistical inflationary correlators. Finally we draw the conclusions in \cref{sec:Conclusions}. Throughout this paper we will use   a Lorentzian metric $g_{\mu\nu}$  with signature $(-, +,+,+)$; greek indices $\alpha,\beta,\gamma,\ldots$ will denote space-time coordinates, while latin indices $i,j,k,\ldots$ denote spatial coordinates.

%%%%%%%%%%%%%%%%%%%%%%%%%%%%%%%%%%%%%%%%%%%%%%%%%%%%%%%%%%%%%%%%%
\section{ $p-$forms coupled to a scalar in $D$ dimensions}\label{sec:pform}
%%%%%%%%%%%%%%%%%%%%%%%%%%%%%%%%%%%%%%%%%%%%%%%%%%%%%%%%%%%%%%%%%

In this section, we recall the basic definitions of $p$-forms and establish our notation. Our goal is to construct a general Lagrangian compatible with gauge symmetry and   allowing for couplings between different arbitrary $p$-forms of different rank. Having in mind applications to inflationary physics, we also concede for kinetic coupling with a scalar field $\phi$ which we can see  as a $0-$form whose dynamics is introduced through a field strength $\partial_{\mu}\phi$. So, the Lagrangian that we are going to consider is the following
\begin{equation}
{\cal L}=  {\cal L}_{\phi} + {\cal L}_{p}(\phi, A_{p}),
\end{equation}
where 
%\begin{equation}\label{eq:Kphi}
${\cal L}_{\phi}  = {\cal L}_{\phi} (\phi, K)$
%\end{equation}
is an arbitrary function of the field and the kinetic term $K \equiv \partial_{\mu}\phi \partial^{\mu}\phi$. In the canonical case we have 
\begin{equation}\label{eq:Lph}
{\cal L}_{\phi}  =- \frac{1}{2}\partial_{\mu}\phi \partial^{\mu}\phi -  V(\phi),
\end{equation}
being $V(\phi)$ a potential suitable for the specific case of study, for instance, in the case of  inflationary physics, the potential $V(\phi)$ should be able to drive slow roll inflationary evolution. $A_{(p)}$ is a $p-$form
\begin{equation}\label{eq:pf}
A_{(p)} = \frac{1}{p!} A_{(p) \mu_1 \mu_2\cdots \mu_{p}}{\rmd}x^{\mu_1}\wedge{\rmd}x^{\mu_2}\cdots \wedge {\rmd}x^{\mu_{p}},
\end{equation}
 where the $A_{(p)  \mu_1 \mu_2\cdots \mu_{p}}$ are taken totally anti-symmetric and $\wedge$ representing the usual ``wedge'' product. To avoid further confusion with indices in a specific dimension, we will use a subscript $p$ between parenthesis to denote the order of the $p-$form.
The Lagrangian ${\cal L}_{p}(\phi, A_{p})$ will include all the possible $p-$forms in a given $D-$dimensional spacetime. The highest rank of a $p-$form in $D$ dimensions is obviously $D$,  which is proportional to the $D-$dimensional volume element, this is:
\begin{equation}
A_D \propto \sqrt{-g} \epsilon_{1\cdots D}  {\rmd}x^{{1}} \wedge \cdots \wedge {\rmd}x^{{D}},
\end{equation}
where $ \epsilon_{1 \cdots D}$ is the Levi-Civita tensor. This term can be absorbed as a cosmological constant term or can be seen as a redefinition of the vacuum of the potential \cite{Koivisto:2009sd,Ito:2015sxj,Mulryne:2012ax,Kumar:2016tdn}, it has no dynamics, and for this reason, we shall only consider up to $A_{D-1}$ in the following. Given a $p$-form $A_{(p) \mu_1,\mu_2\cdots \mu_p}$, its dynamics is introduced through the field strength $F_{(p)}$, which is a $(p+1)$-form defined as the exterior derivative of the $p$-form:\footnote{Notice that we have used the subscript $p$ as a label for the $p$-form. The field strength has the same label, but we emphasize that $F_{(p)}$ is a $(p+1)$-form. } 
\begin{equation}\label{eq:FS}
F_{(p)} = 
\frac{1}{p!}\nabla_{[\mu_1}A_{(p)  \mu_2\mu_3\cdots\mu_{p+1}]} {\rmd}x^{\mu_1}\wedge{\rmd}x^{\mu_2}\cdots \wedge {\rmd}x^{\mu_{p+1}}.
\end{equation} 
Here, it is important to mention that, although we define the field strength with covariant derivatives, in a spacetime endowed with a symmetric connection, without torsion, the antisymmetrization in \cref{eq:FS} will transform all covariant derivatives into ordinary partial derivatives. Having said that, applying the definition \cref{eq:pf} for the field strength, their components  are computed as
\begin{equation}\label{FScomponents}
F_{(p)  \mu_1\mu_2\cdots \mu_{p+1}} \equiv (p+1) \partial_{[\mu_1} A_{(p) \, \mu_2 \mu_3\cdots \mu_{p+1}]}.
\end{equation}
Joint with this, we define the Hodge dual of  a $p$-form $B_p$ as a  $(D-p)-$form in the following way
\begin{equation}
\star {B}_{(p)} = \frac{1}{p!}\frac{1}{(D-p)!}\eta_{\mu_1 \cdots \mu_{p} \nu_{1}\cdots \nu_{D-p}}B_{(p)}^{\, \, \mu_{1}\cdots  \mu_{p}  } {\rmd}x^{\nu_{1}} \wedge\cdots \wedge {\rmd}x^{\nu_{D-p}},
\end{equation}
with $\eta_{\mu_1 \cdots \mu_{D}} = \sqrt{-g} \epsilon_{\mu_1 \cdots \mu_{D} }$. With this definition the Hodge dual of the field strength $F_{(p)}$ reads
\begin{equation}\label{eq:HDu}
\star {F}_{(p)} = \frac{1}{(p+1)!}\frac{1}{(D-p-1)!}\eta_{\mu_1 \cdots \mu_{p+1} \nu_{1}\cdots \nu_{D-p-1}} F_{(p)}^{\,  \, \mu_{1}\cdots  \mu_{p+1}  } {\rmd}x^{\nu_{1}} \wedge\cdots \wedge {\rmd}x^{\nu_{D-p-1}}.%\\
\end{equation}
Denoting the components of the Hodge dual as $\tilde{F}_{(p) \nu_{1}\cdots   \nu_{D-p-1}}$, and using again \cref{eq:pf} for a $(D-(p+1))-$form we find
\begin{equation}\label{eq:HDuc}
\tilde{ F}_{(p)  \nu_{1}\cdots   \nu_{D-p-1} }= \frac{\sqrt{-g} }{(p+1)! } \epsilon_{\mu_1 \cdots \mu_{p+1} \nu_{1}\cdots \nu_{D-p-1}}  
F_{(p)}^{\, \, \mu_{1} \cdots \mu_{p+1}}.  
\end{equation}
Along with the previous definitions, we will also use the notation for the wedge product of any two forms
\begin{equation}\label{eq:wedge}
(A_{(r)}\wedge B_{(q)})_{\mu_{1}\cdots \mu_{r} \, \nu_{1}\cdots \nu_{q}} = \frac{(p+r)! }{p!r! } A_{(r)[\mu_{1}\cdots \mu_{r} }B_{(q)\nu_{1}\cdots \nu_{q}]}.
\end{equation}

%%%%%%%%%%%%%%%
\subsection{Gauge invariance and minimal coupling to gravity}\label{sec:gauge}
%%%%%%%%%%%%%%%
The field strength is endowed with symmetry under the redefinitions of the $p-$form
\begin{equation} \label{eq:gaugesymm}
A_{(p) {\mu_1\cdots \mu_p }} \rightarrow  A_{(p) {\mu_1\cdots \mu_p }} + \partial_{[ \mu_1} \xi_{(p-1) \mu_2 \cdots \mu_p ]}, 
\end{equation}
being $\xi_{(p-1) \mu_2 \ldots \mu_p}$ a $(p-1)$-form. This  is the usual form to state that a model is %(globaly) 
gauge invariant. It is important to mention that, as we said before, the derivatives in the gauge transformation before involves only ordinary derivatives and the quantity $\xi_{\mu_1\cdots \mu_p}$ is an antisymmetric tensor.  If we restrict our construction to gauge invariant terms, with only first order derivatives in the Lagrangian, the covariant version of this theory do not need to introduce non-minimal couplings to gravity. In theories with higher order derivatives, the covariant version for curved backgrounds requires the inclusion of non-minimal coupling terms in order to avoid that higher than two derivatives terms appearing in the equations of motion for both, the $p-$forms and the gravitational field. The general description to include non-minimal coupling terms for theories with second order derivatives in the Lagrangian is carefully described in \cite{Deffayet:2010zh}. As we restrict ourselves to first order derivatives, gauge invariant combinations in the Lagrangian, we will not need to care about non-minimal coupling with gravity. 

%%%%%%%%%%%%%%%
\subsection{General procedure}\label{sec:Genproc}
%%%%%%%%%%%%%%%
The Lagrangian that we are going to formulate will be built out from the appropriate combinations of the $p-$forms field strengths  and their duals. To start with, in $D$-dimensions, we endow a $p-$form with dynamics via a term like
\begin{equation}\label{eq:SAD}
S_{(p)} = - \frac{1}{2} \int F_{(p)} \wedge \star F_{(p)}  \equiv -\frac{1}{2(p+1)!} \int {\rmd}^{D}x \sqrt{-g} F_{(p)}^2,  %\equiv -\frac{1}{2(p+1)!} \int {\rmd}^{D}x \sqrt{-g} F_{(p)  \mu_1\cdots \mu_{p+1}} F_{(p)}^{\, \, \mu_1\cdots \mu_{p+1}} , 
\end{equation}
where 
\begin{equation}\label{F2}
F_{(p)}^2\equiv F_{(p)  \mu_1\mu_2\ldots \mu_{p+1}}F_{(p)}{}^{\mu_1\mu_2\ldots \mu_{p+1}}, 
\end{equation}
which is a generalization of the usual Maxwell term for the $1-$form ${\cal L}_{M}=-\frac{1}{4}F_{\mu\nu}F^{\mu\nu}$. Now, we will consider couplings as
\begin{equation}\label{eq:fphi}
{\cal L}_{{\rm int}} \propto  f_{p}(\phi)  F_{(p)} \wedge \star F_{(p)},
\end{equation}
where $f_{p}(\phi)$ is a function of the scalar field only. We do not consider couplings of the form $f_{p}(\phi, K)$ where $K= \partial_{\mu}\phi \partial^{\mu} \phi$, since, for the $A_{\mu_1}$ case, this leads to configurations in which the Hamiltonian of the theory is not bounded by below, which renders the theory unstable  \cite{Fleury:2014qfa}. It is possible to find, however, regions in the parameter space, with appropriate initial conditions, in which the theory behaves stably \cite{Holland:2017cza}, but, on general grounds we will not consider such couplings here, neither contractions of $p-$forms with $\partial_{\mu}\phi$ since this leads to non-causal equations of motion in the case of a $1-$form \cite{Fleury:2014qfa}.  Couplings like \cref{eq:fphi} have been extensively studied in the literature in the context of inflationary physics \cite{Dimopoulos:2009am, Watanabe:2010bu, Bartolo:2012sd, Biagetti:2013qqa, Shiraishi:2013vja, Abolhasani:2013zya, Lyth:2013kah, Shiraishi:2013oqa, Chen:2014eua, Almeida:2014ava, Fujita:2017lfu}. The main motivation for the introduction of those couplings is that they allow the possibility for the massless perturbations of the antisymmetric tensors to leave some imprint on the inflationary correlators. It also has some impact on the dynamics of the inflationary curvature and tensor perturbations. Without the coupling, the signatures of the $p-$forms would be totally washed out by the inflationary dynamics.   \\
Aside of the quadratic term in \cref{eq:SAD}, we shall also consider general mixing between $p-$forms of different rank. In $D-$dimensions we have the following general form to couple $p-$forms of different rank:
\begin{equation}\label{eq:mixing}
%{\cal L} &\propto & \eta^{\mu_{1}\cdots \mu_{D}} \eta^{\nu_{1}\cdots \nu_{D}} F_{(p)}{}_{\mu_{1}\cdots }F_{(q)}{}_{\nu_{1}\cdots }\\
{\cal L}_{{\rm mixing}} = g_{p_{1}p_{2} \cdots p_{r}}(\phi)  X_{(p_{1})}\wedge \cdots \wedge X_{(p_{r})} =   g_{p_{1}p_{2} \cdots p_{r}}(\phi)  \eta^{\mu_{1}\cdots \mu_{D}} X_{(p_{1})}{}_{\mu_{1}\cdots \mu_{p_{1}+1} } \cdots X_{(p_{r})}{}_{\mu_{p_r +1}\cdots \mu_{D} }, 
\end{equation} 
where $g_{p_{1}p_{2} \cdots p_{r}}(\phi)$ as before, is a coupling function depending only of the scalar field, $X_{(p_i)}$ can be either the field strength of a $p-$form or its Hodge dual, and the ranks of the $p-$forms involved in the product is such that $(p_1 +1) +\cdots + (p_r +1) = D$, when only field strengths are involved, and $(p_1 +1) +\cdots+ D-(p_{i}+1)\cdots+ (p_r +1) = D$ if the Hodge dual of the $p_i-$form is involved. It is straightforward to see that with the expression \cref{eq:mixing} we can write all the possible combinations of forms 
constructed as  contractions of the field strengths with the appropriate number of indices in the covariant and contravariant factors. {We could also include explicit couplings with the $p-$forms but we will exclude those couplings because they break the gauge symmetry  reflected by Eq. \eqref{eq:gaugesymm}. Nevertheless, there are some couplings that include manifestly the $p-$forms and still retain the gauge symmetry. We will list this cases in the next subsection.} 
Restricting to the field strengths, for instance, we can combine different $p-$forms in the following way:
\begin{equation}
F_{(p)}F^{(n)}F^{(m)} \equiv F_{(p)  \mu_1\mu_2\ldots \mu_{p+1}}F_{(n)}^{\; \, \mu_1\mu_2\ldots \mu_{n+1}}F_{(m)}^{\; \, \mu_{n+2}\ldots \mu_{p+1}}, 
\end{equation}
such that the condition $p = n + m +1$ holds. Indeed, we can contract a $p-$form with a number $i$ of $n_i-$forms satisfying
\begin{equation}
F_{(p)}F^{(n_1)}\cdots F^{(n_i)} \equiv F_{(p)  \mu_1\mu_2\ldots \mu_{p+1}}F_{(n_1)}^{\; \, \mu_1\mu_2\ldots \mu_{n_1+1}}\cdots F_{(n_i)}^{\; \, \mu_{s}\ldots \mu_{p+1}},
\end{equation}
with $p \equiv n_1 + \cdots + n_i +i-1$. In an analogous way,  including the Hodge duals, such that  the rank fullfils $p \leq D$, we can also construct gauge invariant terms such as
\begin{equation}
\tilde{F}_{(p)}F^{(n)}F^{(m)} \equiv \tilde{F}_{(p)  \mu_1\mu_2\ldots \mu_{D-(p+1)}}F_{(n)}^{\; \, \mu_1\mu_2\ldots \mu_{n+1}}F_{(m)}^{\; \, \mu_1\mu_2\ldots \mu_{m+1}},
\end{equation}
taking into account  that $D-(p+1) = n + m+2$ holds. And so on and so forth. It is clear that all the possible combinations strongly depend on the dimension of the spacetime and for this reason, such possibilities can only be listed when we are working on a specific dimension. In the next section, we will focus on the four dimensional case and we describe with detail all the possibilities. \\

Finally, we add gravity to the system. As we said before, gauge invariance and the antisymmetric structure of the $p-$forms avoids us from including non-minimal coupling terms, so,  we can include gravity in a minimal way to the system by just adding tan  Einstein-Hilbert like term in $D$ dimensions in the Lagrangian:
%Generalizing this, and including curvature, we can extend the result in $D-$dimensions as
\begin{equation}\label{eq:STP}
S = \int {\rmd}^{D}x \sqrt{\bar{g}}\left( \frac{\bar{M}_{\rm{p}}^{D-2}}{2}\bar{R} + {\cal L}_{\phi} +{\cal L}_{p}\right ), %\quad \cal{L}_{p} =  \sum_{n=1}^{D-1} \frac{f_{n}(\phi)}{2(n+1)!}F_{(n)}^2 ,
\end{equation}
where
\begin{align}\nonumber
{\cal L}_{p} &=-\frac{1}{2}  \sum_{n=1}^{D-1}  f_{n}(\phi)  F_{(n)} \wedge \star F_{(n)}  +  \sum_{(p_{1}p_{2} \cdots p_{r})} g_{p_{1}p_{2} \cdots p_{r}}(\phi)  X_{(p_{1})}\wedge \cdots \wedge X_{(p_{r})}, \\ \label{eq:Lpint}
&= -\frac{1}{2} \sum_{n=1}^{D-1} \frac{f_{n}(\phi)}{(n+1)!}F_{(n)}^2 +  \sum_{(p_{1}p_{2} \cdots p_{r})} g_{p_{1}p_{2} \cdots p_{r}}(\phi)  X_{(p_{1})}\wedge \cdots \wedge X_{(p_{r})},
\end{align}
where $(p_{1}p_{2} \cdots p_{r})$ is a shorthand notation to express all the possible combinations of field strengths and duals for a given dimension. $\bar{g}, \ \bar{R}$ and $\bar{M}_{\rm{p}}$ are the determinant of the metric, the Ricci scalar and the Planck mass in $D$ dimensions, respectively.

%%%%%%%%%%%%%%%
\subsection{Topologic terms}\label{topological}
%%%%%%%%%%%%%%%
The procedure that we outlined before allows us to include topological terms. Although such terms are not relevant for the dynamics of the system, they will play  a non-trivial dynamical role   when they couple to a scalar field. One of the most widely studied case in even dimensions $D=2(p+1)$ is the Chern-Pontryagin density or $\theta-$term, which is written as:
\begin{equation}\label{eq:SADT}
S_{\rm CP} = - \frac{1}{2} \int F_{(p)} \wedge  F_{(p)}  \quad \mbox{with}\quad D=2(p+1).
\end{equation}
According with the definition in \cref{eq:wedge}, we have
\begin{equation}\label{eq:SADT1}
S_{\rm CP} =  %-\frac{1}{2(p+1)!} \int {\rmd}^{D}x \sqrt{-g} \eta^{\mu_{1}\cdots \mu_{D}} {F}_{(p)}{}_{\mu_1\cdots \mu_{p+1}} {F}_{(p)}{}_{\mu_{p+2}\cdots \mu_{D}}
  -\frac{1}{2(p+1)!} \int {\rmd}^{D}x \sqrt{-g}  \tilde{F}_{(p)}{}_{\mu_1\cdots \mu_{p+1}} {F}_{(p)}{}^{\mu_1\cdots \mu_{p+1}}.
\end{equation}
This theory is topological as it is manifestly independent of the metric, so, this particular term does not modify the structure of the gravitational field equations because its energy-momentum tensor vanishes. Nevertheless, once it is coupled to a scalar field,
\begin{equation}\label{eq:SCSphi}
S_{\phi {\rm CP}} =  \int   g_1(\phi) F_{(p)} \wedge  F_{(p)}, 
\end{equation}
it becomes relevant for the dynamics of the scalar and the $p-$form field. In this case, it could leave an imprint on the gravitational field because of the scalar field coupling.  This term has been extensively studied for the case of a vector field  in the context of inflation \cite{Sorbo:2011rz, Dimopoulos:2012av, Anber:2012du, Bartolo:2014hwa, Caprini:2014mja, Bartolo:2015dga, Namba:2015gja, Shiraishi:2016yun, Caprini:2017vnn, Almeida:2017lrq, Almeida:2018pir}, in particular, it has been used to provide a mechanism to seed chiral gravitational waves \cite{Anber:2012du,Caprini:2014mja,Caprini:2017vnn}.  \\

At this point, we have to stress out that, with the procedure outlined before, by including couplings of the form \eqref{eq:mixing}, we have not considered other topological terms which are gauge invariant under the transformation given in \cref{eq:gaugesymm}. Depending on the dimensionality of the spacetime, we can add two more terms to the list. For odd dimensions we have the Chern-Simons invariant \cite{Chern:1974ft}
\begin{equation}
S_{\phi {\rm CS}} =  \int   g_2(\phi) A_{(p)} \wedge  F_{(p)},   \quad \mbox{with}\quad D=2p+1, 
\end{equation}
which is only gauge invariant when the coupling $g_2$ is a constant. For even dimensions, we have the so called $BF-$theories \cite{Blau:1989dh,Horowitz:1989ng,Blau:1989bq, Birmingham:1991ty} which couples a $p-$form with the field strength of a $(p-1)-$form. This theory is  usually written in the following way
\begin{equation}\label{eq:BFD}
S_{\phi {\rm BF}} =  \int  g_3(\phi)    A_{(p)} \wedge  F_{(D-p-1)}, % = g_3(\phi)  \int {\rmd}^{D}x \sqrt{-g} \eta^{\mu_{1}\cdots \mu_{D}} A_{(p)}{}_{\mu_{1}\cdots\mu_{p}} F_{(D-p-1)}{}_{\mu_{p+1}\cdots\mu_{D}} , 
\end{equation}
which, in the decoupled case, produces the equations of motion ${\rmd}A_{(D-p-1)}={\rmd}B_{(p)}=0$, where ${\rmd}$ is the exterior derivative of the form $\epsilon^{\mu_1 \mu_2 \mu_3 \mu_4}\partial_{\mu_1}$. There exists another combination of the form $A_{(D-p-1)} \wedge  F_{(p)}$ but it is easy to see that, after integration by parts,  this is equivalent to \cref{eq:BFD} and produces the same dynamics. The better studied case is the four dimensional case, which couples a $2-$form $B_{(2)}$ and the field strength of a $1-$form $F_{(1)}$ 
\begin{equation}
S_{\phi {\rm BF}}  =   \int   g_3(\phi) B_{(2)} \wedge  F_{(1)}  =   \int {\rmd}^{4}x \sqrt{-g} g_3(\phi) \eta^{\mu_{1}\cdots \mu_{4}} B_{(2)}{}_{\mu_{1}\mu_{2}} F_{(1)}{}_{\mu_{1}\mu_{2}}.
\end{equation} 
Notice that when we couple the BF term with the scalar field, the action is not  gauge invariant anymore under the transformations in  \cref{eq:gaugesymm}. However, the theory is still symmetric under the next transformations:
\begin{equation}
A_{(1)}{}_{\mu} \rightarrow A_{(1)}{}_{\mu} + \partial_{\mu} \xi \quad \mbox{and} \quad B_{(2)}{}_{\mu \nu} \rightarrow B_{(2)}{}_{\mu \nu} + \frac{1}{g_3 (\phi)} \partial_{[\mu} \xi_{\nu ]}.
\end{equation}
In the presence of terms like $F_{(1)} \wedge \star F_{(1)}$ and $F_{(2)} \wedge \star F_{(2)}$, the theory is not invariant under the previous transformation unless the coupling $g_3$ is constant. 
Of course, all the previous quantities, although expressed for an Abelian symmetry group, can be extended straightforwardly for non-Abelian internal symmetry groups. \\

An interesting feature that we want to highlight here is the fact that some of the terms listed before break parity symmetry. As mentioned previously, in the context of inflation,  the parity violating nature of \cref{eq:SCSphi}, have been exploited to source chiral gravitational waves and to study parity violating signals in the inflationary correlators \cite{Sorbo:2011rz, Anber:2012du}. For instance, for the three point correlator, it was found in \cite{Bartolo:2014hwa,Bartolo:2015dga} that the term in \cref{eq:SCSphi} is related to the presence of angular dependencies with odd multipole terms in the correlation functions.  The idea that we want to promote here is that such parity violating features are generic from topological terms like the ones that we discussed in this section, and, in this way, the presence of parity violating signatures in the correlation functions can be related to global, topological features of the spacetime. In particular, we consider that  BF-like models could be of potential interest for the discussion of chiral gravitational waves and parity odd signals in the statistical distribution of the inflationary perturbations. We expect to come back to this issues elsewhere.

%%%%%%%%%%%%%%%%%%%%%%%%%%%%%%%%%%%%%%%%%%%%%%%%%%%%%%%%%%%%%%%%%
\section{$p-$forms in four dimensions}\label{sec:4d}
%%%%%%%%%%%%%%%%%%%%%%%%%%%%%%%%%%%%%%%%%%%%%%%%%%%%%%%%%%%%%%%%%
In this section, we work in a four dimensional background. According to the discussion in the previous section, we will have three $p-$forms in this case. Following the definition of \cref{eq:pf}:
\begin{align}
A_{(1)} = A_{(1) \, \mu_{1}   } {\rmd}x^{\mu_{1}},  \quad
A_{(2)} = \frac{1}{2}A_{(2) \, \mu_{1} \mu_{2}  } {\rmd}x^{\mu_{1}} \wedge {\rmd}x^{\mu_{2}}, \quad \mbox{and}  \quad
A_{(3)} = \frac{1}{6}A_{(3) \, \mu_{1} \mu_{2} \mu_{3}  } {\rmd}x^{\mu_{1}} \wedge {\rmd}x^{\mu_{2}} \wedge {\rmd}x^{\mu_{3}}. \quad% \\
%A_{(4)} &=& \frac{1}{24}A_{(4) \, \mu_{1} \mu_{2} \mu_{3} \mu_{4} } {\rmd}x^{\mu_{1}} \wedge {\rmd}x^{\mu_{2}} \wedge {\rmd}x^{\mu_{3}} \wedge {\rmd}x^{\mu_{4}},
\end{align}
As we said before, in four dimensions the list stops here since the 4-form is proportional to the volume element, which is proportional to a cosmological constant term. Moreover, there is no associated field strength and consequently, this field is non dynamical and for this reason, we do not consider it here\footnote{Nevertheless, see \cite{Aydemir:2010tna} for an attempt to construct a model with a propagating degree of freedom out of a $4-$form.}.

Associated with them, from \cref{FScomponents}, we have the  field strengths 
\begin{align}
 F_{(1)  \mu_{1}  \mu_{2}  } &=& 2 \partial_{[\mu_1}A_{(1)\mu_2]},\quad
F_{(2)  \mu_{1} \mu_{2}  \mu_{3}} = 3 \partial_{[\mu_1}A_{(2) \mu_2 \mu_3]},\quad \mbox{and}\quad
F_{(3) \, \mu_{1} \mu_{2} \mu_{3}  \mu_{4}  } = 4 \partial_{[\mu_1}A_{(3) \mu_2 \mu_3 \mu_4]}.\quad
\end{align}
%\ba
%F_{(1)} &=& F_{(1)  \mu_{1}  \mu_{2}  } {\rmd}x^{\mu_{1}}  \wedge {\rmd}x^{\mu_{2}} = \partial_{[\mu_1}A_{(1)\mu_2]} {\rmd}x^{\mu_{1}}  \wedge {\rmd}x^{\mu_{2}},  \\[2mm]
%F_{(2)} &=& \frac{1}{2}F_{(2)  \mu_{1} \mu_{2}  \mu_{3}} {\rmd}x^{\mu_{1}} \wedge {\rmd}x^{\mu_{2}} \wedge {\rmd}x^{\mu_{3}} =  \frac{1}{2}\partial_{[\mu_1}A_{(2) \mu_2 \mu_3]}  {\rmd}x^{\mu_{1}} \wedge {\rmd}x^{\mu_{2}} \wedge {\rmd}x^{\mu_{3}}, \\[2mm]
%F_{(3)} &=& \frac{1}{6}F_{(3) \, \mu_{1} \mu_{2} \mu_{3}  \mu_{4}  } {\rmd}x^{\mu_{1}} \wedge {\rmd}x^{\mu_{2}} \wedge {\rmd}x^{\mu_{3}} \wedge {\rmd}x^{\mu_{4}} = \frac{1}{6}\partial_{[\mu_1}A_{(3) \mu_2 \mu_3 \mu_4]}  {\rmd}x^{\mu_{1}} \wedge {\rmd}x^{\mu_{2}} \wedge {\rmd}x^{\mu_{3}} \wedge {\rmd}x^{\mu_{4}}.
%\ea
Besides, we also have the Hodge duals, which, according with \cref{eq:HDuc}, are:
\begin{align}
\tilde{ F}_{ (1)  \mu_{1}  \mu_{2} } &= \frac{ \sqrt{-g}}{2!} \epsilon_{ \mu_{1} \mu_{2} \mu_{3} \mu_{4} }F_{(1)}{}^{\mu_{3} \mu_{4}},  \\[1mm]
\tilde{ F}_{ (2) \mu_{1}  } &= \frac{\sqrt{-g}}{3!} \epsilon_{ \mu_{1} \mu_{2} \mu_{3} \mu_{4} } F_{(2)}{}^{ \mu_{2} \mu_{3} \mu_{4} },\\[1mm]
\tilde{ F}_{ (3) } &= \frac{\sqrt{-g}}{4!} \epsilon_{ \mu_{1} \mu_{2} \mu_{3} \mu_{4} } F_{(3)}{}^{ \mu_{1} \mu_{2} \mu_{3} \mu_{4} }. 
\end{align}
Then, we will construct the Lagrangian with the field strengths $F_{(1)}$, $F_{(2)}$ and $F_{(3)}$, their duals, and with  $A_{(1)}$, $A_{(2)}$ and $A_{(3)}$ whenever gauge invariance is respected. This is, aside from the scalar field couplings, we have the following building blocks:
\begin{align}
A_{(1)}{}_{ \mu_1}, \quad A_{(2)}{}_{ \mu_1 \mu_2}, \quad   A_{(3)}{}_{ \mu_1 \mu_2 \mu_3}, \quad F_{(1)}{}_{ \mu_1 \mu_2},  \quad \tilde{F}_{(1)}{}_{ \mu_1 \mu_2}, \quad  F_{(2)}{}_{ \mu_1 \mu_2 \mu_3}, \quad  \tilde{F}_{(2)}{}_{ \mu_1}, \quad  F_{(3)}{}_{ \mu_1 \mu_2 \mu_3 \mu_4}, \quad \tilde{F}_{(3)},\quad
\end{align}
and we will construct our Lagrangian as the various contractions build up from these, having always in mind that the presence of $A_{(p)}$ terms is restricted by gauge invariance. First, let us write the Maxwell like terms $F_{(p)}\wedge \star F_{(p)}$, with the proper coefficients coming from the first term of \cref{eq:Lpint}. This is:
%\ba\label{eq:lphiAp} \nonumber
 %{\cal L}_{p}(\phi, A_p) & =&  f_1(\phi)  F_{(1)  \mu_{1} \mu_{2} }  F_{(1)}{}^{ \mu_{1} \mu_{2} } +  f_2(\phi)  F_{(2)  \mu_{1} \mu_{2}  \mu_{3}  }  F_{(2)}{}^{  \mu_{1} \mu_{2}  \mu_{3} }  +  f_3(\phi)  F_{(3)  \mu_{1} \mu_{2}  \mu_{3}  \mu_{4}  }  F_{(3)}{}^{  \mu_{1} \mu_{2}  \mu_{3}  \mu_{4} } \\[2mm]
%  +    f_4(\phi,K)  (F_{(2) \, \mu_{1} \mu_{2}  \mu_{3}  } F_{(1)}{}^{  \mu_{1} \mu_{2}   } \partial^{\mu_{3}} \phi +\mbox{perms.} ) +  f_5(\phi,K)  ( \partial^{\mu_{1}} \phi  F_{(1)  \mu_{1} \mu_{2}  } F_{(1)}{}^{ \mu_{2} \mu_{3} } \partial_{\mu_{3}} \phi +\mbox{perms.} ) \\[2mm]
 % &+&  f_4(\phi)  F_{(3)  \mu_{1} \mu_{2}  \mu_{3}  \mu_{4} } F_{(1)}{}^{  \mu_{1} \mu_{2}   } F_{(1)}{}^{  \mu_{3} \mu_{4}   }  %+  f_7(\phi,K)  (    F_{(3)  \mu_{1} \mu_{2}  \mu_{3} \mu_{4}  } F_{(2)}{}^{ \mu_{1} \mu_{2}  \mu_{3}} \partial_{\mu_{4}} \phi +\mbox{perms.} ),
%\ea
\begin{equation}\label{eq:lphiAp1} 
 {\cal L}^{\rm M}_{p}(\phi, A_p) = -\frac{f_1(\phi)}{4}  F_{(1) \mu_{1} \mu_{2} }  F_{(1)}{}^{ \mu_{1} \mu_{2}  } - \frac{f_2(\phi) }{12}  F_{(2)  \mu_{1} \mu_{2}  \mu_{3}}  F_{(2)}{}^{ \mu_{1} \mu_{2}  \mu_{3} } - \frac{f_3(\phi)}{48}  F_{(3) \, \mu_{1} \mu_{2}  \mu_{3}  \mu_{4}  }  F_{(3)}{}^{  \mu_{1} \mu_{2}  \mu_{3}  \mu_{4} }.
\end{equation}
Now, we formulate the mixing terms following the prescription in \cref{eq:mixing}. The only possible contractions that we can form according to \cref{eq:mixing} are the following:
\begin{align}\label{eq:lphiAp2} 
 {\cal L}^{\rm mixing}_{p}(\phi, A_p) = -\frac{g_1(\phi)}{4}  F_{(1) \mu_{1} \mu_{2} }  \tilde{F}_{(1)}{}^{ \mu_{1} \mu_{2}  } - \frac{g_2(\phi) }{2}A_{(2)}{}_{\mu_1 \nu_2}  \tilde{F}_{(1)}{}^{ \mu_{1} \mu_{2} }  - {g_3(\phi)}  \tilde{F}_{(3) }.
\end{align}
Other possible cubic terms, mixed contractions (in the abelian case that we are dealing here with), are trivially zero due to antisymmetrization. Alternative combinations can be seen to be equivalent to the combinations already present in \cref{eq:lphiAp1} and  \cref{eq:lphiAp2}, as example:
\begin{align}
&& F_{(3)}{}^{  \mu_{1} \mu_{2} \mu_{3} \mu_{4}} F_{(1)}{}_{\mu_{1}\mu_{2}}F_{(1)}{}_{\mu_{3}\mu_{4}} \propto F_{(1)}{}_{\mu_{1}\mu_{2}}\tilde{F}_{(1)}{}^{\mu_{1}\mu_{2}}, \\
&& F_{(3)}{}^{  \mu_{1} \mu_{2} \mu_{3} \mu_{4}} F_{(1)}{}_{\mu_{1}\mu_{2}}\tilde{F}_{(1)}{}_{\mu_{3}\mu_{4}} \propto F_{(1)}{}_{\mu_{1}\mu_{2}}{F}_{(1)}{}^{\mu_{1}\mu_{2}}.
\end{align}
Combinations such as $F_{(2)}{}^{  \mu_{1} \mu_{2} \mu_{3} }  \tilde{F}_{(2)}{}_{ \mu_{1}} \tilde{F}_{(1)}{}_{ \mu_{2} \mu_{3}  }$ and $F_{(2)}{}^{  \mu_{1} \mu_{2} \mu_{3} }  \tilde{F}_{(2)}{}_{ \mu_{1}} F_{(1)}{}_{\mu_{2} \mu_{3} }$  that are apparently nonvanishing, can be proved to be identically zero when we write them as  in \cref{eq:mixing}. This is:
\begin{align}
&& F_{(2)}{}^{  \mu_{1} \mu_{2} \mu_{3} }  \tilde{F}_{(2)}{}_{ \mu_{1}} \tilde{F}_{(1)}{}_{ \mu_{2} \mu_{3} } \propto \eta^{\sigma \mu_1 \mu_2 \mu_3} \tilde{F}_{(2)}{}_{\sigma} \tilde{F}_{(2)}{}_{\mu_1} \tilde{F}_{(1)}{}_{\mu_2 \mu_3} =0, \\
&& F_{(2)}{}^{  \mu_{1} \mu_{2} \mu_{3} }  \tilde{F}_{(2)}{}_{ \mu_{1}} F_{(1)}{}_{\mu_{2} \mu_{3} } \propto \eta^{\sigma \mu_1 \mu_2 \mu_3} \tilde{F}_{(2)}{}_{\sigma} \tilde{F}_{(2)}{}_{\mu_1} {F}_{(1)}{}_{\mu_2 \mu_3} =0,
\end{align}
where the last equality is a consequence of antisymmetrization over the indices $\sigma, \mu_1$. \\

The first term in \cref{eq:lphiAp2}, as discussed in subsection \ref{topological}, is topological and parity violating. The second one, as we will see in section \ref{sec:12form} can be absorbed in a parity conserving massive vector field model. Despite its appearance, the last term $\tilde{F}_{(3)}$ is not parity violating since it consists of two $\epsilon$ symbols contracted, one from the Hodge dual definition and the other one from the field strength $4-$form. This term is not topological either and it has a nontrivial contribution to the dynamics of the coupled system. To understand better the role of the $3-$form in the dynamics of the system, it is useful to isolate the part of the Lagrangian involving only the scalar and the $3-$form:
\begin{equation}\label{eq:lphiA3} 
 {\cal L}_{\phi A_{(3)}} = -\frac{1}{2}  \partial_{\mu}\phi \partial^{\mu}\phi - V(\phi)  - \frac{f_3(\phi)}{48} F_{(3)}^{2 } - {g_3(\phi)} \tilde{F}_{(3) } .
\end{equation}
The structure of the action of the $3-$form is very particular, and, as  pointed out in \cite{Brown:1987dd,Brown:1988kg,Duncan:1989ug,Duff:1989ah} (and more recently reviewed and generalized in \cite{Farakos:2017jme}), in order to have well defined variation for general field configurations, it is necessary to add the boundary term $ \partial_{\mu_{1}} \left[(g_3 - f_3 \tilde{F}_{(3)})  \epsilon^{ \mu_{1} \mu_{2} \mu_{3} \mu_{4} } A_{(3) \, \mu_{2} \mu_{3} \mu_{4}} /4!\sqrt{-g}   \right]$ to the action. So, the full action reads 
\begin{equation}\label{eq:lphiA3BT} 
 {\cal L}_{\phi A_{(3)}}= -\frac{1}{2}  \partial_{\mu}\phi \partial^{\mu}\phi - V(\phi)  - \frac{f_3(\phi)}{48} F_{(3)}^{2 } - {g_3(\phi)} \tilde{F}_{(3) } + \partial_{\mu_{1}} \left[(g_3 - f_3 \tilde{F}_{(3)}) \frac{ \epsilon^{ \mu_{1} \mu_{2} \mu_{3} \mu_{4} } }{4! \sqrt{-g} }A_{(3) \, \mu_{2} \mu_{3} \mu_{4}}  \right].
\end{equation}
The equations of motion derived from the previous action are:
\begin{align}
\Box \phi -V_{,\phi}-\frac{f_{3,\phi}}{48} F_{(3)}^2   -  \frac{g_{3,\phi}}{24}   \frac{\epsilon^{\mu_{1} \mu_{2} \mu_{3} \mu_{4}}}{\sqrt{-g}} {F}_{(3)\mu_{1} \mu_{2} \mu_{3} \mu_{4} } &= 0, \label{eq:EA3phi1}\\[1mm]
\nabla^{\mu}\left[f_3 F_{(3) \mu_{1} \mu_{2} \mu_{3} \mu_{4}}  + g_{3} \sqrt{-g} \epsilon_{\mu_{1} \mu_{2} \mu_{3} \mu_{4} } \right] & =0,  \label{eq:EA3phi2}
\end{align}
 where we use the notation $V_{,\phi} \equiv \rm{d} V / \rm{d}\phi$. The equation of the $3-$form can be integrated exactly by using the fact that $ {F}_{(3)\mu_{1} \mu_{2} \mu_{3} \mu_{4} } $ is proportional to the four dimensional volume element: 
\begin{equation}\label{F4X} 
 {F}_{(3)\mu_{1} \mu_{2} \mu_{3} \mu_{4} }  = X(x^{\mu}) \sqrt{-g}  \epsilon_{\mu_{1} \mu_{2} \mu_{3} \mu_{4} },
\end{equation} 
 where $X(x^{\mu}) $  is a scalar function. With this, we obtain the solution to \cref{eq:EA3phi2}:
\begin{equation}\label{F4Xsol} 
  X(x^{\mu})  = \frac{c - g_3(\phi)}{f_3(\phi)},
\end{equation} 
where $c$ is an integration constant. Using $F_{(3)}^2 = -4!X^2$ and $\tilde{F}_{(3)}= - X$, and substituting the solution \eqref{F4Xsol} into the action \eqref{eq:lphiA3BT} we obtain
\begin{equation}\label{eq:lphiA3BTint} 
 {\cal L}_{\phi A_{(3)}}= -\frac{1}{2}  \partial_{\mu}\phi \partial^{\mu}\phi - V(\phi)  - \frac{\left( c - g_3(\phi) \right)^2 }{2 f_3 (\phi)}.
\end{equation}
From the previous equation we see that the $3-$form was integrated out and it was absorbed as a potential term for the scalar field, generating an effective potential
\begin{equation}\label{eq:Veff} 
 V_{{\rm eff}} (\phi) = V (\phi)  + \frac{\left( c - g_3(\phi) \right)^2 }{2 f_3 (\phi)}.
\end{equation} 
It is straightforward to see that the equation of motion \eqref{eq:EA3phi1} coincides with the equation of motion derived from the effective Lagrangian \eqref{eq:lphiA3BTint}. 
Interesting applications of this mechanism to dark energy in the string theory landscape and to chaotic inflation were considered  in \cite{Kaloper:2008qs,Kaloper:2008fb}. Models involving the mixing of multiple scalars with $3-$forms were also studied in detail in \cite{Bielleman:2015ina, Ibanez:2015fcv, Valenzuela:2016yny,Farakos:2017jme}. \\
With the previous results we see that the role of the $3-$form in the dynamics of the system is to contribute with an induced potential, so, the complete system of interacting $p-$forms is reduced to a system composed by a scalar field, a $1-$form and a $2-$form. The $3-$form was integrated out and absorbed by the scalar field. Thus, the complete Lagrangian for the $p-$forms is reduced to:
%\begin{multline}\label{eq:lphiAp3}
% {\cal L}_{p}(\phi,A_p) = -\frac{f_1(\phi)}{4}  F_{(1) \mu_{1} \mu_{2} }  F_{(1)}{}^{ \mu_{1} \mu_{2}  } - \frac{f_2(\phi) }{12}  F_{(2)  \mu_{1} \mu_{2}  \mu_{3}}  F_{(2)}{}^{ \mu_{1} \mu_{2}  \mu_{3} } - \frac{f_3(\phi)}{48}  F_{(3) \, \mu_{1} \mu_{2}  \mu_{3}  \mu_{4}  }  F_{(3)}{}^{  \mu_{1} \mu_{2}  \mu_{3}  \mu_{4} }  \\
%-  \frac{f_4(\phi)}{24}  \tilde{F}_{(3) } 
%- \frac{g_1(\phi)}{4}  F_{(1) \mu_{1} \mu_{2} }  \tilde{F}_{(1)}{}^{ \mu_{1} \mu_{2}  } - \frac{g_2(\phi) }{2}A_{(2)}{}_{\mu_1 \nu_2}   \tilde{F}_{(1)}{}^{ \mu_{1} \mu_{2}  } ,
%\end{multline}
\begin{align}\label{eq:lphiAp4}
 {\cal L}_{p}(\phi,A_p) = -\frac{1}{2} \sum_{n=1}^{2} \frac{f_{n}(\phi)}{(n+1)!}F_{(n)}^2 %- \frac{f_4(\phi)}{24}  \tilde{F}_{(3) } 
 - \frac{g_1(\phi)}{4}  F_{(1) \mu_{1} \mu_{2} }  \tilde{F}_{(1)}{}^{ \mu_{1} \mu_{2}  } - \frac{g_2(\phi) }{2}A_{(2)}{}_{\mu_1 \nu_2}  \tilde{F}_{(1)}{}^{ \mu_{1} \mu_{2}  },
\end{align}
where we used the shorthand notation for the Maxwell like terms \cref{F2}. 

Remarkably, although at first sight we could expect more available non-trivial mixing combinations, the list reduces to the terms shown in \cref{eq:lphiAp4}. The only term which actually mixes $p-$forms of different rank is the $B\wedge F$ term, which, as we pointed out before, preserves gauge invariance only when the coupling function $g_2$ is a constant.  The Lagrangian consists only of two dynamical and two topological terms. As we said before, including gravity can be done in a minimal way by just adding the Einstein-Hilbert Lagrangian, so, the total Lagrangian for the coupled system, including the scalar field is 
\begin{equation} \label{eq:LT}
S_p = \int {\rmd}^4 x \sqrt{-g}{\cal L}_{T}, \quad {\cal L}_{T} = \left( \frac{M_{\rm{p}}^2}{2}R + {\cal L}_{\phi} + {\cal L}_{p}(\phi, A_p)\right ).
\end{equation}

Even if in this work we only deal with abelian gauge fields, all the arguments and procedures can be generalized straightforwardly to the case of non-Abelian gauge symmetry groups. Certainly, much more combinations will appear if we consider non-abelian groups or multiple $p-$forms of the same rank. As an example, in the presence of non-abelian group, cubic non-vanishing terms will appear, terms like
\begin{equation}\label{cubic}
f_{abc} {\bf F}^{(a)}{}_{\mu\alpha} {\bf F}^{(b)}{}^{\alpha}{}_{\beta} {\bf F}^{(c)}{}^{\beta \mu} \quad \mbox{and} \quad f_{abc} {\bf F}^{(a)}{}_{\mu\alpha} {\bf F}^{(b)}{}^{\alpha}{}_{\beta} {\bf \tilde{F}}^{(c)}{}^{\beta \mu},
\end{equation}
where $f_{abc}$ are the structure constants of the gauge group and ${\bf F}^{(a)}{}_{\mu\alpha} = \partial_{[\mu} {\bf A}^{(a)}_{\nu]} + f^{a}{}_{bc} \left[  {\bf A}^{(b)}_{\mu} , {\bf A}^{(c)}_{\nu}\right]$. 
Some four dimensional models have been considered in the recent literature in the context of inflation. For an extensive review of inflationary models with non-abelian symmetry groups, see \cite{Dimastrogiovanni:2010sm, Maleknejad:2012fw}. From those models, it is important to highlight the Gauge-flation model \cite{Maleknejad:2011jw, Maleknejad:2011sq} which involves a dimension eight operator $( {\rm Tr}[{\bf F}^{(a)}_{\mu\nu} \tilde{{\bf F}}^{(a)}{}^{\mu\nu}])^2 $ as the cause of the inflationary expansion and the closely related one, the Chromo-natural inflation model \cite{Adshead:2012kp}, which involves a coupling between a scalar, inflaton field and a Chern-Pontryagin term $\phi   {\rm Tr}[{\bf F}^{(a)}_{\mu\nu} \tilde{{\bf F}}^{(a)}{}^{\mu\nu}] $.  Those models, on their original version, have serious challenges when confronted with current CMB data \cite{Namba:2013kia}, and in an attempt to correct the observational discrepancies,  massive versions of them have been proposed, so, we can find the massive Gauge-flation \cite{Nieto:2016gnp} and the Higgsed chromo-natural model \cite{Adshead:2016omu}, and in a more theoretical context, a generalized Proca model with non-abelian gauge groups \cite{Allys:2016kbq}. Models including the presence of cubic terms like the ones in \cref{cubic} and its relevance for gravitational leptogenesis and the production of chiral gravitational waves were considered in \cite{Maleknejad:2014wsa, Maleknejad:2016dci}.  The models listed before and models related to them, have appealing phenomenological features and provide interesting connections with particle physics. \\ 

In the next subsection we will write the equations of motion and the energy-momentum tensor for the coupled system. 

%%%%%%%%%%%%%%%%%%%%
\subsection{Equations of motion and energy momentum tensor} \label{sec:EOM}
%%%%%%%%%%%%%%%%%%%%
For obtaining the energy-momentum tensor of the $p$-form Lagrangian \cref{eq:lphiAp4}, we take into account that the topological terms $F_{(1)}\wedge F_{(1)}$ and $A_{(2)}\wedge F_{(1)}$  do not contribute to the energy-momentum tensor as they are metric independent. The result from a direct calculation is
\begin{align} \label{eq:EMTp}
T_{\alpha\beta}^{(p)} &= - \frac{2}{\sqrt{-g}}\frac{\delta (\sqrt{-g} {\cal L}_{p})}{\delta g^{\alpha\beta}}, \nonumber \\  
&= \sum_{n=1}^{2} \frac{f_n(\phi)}{(n+1)!}\left[ (n+1) F_{(n)  \alpha \mu_2\cdots\mu_{n+1}}F_{(n)  \beta}{}^{ \mu_2 \cdots \mu_{n+1}} - \frac{1}{2} g_{\alpha\beta} F_{(n)}^2 \right], \nonumber \\  % -\frac{f_4(\phi)}{24}  \tilde{F}_{(3) } g_{\alpha\beta},
&= f_1(\phi) \left( F_{(1)}{}_{ \alpha \mu_2} F_{(1)  \beta}{}^{ \mu_2 } -\frac{1}{4} g_{\alpha\beta} F_{(1)}^2 \right) + f_2(\phi) \left( \frac{1}{2}F_{(2)}{}_{ \alpha \mu_2 \mu_3} F_{(2)  \beta}{}^{ \mu_2 \mu_3} -\frac{1}{12} g_{\alpha\beta} F_{(2)}^2 \right). %\nonumber \\
%&+ f_3(\phi) \left( \frac{1}{6}F_{(3)}{}_{ \alpha \mu_2 \mu_3 \mu_4} F_{(3)  \beta}{}^{ \mu_2 \mu_3 \mu_4} -\frac{1}{48} g_{\alpha\beta} F_{(3)}^2 \right) - \frac{f_4(\phi)}{24} \tilde{F}_{(3) } g_{\alpha\beta}.
\end{align}
Additionally, we have the energy momentum tensor for the scalar field
\begin{equation}\label{eq:Temph}
T_{\alpha\beta}^{(\phi)} = \partial_{\alpha}\phi \partial_{\beta}\phi - \frac{1}{2} g_{\alpha\beta} \partial_{\sigma}\phi \partial^{\sigma}\phi - g_{\alpha\beta}V_{{\rm eff}}(\phi),
\end{equation}
so, the Einstein equations are written as
\begin{equation}\label{eq:EinsM}
R_{\alpha\beta} - \frac{1}{2}R g_{\alpha\beta} = 8 \pi G \left(T_{\alpha\beta}^{(\phi)} +  T_{\alpha\beta}^{(p)} \right),
\end{equation} 
where we split the metric variation of the Lagrangian in two parts:  $T_{\alpha\beta}^{(\phi)}$ the energy-momentum tensor for the scalar field (\cref{eq:Temph}) and $T_{\alpha\beta}^{(p)}$ the associated with the $p$-forms (\cref{eq:EMTp}). We also set $M_{{\rm pl}}^{-1} =8 \pi G $. \\

The equations of motion of the scalar field and the $p$-form are obtained via variation of the Lagrangian with respect to $\phi, A_{p}$ 
\begin{equation}
\mathcal{E}_{\phi } = \frac{1}{\sqrt{-g}}\frac{\delta (\sqrt{-g}{\cal L}_T) }{\delta \phi} = 0, \quad \mathcal{E}_{(p)  \mu_1\cdots \mu_{p}} = \frac{1}{\sqrt{-g}}\frac{\delta (\sqrt{-g}{\cal L}_p) }{\delta A_{(p)}^{\; \mu_1\mu_2\cdots\mu_p}} = 0.
\end{equation}
Explicitly, we have, for the scalar field:
\begin{equation}
 \square \phi - V_{{\rm eff}}{}_{,\phi} +    \left(   \frac{f_{1,\phi}}{4}F_{(1)}^2 + \frac{f_{2,\phi}}{12}  F_{(2)}^2   + \frac{g_{1,\phi}}{4} F_{(1)}\wedge {F}_{(1)}  % +   \frac{f_{3,\phi}}{48} F_{(3)}^2  +  \frac{f_{4,\phi}}{24}  \tilde{F}_{(3)} + \frac{g_{2,\phi}}{2} A_{(2)}\wedge {F}_{(1)}
  \right) = 0, 
 \end{equation}
and for the $p-$forms
\begin{align}
 \nabla^{\mu}\left(f_1(\phi) F_{(1)}{}_{\mu\nu} + g_1(\phi) \tilde{F}_{(1)}{}_{\mu\nu} +  g_2  \tilde{A}_{(2)}{}_{\mu\nu}  \right) & =0, \label{eq:EA1}\\[1mm]
 \nabla^{\mu}\left(f_2(\phi) F_{(2)}{}_{\mu\nu \alpha}  \right)  + \frac{g_2}{2}  \tilde{F}_{(1)}{}_{\nu\alpha} & =0, \label{eq:EA2} %\\[1mm]
%\nabla^{\mu}\left(f_3(\phi) F_{(3)}{}_{\mu\nu \alpha \beta}   + f_4(\phi) \sqrt{-g} \epsilon_{\mu\nu \alpha \beta}  \right)  & =0. \label{eq:EA3}
\end{align}
These equations are supplemented with the Bianchi identities:
\begin{equation}
 \nabla^{\mu} \tilde{F}_{(1)}{}_{\mu \nu}= 0,\quad   \nabla^{\mu} \tilde{F}_{(2)}{}_{\mu }= 0.
\end{equation}
Next, we will deal with each field in  detail. For concrete reference, we will write the equations of motion for a Friedmann-Lema\^itre-Robertson-Walker (FLRW) solution in conformal time:
\begin{equation}\label{FLtau}
 {\rmd}s^2 = a(\tau)^2\left[ -{\rmd}\tau^2 + {\rmd}{\vc{x}}^2 \right].
\end{equation}

%%%%%%%%%%%%%
\subsubsection{Equations for the coupled $1-$form and $2-$form system}\label{sec:12form}
%%%%%%%%%%%%%
The previous system of equations describe an interesting interacting system when the coupling functions are different from zero, otherwise, all the $p-$forms evolve independently. For definiteness, we will introduce some simplifying assumptions. First of all, to untangle notation, in this and in the next subsection, we will use $A$ for the $1-$form, $F$ for its field strength, $B$ for the $2-$form and $H$ for its field strength, as it is commonly used in the literature (see for instance \cite{Ito:2015sxj}). In order to retain gauge invariance in the coupled system, we have to demand that $g_2$ is a constant coupling.  Moreover, we will assume that aside of $g_2$, all the coupling functions are only time dependent which is consistent with a homogeneous background. 
The e.o.m for the 1-form and for the $2-$form \cref{eq:EA1,eq:EA2}, can be written as
\begin{align}\label{A1eq}
\nabla^{\mu}\left(f_1(\phi) F_{\mu\nu} + g_1(\phi) \tilde{F}_{\mu\nu} +  g_2\tilde{B}_{\mu\nu}  \right) & =0,\\[1mm] \label{A2eq}
\nabla^{\mu}\left(f_2(\phi) H_{\mu\nu \alpha}  \right)  + \frac{g_2}{2}  \tilde{F}_{\nu\alpha} & =0, 
\end{align}
And in the following, we describe the system in terms of the  four dimensional components of the fields. To start with, we notice that the components $A_0, B_{0i}$ are non dynamical as its time derivative do not appear in the Lagrangian, so, we will use the gauge freedom and set them to zero. Moreover, we will also use the gauge freedom to set a null divergence of the fields, this is $\partial_i A_i = \partial_{i}B_{ij}= \partial_j B_{ij}=0.$ 
In components, an using $\nabla^{\mu}\tilde{F}_{\mu\nu}=0$, in the first equation when $\nu=0$ we have
\begin{equation}
f_1 \partial_{i}F_{i0} = \frac{1}{2} g_2 \epsilon_{0ijk}\partial_{i}B_{jk},
\end{equation}
which using the gauge choice $A_{0}=0=\partial_{i}A_{i}=0$ we get
\begin{equation}
\epsilon_{0ijk}\partial_{i}B_{jk} = \frac{1}{3}\epsilon_{0ijk} H_{ijk}=0. 
\end{equation}
For the spacial components $\nu=i$ we get
\begin{equation}
\nabla^{\mu}\left(f_1(\phi) F_{\mu i} + g_1(\phi) \tilde{F}_{\mu i } +  g_2  \tilde{B}_{\mu i }  \right)  =0,
\end{equation}
which, after using the gauge choice gives:
\begin{equation}
f_1 (\nabla^2  - \partial_{\tau}^2)A_i - f_1'  A'_i - g_1' \epsilon_{0ijk}\partial_{j}A_k - g_2 \nabla_{0} \tilde{B}_{0i} = 0, % + g_2 \nabla_{j}\tilde{B}_{ji}  -g_2' \tilde{B}_{0i} =0,
\end{equation}
where we used, according with the gauge choice, that $\tilde{B}_{ij}=0$, and primes $'$ denoting derivatives with respect to conformal time.  Using $\tilde{B}_{0i} =\frac{1}{2}\epsilon_{0ijk}B_{jk} $ we can write
\begin{equation}\label{eqAB}
-f_1\left[ A''_i  -\nabla^2 A_i  + \frac{ f_1'}{f_1}  A'_i  + \frac{ g_1'}{f_1} \epsilon_{0ijk}\partial_{j}A_k \right] =  \frac{g_2}{2} \left[  \partial_{\tau}  -2\frac{ a'}{a} \right] \epsilon_{0ijk}B_{jk}. 
\end{equation}
On the other hand, we can write the equations for the $2-$form as
\begin{align}
\nabla^{j}\left(f_2(\phi) H_{j 0 i}  \right)  + \frac{g_2 }{2}  \epsilon_{0ijk}\partial_{j}A_{k}&=0, \\
\nabla^{\mu}\left(f_2(\phi) H_{\mu ij }  \right)  - \frac{g_2 }{4}  \epsilon_{0ijk}\partial_{0}A_{k}&=0, 
\end{align}
which in the Friedmann metric becomes
\begin{align}
& \frac{f_2}{a^2} \partial_{j} H_{0ij} = - \frac{g_2}{2}  \epsilon_{0ijk}\partial_{j}A_{k}, \\
&  \frac{f_2}{a^2} \left( \partial_{0} H_{0 ij } -  \partial_{k} H_{k ij } + \left(\frac{f'_2}{f_2} - 2 \frac{a'}{a} \right) H_{0 ij }  \right)    = -\frac{g_2}{4}  \epsilon_{0ijk}\partial_{0}A_{k}.
\end{align}
Using
\begin{equation}
H_{0ij} = \partial_0 B_{ij} +  \partial_i B_{j0} +  \partial_j B_{0i}, \qquad
H_{ijk} = \partial_i B_{jk} +  \partial_j B_{ki} +  \partial_k B_{ij}, 
\end{equation}
and the gauge choice mentioned before $B_{0i}=\partial_i B_{ij}= \partial_{j}B_{ij}=0$, we obtain
\begin{align}\label{eqBA1}
& \frac{f_2}{a^2} \partial_{j} B'_{ij} = - \frac{g_2}{2}  \epsilon_{0ijk}\partial_{j}A_{k}=0, \\ \label{eqBA2}
& \frac{f_2}{a^2}\left( B''_{ij} - \nabla^2 B_{ij} + \left( \frac{f'_2}{f_2} - 2\frac{a'}{a} \right)B'_{ij} \right)  = - \frac{g_2}{4}  \epsilon_{0ijk}A'_{k}. 
\end{align}
%The equations \eqref{eqAB}, \eqref{eqBA1}, and \eqref{eqBA2} defines the dynamics of the coupled system in a Friedmann background. 
By doing $g_2=0$ we obtain the uncoupled version of this system
\begin{align}\label{eqg201}
&  A''_i  -\nabla^2 A_i  + \frac{ f_1'}{f_1}  A'_i  + \frac{ g_1'}{f_1} \epsilon_{0ijk}\partial_{j}A_k  =0, \\ \label{eqg202}
&  B''_{ij} - \nabla^2 B_{ij} + \left( \frac{f'_2}{f_2} - 2\frac{a'}{a} \right)B'_{ij}  = 0, 
\end{align}
which (except by the parity violating term due to the coupling $g_1$) has been studied in extent in \cite{Ohashi:2013qba,Ito:2015sxj}. \\
 The \cref{eqAB,eqBA1,eqBA2},  define the dynamics of the coupled system in a Friedmann background. This is a system difficult to solve analytically and numerically, and depends on the particular details of the kinetic coupling functions. It is beyond the scope of this paper to enter into the concrete details of the solution of the coupled system as our main goal here is to describe the general features and possibilities that arise when a general mixing between arbitrary $p-$forms is allowed. Nevertheless, the coupled system involving the $B\wedge F$, has a great potential interest for cosmological applications due to the fact that it constitutes, as we will see next, an interesting mechanism 
%to source parity violating features in the observables, which, is well known to be related to the enhancement of the mechanism for sourcing chiral gravitational waves. This constitutes  a different mechanism  from the most studied case involving a term $F\wedge F$. Aside of this, it is also very interesting to study in more detail the mechanism 
for generation of mass due to the topological coupling in the context of inflationary physics, a mechanism which is different from the Higgs mechanism. We hope to come back to the study of the specific details of this system elsewhere. \\

To conclude this section, we  comment on the mass generation mechanism that we mentioned before which is an interesting possibility offered by the coupled system  \cref{A1eq,A2eq}. First, we notice that the $2-$form  \cref{A2eq} can be solved for $F^{\mu \nu}$  as follows:
\begin{equation}\label{FH}
F^{\mu\nu} = -\frac{1}{3g_2}\nabla^{[\mu} f_2(\phi) \tilde{H}^{\nu ]}. 
\end{equation}
Defining the vector field:
\begin{equation}\label{defvh}
V^{\mu} \equiv \frac{f_2(\phi)}{3g_2} \tilde{H}^{\mu},
\end{equation}
and replacing in \cref{A1eq} we get
\begin{equation}\label{massv}
\nabla^{\mu}\left(f_1(\phi) \nabla_{[\mu} V_{\nu] }+\frac{g_1(\phi)}{2}  \sqrt{-g} \epsilon_{\mu \nu \alpha \beta} \nabla^{[\alpha} V^{\beta] }  \right)  = m^2(\phi) V_{\nu}, 
\end{equation}
where $m^2(\phi) \equiv \frac{3 g_2^2}{2f_2} $ and where we used $\nabla^{\mu}\tilde{B}_{\mu \nu} = \tilde{H}_{\nu}/2$. Equation \eqref{massv} is the equation corresponding to a massive vector field derived from an action of the form
\begin{equation}\label{Smass}
S_{V} = - \frac{1}{4}\int {\rmd}^4x \sqrt{-g} \left[ f_1(\phi) W_{\mu\nu}  W^{\mu\nu}  + g_1(\phi) W_{\mu\nu} \tilde{W}^{\mu\nu} + 2 m^2({\phi}) V_{\mu}V^{\mu} \right],
\end{equation}

with $W_{\mu\nu}  = \nabla_{[\mu} V_{\nu] } $. This action  is not invariant under gauge transformations $V_{\mu}\rightarrow V_{\mu} + \partial_{\mu}\xi$ as the gauge symmetry is broken by the  coupling $f_2(\phi)$. Nevertheless, the theory is indeed invariant under the transformation $\tilde{H}_{\mu}\rightarrow \tilde{H}_{\mu} + \partial_{\mu}\xi$. Actually, the action and the equations of motion can be formulated in terms of the $1-$form $A_{\mu}$ by solving  \cref{FH}. By using \cref{defvh}, we see that \cref{FH} can be solved as $V_{\mu} = A_{\mu} + \partial_{\mu}v$ for a convenient $v$ function. Using the gauge $\nabla_{\nu}\nabla_{\mu}A^{\mu} = -\frac{m^2}{f_1}\partial_{\nu}v$ we obtain
\begin{equation}
\Box A_{\nu} - R_{\nu \mu}A^{\mu} + \frac{\partial^{\mu}f_1}{f_1}(\nabla_{\mu} A_{\nu}- \nabla_{\nu} A_{\mu} ) + \frac{\partial^{\mu}g_1}{f_1}\eta_{\mu\nu\alpha\beta} \nabla^{\alpha} A^{\beta} - \frac{m^2}{f_1} A_{\nu} = 0,
\end{equation}
which is the equation of motion of a Proca vector field with kinetic couplings in a curved background. 
 This peculiarity of obtaining a massive theory from a massless gauge invariant theory through the introduction of a topological coupling term such as $B \wedge F$, is an example of the mechanism %for generating massive guage invariant fields usually dubbed  as 
of  ``topological  mass generation'' described in \cite{Allen:1990gb,Dvali:2005ws}. Generalizations of the  Proca model has been studied with interest in recent literature \cite{Tasinato:2014eka, Heisenberg:2014rta,Allys:2015sht,Jimenez:2016isa,Allys:2016jaq} and, although the mechanism described here differs  from the construction developed in those studies  by the fact that we keep gauge invariance of the fields, it would be interesting to study if some connection can be made with those models at certain limits.
An appealing and interesting relation of these terms with the Julia-Toulouse mechanism in solid state physics \cite{Julia:1979ur}  and the  condensation of topological defects was also studied in \cite{Quevedo:1996uu}.  We refer the interested reader to go through this reference for further details. It is worth mentioning that we can also extrapolate the considerations followed here to the non-abelian case that we discussed before and obtain a non-abelian version of  \cref{Smass}. Notice however,  that here we have a mechanism different from the Higgs-like mechanism for the generation of a mass that was considered in the massive Gauge-flation model \cite{Nieto:2016gnp} and the Higgsed chromo-natural inflation model \cite{Adshead:2016omu} mentioned before.\\

We should also comment that the theory described by \cref{Smass} is consistent with the gauge symmetries and the number of degrees of freedom in the theory described by  \cref{A1eq,A2eq}. The theory described by a $1-$form coupled to a $2-$form consistent with gauge symmetry contains a total of 3 propagating degrees of freedom, two transverse polarizations for the $1-$form and one degree of freedom for the $2-$form. This is precisely the same number of degrees of freedom present in the model \cref{Smass} which contains two transverse and one longitudinal propagating polarization. Equivalently, we could think this theory in terms of a massive $2-$form field with three propagating degrees of freedom.

%%%%%%%%%%%%%
\subsubsection{3-form}\label{sec:3formsec}
%%%%%%%%%%%%%
We saw before that the $3-$form can be integrated out and absorbed as a potential term of the scalar field. Nevertheless, it is instructive to analyze the time evolution of the $3-$form potential and discuss some of its particular features. To this end, let us recall \cref{eq:EA3phi2} and the solution \cref{F4Xsol}. From \cref{F4X}  
we evaluate the component $(0ijk)$ 
\begin{equation} 
F_{(3)}{}_{0ijk} = \sqrt{-g} X(x^{\alpha})\epsilon_{0ijk}= \partial_{0}A_{(3)}{}_{ijk}, \quad \mbox{or} \quad \sqrt{-g} X(x^{\alpha}) = \partial_{0}A_{(3)}, %-3\frac{\partial_{\tau}a}{a} A_{3}{}_{ijk}, 
\end{equation}
 where $A_{(3)} \equiv  A_{(3)}{}_{ijk} $. It is interesting to notice that the solution of this equation is invariant under the transformation $A_{(3)}\rightarrow A_{(3)} + b(x^i)$ where $b(x^i)$ is an arbitrary time-independent function. This invariance reflects the gauge symmetry of the $3-$form which manifests as a shift (space dependent) symmetry.  For an arbitrary background, the solution of this equation is  
\begin{equation}
A_{(3)}(\tau) = \int {\rmd}\tau'  \sqrt{-g } \left( \frac{{c} - g_3(\phi ) }{f_3(\phi )} \right).
\end{equation}
For a Friedmann background, and assuming that the coupling functions depend only on the conformal time, the previous solution reads 
\begin{equation}
A_{(3)}(\tau)  = \int {\rmd}\tau'  a^4(\tau') \left( \frac{{c} - g_3(\tau') }{f_3(\tau')}\right) .
\end{equation}
This solution is valid for any Friedmann cosmology and for any time dependence of the couplings $f_3, g_3$. Given this  result, we realize that, when coupled to a scalar field, the $3-$form potential has a non-trivial time evolution depending on the particular form of the couplings $f_3, g_3$. As mentioned before, the $3-$form is invariant under constant space dependent shifts of the form $A_{(3)}\rightarrow A_{(3)} + b(x^i)$ which also reflects the fact that this field does not propagate in space, and it is not a true physical propagating degree of freedom.  Furthermore, given that  it only depends on time, the $3-$form evolves in a homogeneous way and no statistical anisotropies are seeded by this field. \\

To conclude, we comment about the energy-momentum tensor of the $3-$form.  The $3-$form only enters in the energy-momentum tensor through the effective potential \cref{eq:Veff} in \cref{eq:Temph}: % replacing \cref{F3vol} in the $3-$form part of \cref{eq:EMTp}, the total energy   momentum tensor of the $3-$form is
%we can write the contributions to the energy momentum tensor of the $3-$form as 
%\be
%\frac{f_3(\phi)}{3!} F_{(3)}{}_{ \alpha \mu_2 \mu_3 \mu_{4}}F_{(3)}{}_{\beta}{}^{ \mu_2  \mu_3  \mu_{4}} = -  g_{\alpha \beta} (4!)^2 f_3(\phi)  X^2,\quad  - \frac{f_3(\phi)}{2\times 4!} g_{\alpha\beta} F_{(3)}^2= \frac{(4!)^2}{2}g_{\alpha \beta} f_3(\phi)X^2,
%\ee
%\begin{equation}\label{eq:Emt3}
%T_{\alpha\beta}^{(3)}  = - \left(\frac{(4!)^2}{2}  f_3(\phi)  X^2  -  {f_4(\phi)}  X \right) g_{\alpha\beta},
%\end{equation}
\begin{equation}\label{eq:Emt3}
T_{\alpha\beta}^{(3)}  = - \left(V_{{\rm eff}}(\phi) - V(\phi) \right) g_{\alpha\beta} = - \frac{\left( c - g_3(\phi) \right)^2 }{2 f_3 (\phi)}g_{\alpha\beta},
\end{equation}
which is a cosmological constant term. In the decoupled case, the $3-$form acts as a true constant term, but in the coupled case, the cosmological constant term acquires an evolving behaviour due to the coupling functions $f_3, g_3$ and mimics the effect of a cosmological constant with an exact equation of state (hereafter e.o.s.) parameter $w_{(3)} = -1$. The evolution of this cosmological constant term depends on the specific coupling functions  $f_3, g_3$.  This feature makes the $3-$form useful for applications related to the inflationary period and the late time accelerated inflation of the universe. We discuss some application of this behaviour in \cref{sec:cosmology}.  

%%%%%%%%%%%%%%%%%%%%%%%%%%%%%%%%%%%%%%%%%%%%%%%%%%%%%%%%%%%
\section{Some simple applications to cosmological backgrounds}\label{sec:cosmology}
%%%%%%%%%%%%%%%%%%%%%%%%%%%%%%%%%%%%%%%%%%%%%%%%%%%%%%%%%%%
In this section we  discuss some applications of cosmological interest (at background level) of the model presented here.   
We consider reduced versions of the model for simple choices of the couplings of the $p-$forms and the scalar field. Previous studies of the background evolution of a system involving a $1-$form and a $2-$form fields in the context of inflation, assuming an exponential form for the potential $V(\phi)$ and the kinetic coupling functions $f_{i}(\phi)$, can be found in  \cite{Thorsrud:2012mu,Ito:2015sxj,Obata:2018ilf, Almeida:2019xzt,Ohashi:2013qba, Ohashi:2013mka}. Aside of these, others interesting applications of $p-$forms in the context of inflation and dark energy scenarios have been widely discussed in the recent literature (see {\it e.g.} \cite{Koivisto:2009sd,Koivisto:2009fb,Koivisto:2012xm,Mulryne:2012ax,Kumar:2016tdn,Farakos:2017jme,Almeida:2019xzt,Almeida:2019iqp,Guarnizo:2019mwf}). In the following two subsections we employ dynamical system techniques in order to give statements about the evolution of the degrees of freedom and the physical parameters of the model \cite{Bahamonde:2017ize}.

\subsection{Single scalar field with effective potential induced by a $3-$form}\label{sec:4.1}
Here we concentrate in the effect of the $3-$form field, more precisely, in the induced effective potential obtained in \cref{eq:Veff}. As we discussed before, the presence of $3-$form is consistent with an homogeneous and isotropic evolution, so, the background metric that we use here is the FLRW metric

\begin{equation}
 {\rmd}s^2 = -{\rmd}t^2 + a^{2}(t){\rmd}{\vc{x}}^2. 
 %\qquad a(t) = e^{H t}.
\end{equation}
With this background metric, the Lagrangian \cref{eq:LT} reduces to
\begin{equation}
\mathcal{L}_{T} =  \left[ 3M_{{\rm Pl}}^2\left( \frac{\dot{a}^2}{a^2} + \frac{\ddot{a}}{a} \right)  +  \frac{1}{2} \dot{\phi}^2  - V_{{\rm eff}}(\phi)  %+ \gamma_3 f_1^2 \frac{\dot{A}_{3}^2}{2a^6} - 2\gamma_4 f_1^2 \frac{\dot{A}_3}{a^6} 
\right ],
\end{equation}
and the e.o.m for the scalar field is
\begin{equation}\label{eq:Eomphi}
\ddot{\phi} + V_{{\rm eff}}{}_{,\phi} + 3H \dot{\phi}  %-   \gamma_{3}f_1  f_{1 , \phi} \left(\frac{\dot{A}_{3}^2}{a^6}  - \theta  \frac{\dot{A}_3}{a^6}   \right) 
=0,
\end{equation}
where and overdot means derivative w.r.t cosmic time. Additionally, we will add an energy-momentum for matter as a perfect fluid in the form
\begin{equation}\label{eq:EMF}
T_{\mu\nu}^{x} = (\rho_x  + p_x)u_{\mu} u_{\nu} + p_x g_{\mu\nu}, 
\end{equation}
being $u^{\mu}$ the 4-velocity of the fluid with $u_{\mu}u^{\mu}= -1$, $\rho_x$ the energy density and $p_x$ the pressure. In the following we consider two matter sources, pressureless cold dark matter and radiation with e.o.s. $p_m=0$ and $p_r = \rho_r /3$ respectively. With these definitions, the Friedmann equations coming from \cref{eq:EinsM} reads
\begin{align}
3M_{{\rm Pl}}^2 H^2 & = \frac{1}{2} \dot{\phi}^2 + V_{{\rm eff}}(\phi)  + \rho_m  +  \rho_r , \label{eq:Fri1} \\
 2 M_{{\rm Pl}}^2  \dot{H} & = - \left[\dot{\phi}^2   + \rho_m + \frac{4}{3}\rho_r \right]. \label{eq:Fri2}
\end{align}
It is useful to define the energy density and pressure for each component, namely the scalar field and the $3-$form. From the energy-momentum tensor of the scalar field \cref{eq:Temph}, we can define as usual
\begin{equation}
\rho_{\phi} = \frac{1}{2} \dot{\phi}^2 + V(\phi), \qquad p_{\phi} = \frac{1}{2} \dot{\phi}^2 - V(\phi), \qquad
w_{\phi} = \frac{p_{\phi}}{\rho_{\phi}} = \frac{ \frac{1}{2} \dot{\phi}^2 - V(\phi)}{\frac{1}{2} \dot{\phi}^2 + V(\phi)}.
\end{equation}
While, from \cref{eq:Emt3},  we have 
\begin{equation}\label{eq:eosP}
\rho_{(3)} = - p_{(3)} \equiv \frac{\left( c - g_3(\phi) \right)^2 }{2 f_3 (\phi)}
% \gamma_3 f^2\left(\frac{\dot{A}_{3}^2}{2 a^6}   \right), \quad p_{p}\equiv - \gamma_3f^2\left(\frac{\dot{A}_{3}^2}{2 a^6}  \right), 
\quad \mbox{ so} \quad \quad w_{(3)} \equiv \frac{p_{(3)}}{\rho_{(3)}}=-1,
\end{equation}
for the $3-$form. Furthermore, we will assume $f_3 > 0$ to achieve a strictly positive energy density for the $3-$form. 
Using the definition of \cref{eq:eosP} and  \cref{eq:Veff}, we split the contributions to the effective potential  as
\begin{equation}
V_{\rm{eff}}(\phi) = V(\phi) + \rho_{(3)},
\end{equation}
and we define the dark energy density,  the pressure and its equation of state as
\begin{equation}
\rho_{\rm{DE}} = \frac{\dot{\phi}^2}{2} + V(\phi) + \rho_{(3)}, \quad p_{\rm{DE}} = \frac{\dot{\phi}^2}{2} - V(\phi) - \rho_{(3)}, \quad w_{\rm{DE}} \equiv \frac{p_{\rm{DE}}}{\rho_{\rm{DE}}}. %= \frac{X^2-Y^2 - \Omega_{(3)}}{X^2 + Y^2 + \Omega_{(3)}}.
\end{equation}
Now, to study the dynamics of this system, we define the variables
\begin{equation}\label{eq:params}
X = \frac{1}{\sqrt{6}M_{{\rm pl}}}\frac{\dot{\phi}}{H}, \qquad 
Y = \frac{1}{\sqrt{3}M_{{\rm pl}}}\frac{\sqrt{V}}{H},  \qquad
\Omega_{(3)} = \frac{\rho_{(3)}}{3 M_{{\rm pl}}^2 H^2 }, \qquad  \Omega_r = \frac{\rho_r}{3 M_{{\rm pl}}^2 H^2 }. 
\end{equation}
The first Friedmann equation, \cref{eq:Fri1}, can be written as a constraint in the density parameter $\Omega_m \equiv \frac{\rho_m}{3 M_{{\rm pl}}^2 H^2 }$ as
\begin{equation}\label{eq:DensParam}
\Omega_m = 1 - \Omega_{\rm{DE}} - \Omega_{r} , \qquad 
\Omega_{\rm{DE}} \equiv X^2+ Y^2 + \Omega_{(3)} .
\end{equation}
To gain some insight in the physics, we also define the effective e.o.s. for the system as %and  the equation of state of the as 
\begin{equation}\label{eq:weff}
w_{\rm{eff}}\equiv\frac{p_{m} + p_{r} + p_{\phi} + p_{(3)}}{\rho_{m} + \rho_{r} + \rho_{\phi}+\rho_{(3)}} = w_{m}\Omega_{m} + w_{r}\Omega_{r}+  w_{\phi}\Omega_{\phi} + w_{p}\Omega_{(3)}
= - 1 - \frac{2}{3} \frac{\dot{H}}{H^2},
\end{equation} 
where the ratio $\dot{H}/H^2$ can be computed from \cref{eq:Fri1,eq:Fri2}

\begin{equation}\label{eq:HubRat}
\frac{\dot{H}}{H^2} = - \frac{1}{2}\left(3+3X^2- 3Y^2 -3\Omega_{(3)}+ \Omega_r\right ).
\end{equation}
Now, we notice from \cref{eq:Eomphi} that the evolution of the scalar field depends on the derivative of the effective potential $V_{\rm{eff}}$. Then, in order to solve the system in a closed way, we need to assume a particular form for the potential  $V(\phi)$ and the coupling functions  $f_3(\phi) $ and $g_3(\phi)$. 
Another possibility remains to leave the potential as an additional dynamical variable, but, we prefer here to choose a particular form for the potential and the couplings.  For the sake of simplicity we use exponential functions for them 
\begin{equation}\label{eq:Potentials}
V(\phi) \propto e^{- \lambda\frac{\phi}{M_{{\rm pl}}}}, \qquad f_3(\phi) \propto e^{- \beta\frac{\phi}{M_{{\rm pl}}}}, \qquad g_3(\phi) \propto e^{-\gamma\frac{\phi}{M_{{\rm pl}}}},
\end{equation}
where $\lambda, \beta$ and $\gamma$ are constants. Further, we set the integration constant $c$ in \cref{eq:eosP} to zero. With this setup, the energy density of the $3-$form has a simple form:

\begin{equation}
\rho_{(3)} = \frac{ e^{(\beta-2\gamma) \frac{\phi}{M_{{\rm pl}}}}}{2 } =\frac{ e^{\delta \frac{\phi}{M_{{\rm pl}}}}}{2 } , \quad \delta \equiv \beta - 2\gamma, 
\end{equation}
where we also drop the proportionality constants in \cref{eq:Potentials}; the constant $\delta$ is introduced to simplify further calculatations. We emphasize in the fact that $\delta$ comes directly from the choice of the coupling functions, and its behaviour depends entirely on them. Differentiating w.r.t to $\cal{N}$ each of these parameters, with  ${\rm d}{\cal N} = H{\rm d} t $, and using \cref{eq:Eomphi,eq:HubRat}, we find

\begin{align}
X' &=  \frac{3}{2} X \left(X^2 -Y^2-1 - \Omega_{(3)} + \frac{\Omega_r}{3} \right)+ \sqrt{\frac{3}{2}}\left( \lambda Y^2 - \delta \Omega_{(3)} \right), \label{eq:Dynsi1}\\[1mm]
Y' &= \frac{1}{2}Y \left(3X^2-3 Y^2+3- 3\Omega_{(3)}+ \Omega_r - \sqrt{6}\lambda X \right), \label{eq:Dynsi2}\\[1mm]
\Omega_{(3)}' &=\Omega_{(3)} \left(3X^2-3 Y^2+3 - 3\Omega_{(3)}+ \Omega_r - \sqrt{6} \delta X  \right), \label{eq:Dynsi3} \\[1mm]
\Omega_r ' &= \Omega_r\left( 3X^2-3Y^2 -1 - 3\Omega_{(3)} + \Omega_r\right). \label{eq:Dynsi4}
\end{align}

\begin{center}
\begin{table}[tb]
\small
\centering\begin{tabularx}{\textwidth}{| >{\centering}X | >{\centering}X  | >{\centering}X |  >{\centering}X  | >{\centering}X |  >{\centering}X  | >{\centering}X  | >{\centering}X |}
\hline 
Point & $X$  & $Y$  & $\Omega_{(3)}$ & $\Omega_r$  &  $\Omega_m$ & $\Omega_{\rm{DE}}$ & $w_{\rm{ef}f}$      \tabularnewline
\hline 
\hline 
$\cal{O}$ & 0 & 0  & 0  & 0 & 1 & 0  & 0   \tabularnewline
$\cal{A}_{\pm}$ & $\pm$ 1 & 0  & 0  & 0 & 0  & 1  & 1      \tabularnewline
$\cal{B}$ & 0 & 0 & 0 & 1 & 0 & 0 & $\frac{1}{3}$   \tabularnewline
$\cal{C}$ & $\frac{2 \sqrt{6}}{3\lambda }$ & $ \frac{2 \sqrt{3}}{3 \lambda}$ &  0 & $1-\frac{4}{\lambda ^2}$  & 0  & $\frac{4}{\lambda^2}$ & $\frac{1}{3}$    \tabularnewline
$\cal{D}$ &  0 & $\frac{\sqrt{\delta}}{\sqrt{\delta +\lambda }}$ & $\frac{\lambda }{\delta +\lambda
   }$ & 0 & 0 & 1 & $-1$  \tabularnewline
$\cal{E}$ & $\frac{\sqrt{6}}{2\lambda }$ & $\frac{\sqrt{6}}{2\lambda }$ &  0 & 0  & $1-\frac{3}{\lambda^2}$ &  $\frac{3}{\lambda^2}$ & 0  \tabularnewline
$\cal{F}$ & $\frac{\lambda }{\sqrt{6}}$ & $\sqrt{1-\frac{\lambda ^2}{6}}$ & 0 & 0  & 0 & 1 & $\frac{\lambda ^2-3}{3} $     \tabularnewline
$\cal{G}$ & $\frac{\sqrt{6} \delta}{2(\delta^2-3)}$ & 0 & $-\frac{3 \left(\delta^2-6\right)}{2 \left(\delta^2-3\right)^2}$ & 0  & $1 -\frac{9}{\left(\delta^2-3\right)^2}$ &$\frac{9}{\left(\delta^2-3\right)^2}$ & $\frac{3}{\delta^2-3}$  \tabularnewline
$\cal{H}$ & $\frac{2 \sqrt{6}}{3\delta }$ & 0 & $-\frac{4}{3 \delta^2}$ & $1-\frac{12}{\delta^2}$  & $\frac{32}{3 \delta^2}$ & $\frac{4}{3 \delta^2}$ & $\frac{1}{3}$  \tabularnewline
\hline 
\end{tabularx}\caption{\label{tab:DymSys}Critical points for the dynamical system defined in \cref{eq:Dynsi1,eq:Dynsi2,eq:Dynsi3,eq:Dynsi4}. We also show the expressions for the effective EoS $w_{\rm{eff}}$ and the total dark energy density parameter $\Omega_{\rm{DE}}$.}
\end{table}
\end{center}
The critical points for the system, being the points in which $\frac{{\rmd} X}{{\rmd} \cal{N}}=\frac{{\rmd} Y}{{\rmd} \cal{N}}=\frac{{\rmd} \Omega_{(3)}}{{\rmd} \cal{N}}=\frac{{\rmd} \Omega_{r}}{{\rmd} \cal{N}}=0$, are presented in \cref{tab:DymSys}.  The dynamical system, \cref{eq:Dynsi1,eq:Dynsi2,eq:Dynsi3,eq:Dynsi4}, reveals the invariance under the transformation

\begin{equation}
Y \mapsto - Y, \qquad \text{and} \qquad \Omega_{(3)} \mapsto - \Omega_{(3)},
\end{equation}

this was used in the construction of \cref{tab:DymSys}, since also fixed points with negatives $Y$ and $\Omega_{(3)}$ appeared, but discarded by the previous invariance; in other words, it is enough to consider the dynamics for positive values of $Y$ and $\Omega_{(3)}$. From the exponential form of the potential and the coupling function, another simultaneous invariance is present
\begin{equation}
\lambda \mapsto - \lambda, \qquad \text{\textrm{and}} \qquad \delta \mapsto -\delta,   \qquad \text{\textrm{and}} \qquad X \mapsto -X,
\end{equation}
meaning that the system could be fully described assuming only $\lambda > 0$,  and $\delta > 0$.

\subsubsection{Stability of the fixed points}

The conditions of stability are related with the behavior of the system under small perturbations close to the fixed points \cite{Bahamonde:2017ize,wainwright:1997}. If we denote our set of critical points presented in \cref{tab:DymSys} as $\boldsymbol{x}_0=(X_0, Y_0, \Omega_{(3) \, 0},\Omega_{r \, 0} )$ and consider small perturbations around them, denoted as $\boldsymbol{X} = \{\delta X$, $\delta Y,\delta \Omega_{(3)}, \delta \Omega_r\}$, the linearized system (after perturbing \cref{eq:Dynsi1,eq:Dynsi2,eq:Dynsi3} ), could be written as
\begin{equation}\label{eq:StaCond}
\boldsymbol{X}'  = \cal{M} \boldsymbol{X}, 
\end{equation}

being $ \cal{M}$ a  Jacobian matrix. The stability criteria refers to the relative signs of the eigenvalues for the Jacobian matrix $\cal{M}$, evaluated at the fixed points. Denoting the eigenvalues by $\lambda_i$,  and assuming them in general to be complex, the stability criteria can be summarized as

\begin{enumerate}[\itshape(i)]
\item \textit{Stable point}: The critical point $\boldsymbol{x_0}$ is a stable point if all the eigenvalues $\lambda_i$ are real and negative.
\item \textit{Unstable point}: If the eigenvalues $\lambda_i$ are real and positive, $\boldsymbol{x}_0$ is a unstable point. 
\item \textit{Saddle point}: Again, if the eigenvaues $\lambda_i$ are real, and \textit{any} of these $\lambda_i$ is negative, $\boldsymbol{x}_0$ is a saddle point.
\item \textit{Stable spiral}: if the eigenvalues $\lambda_i$ are complex with negative real parts, the point $\boldsymbol{x}_0$ is called a stable spiral.
\end{enumerate}

Applying these criteria we found  general statements for the behaviour of the critcal points of \cref{tab:DymSys}.

\begin{itemize}
\item \textit{Point $\cal{O}$}: This is the origin of the phase space. Since $\Omega_{\rm{DE}}=0$, this point corresponds to a matter dominated universe with $\Omega_m=1$. The eigenvalues for this point are $\left\{3,-\frac{3}{2},\frac{3}{2},-1\right\}$, corresponding thus to a saddle
point. Since the kinetic and potential energy of the scalar field and the coupled $3-$form are zero, this point has no physical importance. 

\item \textit{Points $\cal{A}_{\pm}$}:   In these two points the universe is dominated by the scalar field kinetic energy in which
$w_{\rm{eff}} = 1$ corresponding to stiff matter, with no acceleration. The eigenvalues are
\begin{equation}
\left\{3,2,6 \mp \sqrt{6} \delta ,3 \mp \frac{\sqrt{6}}{2} \lambda \right\}.
\end{equation}
Thus, they never are stable points. They are unstable if   $\delta < \sqrt{6}$ $(\beta < 2\gamma + \sqrt{6})$, $\lambda < \sqrt{6}$, and  $\delta > - \sqrt{6}$ $(\beta > 2\gamma - \sqrt{6})$, $\lambda > - \sqrt{6}$ for $\cal{A}_{+}$ and $\cal{A}_{-}$, respectively. Similarly, $\cal{A}_{+}$ is a saddle point if $\delta > \sqrt{6}$ $(\beta > 2\gamma + \sqrt{6})$ or $\lambda > \sqrt{6}$, while $\cal{A}_{-}$ becomes saddle for $\delta < - \sqrt{6}$ $(\beta < 2\gamma - \sqrt{6})$, or $\lambda < - \sqrt{6}$. For current observations of accelerated expansion,
a stiff matter fluid is not relevant, and only have importance at early times.
\item \textit{Point $\cal{B}$}: This point matches a pure raditation dominance. The eigenvalues are $\{4,2,-1,1\}$, corresponding to a saddle point.

\item \textit{Point $\cal{C}$}: This point corresponds also a radiation dominance with $\Omega_{\rm{DE}} = \frac{4}{\lambda^2}$. From Big Bang Nucleosynthesis (BBN) bounds \cite{Bean:2001wt,Ohashi:2009xw}, we have $\Omega_{\rm{DE}} < 0.045$ translating into $\lambda > 9.2$.  The eigenvalues read
\begin{equation}
\left\{1,4 - \frac{4\delta}{\lambda },-\frac{1}{2} \pm \frac{\sqrt{64-15 \lambda ^2}}{2 \lambda } \right\},
\end{equation}
which under the previous bound give rise to a saddle point. 
\item \textit{Point $\cal{D}$}: Dark energy dominated point. This point provides accelerated expansion ($w_{\rm{eff}}=-1$) with $\Omega_{\rm{DE}}=1$. For the existence of this point we should set $\lambda>0$ and $\delta > 0$ $(\beta > 2\gamma)$. The eigenvalues are 
\begin{equation}
\left\{ -4,  -3, -\frac{3}{2} \pm \frac{\sqrt{3} \sqrt{(\delta +\lambda ) \left(3\lambda + \delta(3+4\lambda(\delta-\lambda))\right)}}{2 (\delta +\lambda )} \right\}.
\end{equation}
Under the bound $\lambda > 9.2$, three of the eigenvalues are negative and one is positive, becoming a saddle point.  
\item \textit{Point $\cal{E}$}:  This point corresponds to a matter dominated point with $w_{\rm{eff}}=0$, and $\Omega_{\rm{DE}} = \frac{3}{\lambda^2}$. Under the CMB bound for the dark energy density parameter $\Omega_{\rm{DE}} < 0.02$ \cite{Ade:2015rim}, we restrict $\lambda$ to be $\lambda > 12$. The eigenvalues are
\begin{equation}
\left\{-1,3-\frac{3 \delta}{\lambda }, - \frac{3}{4} \pm \frac{3 \sqrt{24-7 \lambda ^2}}{4 \lambda } \right\}.
\end{equation}
Under the previous bound for $\lambda$ this point becomes saddle. 

\item \textit{Point $\cal{F}$}:  This point could give cosmic acceleration with $w_{\rm{eff}} <-\frac{1}{3}$, translating into $\lambda^2 < 2 $. From \cref{tab:DymSys} this point exists for $\lambda^2 < 6$. The eigenvalues are
\begin{equation}
\left\{\lambda  (\lambda-\delta),\frac{1}{2} \left(\lambda ^2-6\right),\lambda ^2-4,\lambda ^2-3\right\}.
\end{equation}
It could be stable if $\lambda^2< 3$ with two options depending the sign of $\delta$: if $\delta >0$, $0 < \lambda < \delta$, otherwise $\delta < \lambda < 0$.  

\item \textit{Point $\cal{G}$}:  This point could give cosmic acceleration with $\Omega_{\rm{DE}} = \frac{9}{(\delta^2-3)^2}$ and exists  for $\delta \neq \pm \sqrt{3}$. From the BBN bound $\Omega_{\rm{DE}} = 0.045$ we have $\delta = \pm 4.14$. The eigenvalues are
\begin{equation}
\left\{ \frac{3}{4} \pm W,\frac{9}{\delta^2-3}-1, \frac{3 \delta (\delta -\lambda )}{2 \left(\delta^2-3\right)}\right\},
\end{equation}
being
\begin{equation}
W = \sqrt{-\frac{\left(7 \delta^2-48\right)
   \delta^2}{\left(\delta^2-3\right)^2}}+\frac{9}{\delta^2-3}-1. 
\end{equation}
The function $W$ has two asymptotes in $\delta = \pm \sqrt{3}$, and inside this range becomes negative.  For $\delta < - 4\sqrt{\frac{3}{7}}$ or $\delta >4\sqrt{\frac{3}{7}}$ the term inside the radical is negative and the whole function is complex; in addition we have the limiting value of $-1+i\sqrt{7}$ for $W$ when $\delta \rightarrow \pm \infty$. Thus, the first eigenvalue ($\frac{3}{4}+W$) is negative when $\delta$ ranges the interval $(-\sqrt{3},\sqrt{3})$, while the second eigenvalue ($\frac{3}{4}-W$) is negative in the interval $[-4\sqrt{3/7},-\sqrt{3}) \cup (\sqrt{3},4\sqrt{3/7}]$. The third eigenvalue is positive inside $\delta \in (-\sqrt{3},\sqrt{3})$. Finally, for $\lambda >0$ there are three options for the fourth eigenvalue to be negative: For $\lambda \in (0,\sqrt{3})$, $\delta \in(-\sqrt{3},0)$ or $\delta \in (\lambda,\sqrt{3})$,
   for $\lambda=\sqrt{3}$, $\delta \in (-\sqrt{3},0)$, and for $\lambda>\sqrt{3}$, $\delta\in(-\sqrt{3},0)$ or $\delta \in(\sqrt{3},\lambda)$. 
\item \textit{Point $\cal{H}$}:  This point corresponds to radiation domination with $\Omega_r = 1- \frac{12}{\delta^2}$. The eigenvalues are
\begin{equation}
\left\{ \frac{1-Z^{2/3}}{\sqrt[3]{Z}},\frac{Z^{2/3}-1 \pm i \sqrt{3}( Z^{2/3} +1)}{2
   \sqrt[3]{Z}}, 2-\frac{2 \lambda }{\delta
   }\right\},
\end{equation}
with
\begin{equation}
Z = \frac{24}{\delta ^2}+\sqrt{5-\frac{96 \left(\delta ^2-6\right)}{\delta ^4}}-2. 
\end{equation}
We check numerically the behaviour of $Z$ in function of $\delta$. Besides the asymptotic nature at $\delta=0$, $Z$ never gets negative values. In fact, the asymptotic value of $-2+\sqrt{5}$ is reached when $\delta \rightarrow \pm \infty$. For $Z>0$, the first eigenvalue is always negative, while is positive for $Z$ ranging $[0,1]$; conversely, the real part of the two complex eigenvalues is positive for $Z>1$, and negative for $Z \in [0,1]$. Finally, the third eigenvalue is positive for $\delta > 0$ or $\delta > \lambda$ being $\lambda$ positive, while is negative with $\delta$ ranging $[0,1]$ assuming $\lambda >0$. 
\end{itemize}

The tipical viable cosmological dynamics corresponds to a transition between a radiation epoch ($w_{\rm{eff}} \simeq \frac{1}{3}, \Omega_r \simeq 1$) to a matter dominated epoch ($w_{\rm{eff}} \simeq 0, \Omega_m \simeq 1$), and finally an accelerated expansion epoch ($w_{\rm{eff}} \lesssim -\frac{1}{3}, \Omega_{\rm{DE}} \simeq 1$). Such transition  can be made as

\begin{equation}
\overbrace{\{ \cal{B}, \cal{C}, \cal{H} \}}^{\text{Radiation}} \quad \longrightarrow\quad \overbrace{\{ \cal{O}, \cal{E} \}}^{\text{Matter}} \quad \longrightarrow\quad \overbrace{\{ \cal{D}, \cal{F}, \cal{G} \}}^{\text{Dark Energy}} .
\end{equation}

For the radiation domination we observe that point $\cal{H}$ can not realize $\Omega_r \simeq 1$, except for $\delta \rightarrow \infty$, implying an unbounded energy density for the $3-$form. Moreover, the point $\cal{B}$ has more positive eigenvalues than point $\cal{C}$, this suggest that the solutions approach the saddle point $\cal{C}$ during the radiation era. The matter domination can be realized by point $\cal{E}$, noticing that point $\cal{O}$ has two negative eigenvalues while $\cal{E}$ has one negative, and two complex eigenvalues with negative real parts when the CMB bound ($\lambda > 12$) is considered. Finally, for the CMB bound for $\lambda$ the point $\cal{F}$ can not realize the accelerated expansion due to the existence condition $\lambda^2 <6 $; under the BBN $\delta = \pm 4.14$ the point $\cal{G}$ has two negative eigenvalues and two complex eigenvalues with positive real parts, nevertheless, point $\cal{D}$ under this bounds has two complex numbers with negative real parts, and two negative eigenvalues, becoming a stable spiral. In summary, we can produce a viable cosmological evolution by the sequence

\begin{equation}
\overbrace{\cal{C}}^{\text{Radiation}} \quad \longrightarrow\quad \overbrace{\cal{E}}^{\text{Matter}} \quad \longrightarrow\quad \overbrace{ \cal{D}}^{\text{Dark Energy}}. 
\end{equation}

From the definitions of the critical points $\cal{C}$ and $\cal{E}$, we realize that the transition from radiation to matter dominance can be made without the existence of the $3-$form, i.e., only with the quintessence field $\phi$. The $3-$form induced potential is only relevant for the dark energy dominated epoch.

\subsection{Anisotropic dark energy: coupled 1-form with a scalar field}
In this section, we briefly discuss some aspects of a model for dark energy built with a $1-$form coupled to a scalar field.  The particular model that we will study is described by the action 
\begin{equation}
S= \int {\rmd}^4 x \sqrt{-g}\left( \frac{M_{\rm{p}}^2}{2}R + {\cal L}_{(1)}+{\cal L}_\phi + {\cal L}_{x} \right),
\end{equation}
where
\begin{equation}
{\cal L}_{(1)} = -\frac{1}{4}f_1(\phi)F_{(1)\mu\nu}F_{(1)}^{\mu\nu},
\end{equation}
  ${\cal L}_\phi$ is give by eq. (\ref{eq:Lph})  and ${\cal L}_{x}$ stands for the perfect fluid of matter and/or radiation.  This model has some interest since it can be responsible for a late time anisotropic accelerated expansion. It was previously studied in ref. \cite{Thorsrud:2012mu}, but adding an explicit coupling between matter and the scalar field. In this reference, the authors use the dynamical system approach to show that the kinetic coupling can produce anisotropic scaling solutions which can be interpreted as a matter and  dark energy dominated epochs. For this reason, in this section  we only focus on the kinetic coupling to call the attention of  a particular behaviour of the equation of state that was not mentioned in \cite{Thorsrud:2012mu}, and that is interesting from the observational point of view.

We will use the gauge freedom $A_0 = \partial^{i}A_{i}=0$, to choose the vector field along the $x$ direction $A_{(1)}=A_{1}(t)\rm{d}x$. Due to the rotational symmetry in the $yz$ plane  we will use a Bianchi I metric:
\begin{equation}
\rm{d}s^2 = - \rm{d}t^2 + e^{2\alpha(t)}\left[e^{- 4\sigma(t)}\rm{d}x^2 + e^{2\sigma(t)}(\rm{d}y^2 + \rm{d}z^2) \right],
\end{equation}
being $e^{\alpha} \equiv a$ with $a$ the scale factor, and $\sigma$ the spatial shear.  In this configuration and from \cref{eq:EA1},  the e.o.m.  for the field $A_{1 \mu_1}$ leads to
\begin{equation}\label{EOMVF}
\ddot{A}_{1} + \left[ 2 \frac{f_1'}{f_1}\dot{\phi} + \dot{\alpha} + 4 \dot{\sigma} \right]\dot{A}_{1} = 0,  \qquad \Longrightarrow \qquad \dot{A}_{1} = \tilde{p}_{1} f_1(\phi)^{-2}e ^{-\alpha - 4 \sigma},
\end{equation}
being $\tilde{p}_{1} $ an integration constant. Friedmann equations and the e.o.m for the scalar field can be written as
\begin{align}
 \dot{\alpha}^2 
& = \dot{\sigma}^2 + \frac{1}{3M_{\rm pl}^2} \left[ \rho_m +  \rho_r + \frac{1}{2} \dot{\phi}^2+V(\phi)+\rho_{(1)} \right]\,, \label{eq:Hmod4}\\
 \ddot{\alpha} 
& =-3\dot{\alpha}^2 + \frac{1}{M_{\rm pl}^2} \left[  \frac{\rho_m}{2}+ \frac{\rho_r}{3}  + V(\phi)+\frac{1}{3} \rho_{(1)} \right]\,,\label{eq:Hmod5}\\
 \ddot{\sigma}
& =-3 \dot{\alpha}\dot{\sigma} +\frac{2}{3M_{\rm pl}^2}  \rho_{(1)}\,, \label{eq:Hmod6}\\
\ddot{\phi} & = -3 \dot{\alpha} \dot{\phi}-V_{,\phi}
+2\frac{f_{1,\phi}}{f_1} \rho_{(1)},
\end{align}
where we have defined the energy density of the $1-$form field as
\begin{equation}
\rho_{(1)}=\frac{f_1^2}{2} e^{-2\alpha+4\sigma} \dot{A}_1^2\,.
\end{equation}
By using the parameters defined in \cref{eq:params}  and two more related to the anisotropy and energy density of the $1-$form:
\begin{equation}
\Sigma = \frac{\dot{\sigma}}{H}, \qquad \Omega_{(1)} = \frac{\rho_{(1)}}{ 3 H^2 M_{\rm pl}^2}.
\end{equation}
As in the previous subsection, we can define the following set of quantities
\begin{equation}
\Omega_{\rm{DE}} = X^2 + Y^2 + \Sigma^2 + \Omega_{(1)},
\end{equation}
\begin{equation}
\rho_{\rm{DE}} = \frac{\dot{\phi}}{2} + V(\phi) + \rho_{(1)} + 3 M_{\rm pl}^2 H^2 \Sigma^2, \quad p_{\rm{DE}} = \frac{\dot{\phi}}{2} - V(\phi) + \frac{\rho_{(1)}}{3} + 3 M_{\rm pl}^2 H^2 \Sigma^2,   
\end{equation}
\begin{equation}
w_{\rm{DE}} \equiv \frac{p_{\rm{DE}}}{\rho_{\rm{DE}}} = \frac{3(X^2 - Y^2 + \Sigma^2)+\Omega_{(1)}}{3(X^2+Y^2+\Sigma^2 + \Omega_{(1)})},
\end{equation}
where $\rho_{\rm{DE}}$, $p_{\rm{DE}}$, $\Omega_{\rm{DE}}$ and $w_{\rm{DE}}$ are the energy density, the pressure, the density parameter and the equation of state related to dark energy component, respectively. The effective equation of state $w_{\rm{eff}}$ has a simple form:
\begin{equation}
w_{\rm{eff}} = X^2 - Y^2 + \Sigma^2  + \frac{\Omega_{(1)}}{3} + \frac{\Omega_r}{3}.  
\end{equation}
Again, we  fix the potential and the coupling $f_1$ with exponential forms as in \cref{eq:Potentials}:
\begin{equation}\label{eq:Potentials2}
V(\phi) \propto e^{- \lambda\frac{\phi}{M_{{\rm pl}}}}, \qquad f_1(\phi) \propto e^{- \mu\frac{\phi}{M_{{\rm pl}}}}.
\end{equation}
With the previous definitions the equations (\ref{EOMVF}) to (\ref{eq:Hmod6}) can be converted like an autonomous system as
\begin{align}
X' &=  \frac{3}{2}X\left( X^2 - Y^2 + \Sigma^2 -1  + \frac{\Omega_{(1)}}{3} + \frac{\Omega_r}{3}\right) + \frac{\sqrt{6}}{2}(\lambda Y^2 - 2\mu \Omega_{(1)}), \label{eq:Dynsi41}\\[1mm]
Y' &= \frac{1}{2}Y \left(3X^2- 3Y^2 + 3\Sigma^2 + 3 + \Omega_{(1)} + \Omega_r  - \sqrt{6}\lambda X  \right), \label{eq:Dynsi42}\\[1mm]
\Sigma' &= \frac{1}{2}\Sigma\left(3 X^2 - 3Y^2 + 3\Sigma^2 - 3 + \Omega_{(1)} +  \Omega_r \right) + 2 \Omega_1, \label{eq:Dynsi43}\\[1mm]
\Omega_{(1)}' &=\Omega _{(1)} \left( 3X^2 - 3Y^2 +3 \Sigma ^2+4 \Sigma-1 +2 \sqrt{6} \mu  X+\Omega _{(1)}+ \Omega _r \right), \label{eq:Dynsi44} \\[1mm]
\Omega_r ' &= \Omega_r \left( 3 X^2 - 3Y^2 + 3\Sigma^2 -1 + \Omega_{(1)} + \Omega_r\right). \label{eq:Dynsi45}
\end{align}

\begin{figure}
\centering
\begin{minipage}{0.44\textwidth}
\centering
\subfloat[]
{\label{fig:1a}\includegraphics[width=\linewidth]{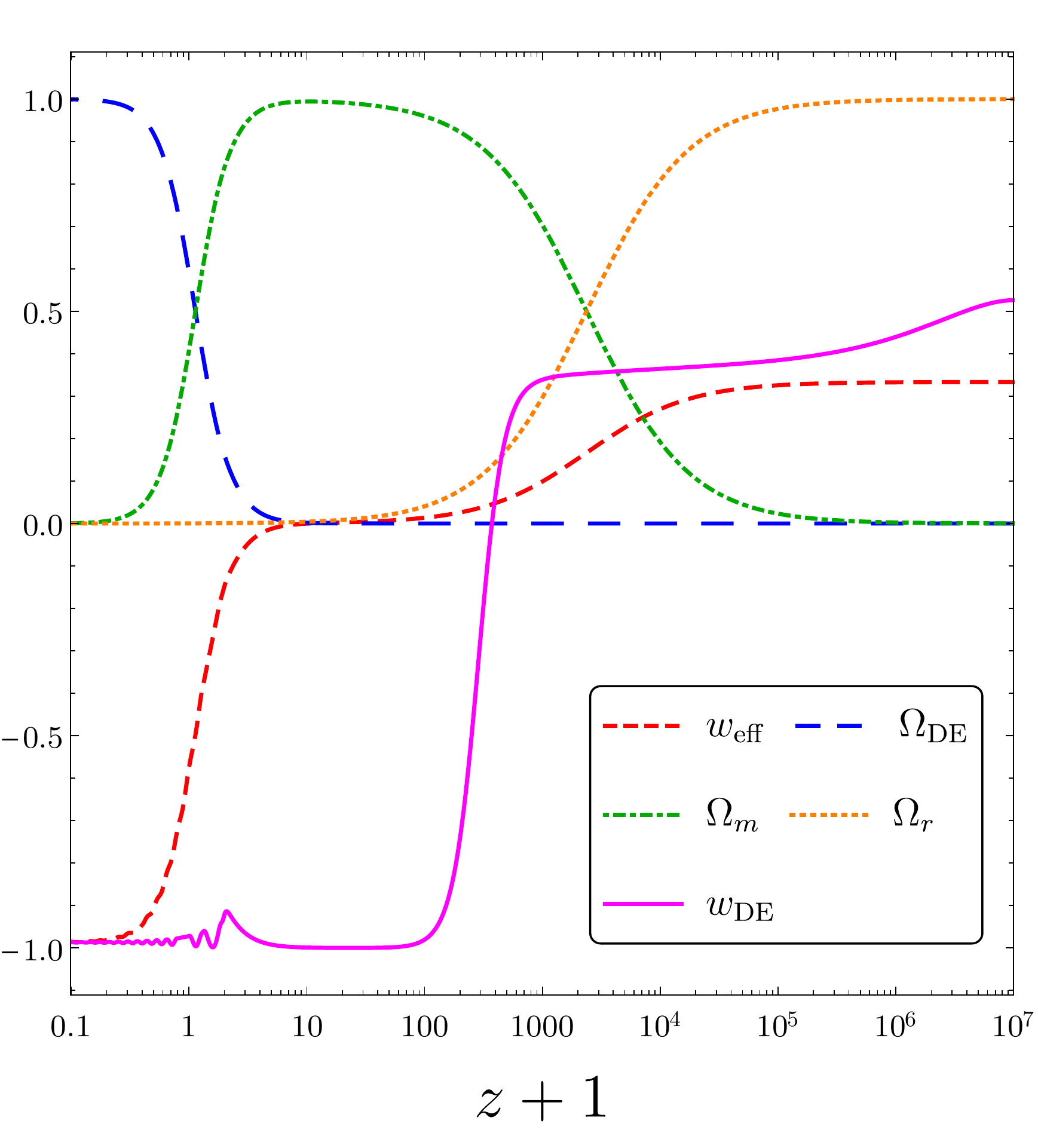}}
\end{minipage}
\begin{minipage}{0.5\textwidth}
\centering
\subfloat[]
{\label{fig:1b}\includegraphics[width=\linewidth]{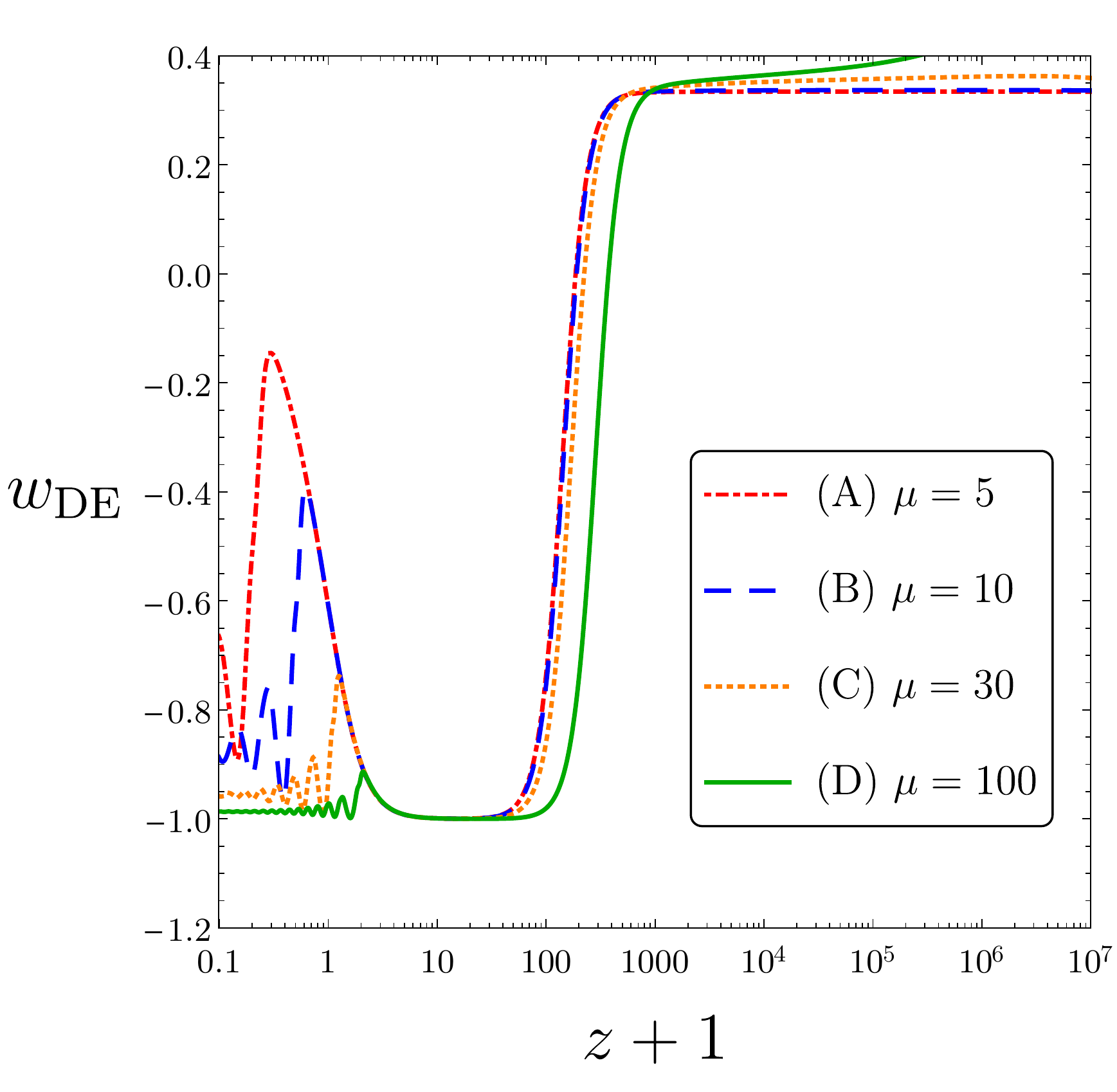}}
\end{minipage}
\caption{\label{fig:fig1} (a) Evolution of $\Omega_{\rm DE}$, $\Omega_r$, $\Omega_m$, 
$w_{\rm DE}$, $w_{\rm eff}$ versus $z+1~(=1/a)$ 
for $\lambda=2$ and $\mu=100$ with 
the initial conditions $X=10^{-13}$, $Y=10^{-14}$, $\Sigma=0$, $\Omega_{(1)}=10^{-5}$, and $\Omega_r=0.99996$ at the redshift $z=7.9 \times 10^7$. 
The present epoch ($z=0$) 
is identified by $\Omega_{\rm DE}=0.68$. (b) Evolution of $w_{\rm DE}$ versus $z+1$ for 
$\lambda=2$ with the same initial values of 
$X$, $Y$, $\Sigma$, and $\Omega_{(1)}$, 
as those in Fig.~\ref{fig:1a}.
Each line corresponds to 
(A) $\mu=5$, (B) $\mu=10$, (C) $\mu=30$, and 
(D) $\mu=100$. 
The initial conditions of radiation density parameter 
are chosen to be 
(A) $\Omega_r=0.99994$ at  $z=5.5 \times 10^{7}$, 
(B) $\Omega_r=0.999951$ at  $z=6.5 \times 10^{7}$, 
(C) $\Omega_r=0.99996$ at  $z=7.9 \times 10^{7}$, and 
(D) $\Omega_r=0.999961$ at  $z=8.3 \times 10^{7}$, 
respectively, 
to realize the value $\Omega_r(z=0) \simeq 10^{-4}$. }
\end{figure}

The previous set of equations allow us to fully describe the background evolution of the universe characterised by this model, from the analysis of the dynamical system, as we made in \cref{sec:4.1}, or by direct numerical integration of the variables. Since our ultimate goal is to illustrate the applications of the coupled $p-$forms formalism, without rigurous treatment, we decide here to implement the second choice.  By fixing the couplings constants to be $\lambda=2$ and $\mu=100$, we integrate the whole system \cref{eq:Dynsi41,eq:Dynsi42,eq:Dynsi43,eq:Dynsi44,eq:Dynsi45} with suitable initial conditions. In \cref{fig:1a}  we show the evolution of the density parameters and the e.o.s. for dark energy. We clearly see a well-behave cosmological dynamics, starting from a radiation-dominated epoch, followed by a matter-dominated era around the redshift $z\simeq 2200$, and afterwards a dark energy-dominated epoch in which the dark energy density $\Omega_{\rm{DE}}$ start to deviate from zero around $z\simeq8$. The shear $\Sigma$, which measure the degree of anisotropy, has interesting features. We check numerically that even if it starts from a null value, it gained shear in two stages: first it decreases from a value of $\simeq 10^{-5}$ at $z\simeq10^{7}$ until a minimum value of order $10^{-8}$ at around $z\simeq1$, and then increases until a value of $0.1$ today. Thus, starting from a nearly isotropic configuration, the system acquires anisotropy through time.  In addition, an interesting feature of this model, that was not mentioned in ref. \cite{Thorsrud:2012mu}, appears: the dark energy e.o.s. acquires a value around $-1$ close to $z\simeq100$, but present an oscillating behaviour over $z\simeq3$ to finally reached again the value $-1$ close to the present time. This characteristic represents an interesting observational signature for distinguishing this scenario from other dark energy models, since one of the current observational interest is to reconstruct the temporal evolution of the equation of state of dark energy. In this sense, we expect that future surveys like Euclid \cite{Amendola:2012ys} can elucidate if models like the one present here are viable or not. In \cref{fig:1b} we present the evolution of the dark energy e.o.s for different values of the coupling $\mu$. We can observe that the amplitude of the oscillations of $w_{\rm{DE}}$ strongly depend on the value of $\mu$, in particular if $\mu\sim 10^2$ then $w_{\rm{DE}}$ approaches to $-1$ at present time, as expected. We plan to make a full analysis for the cosmogical dynamics of coupled $p-$form system in a forthcoming publication. \\

{  Before concluding this section, it is worth to mention the possible impact of this scenario in current issues such as the so called $H_0$ \textit{tension}. This tension appears when comparing the value of the Hubble rate today from local distance indicators or low redshift  (distance ladder method) and high redshift (CMB) observations from the Planck survey; even both methods give precision measurements, they have a significant statistical discrepancy of around $3.3\sigma$ \cite{Riess:2016jrr}. Besides possible explanations coming from statistical analysis and problems when modeling, for instance, the astrophysics behind the supernovae events, models beyond the flat $\Lambda$CDM paradigm had been considered to aliviate this discrepancy \cite{Mortsell:2018mfj,DiValentino:2017rcr,DiValentino:2017iww,Guo:2018ans,Pourtsidou:2016ico,Khosravi:2017hfi}. Most of them use contributions from the dark sector in the form of a dynamical equation of state for dark energy. Since our model reveals an oscillatory behavior as shown in \cref{fig:1b}, it could be 
interesting if, with a proper statistical analysis using current data, we could give some hints in the discussion on this tension. We expect to come back to this issues in a future work. %{\color{red} Another models based in the interaction between $p-$forms and a scalar field also expose oscillations at low redshifts in the equation of state \cite{Almeida:2019iqp,Guarnizo:2019mwf}, and thus they could be subject to this examination.}  
}

%%%%%%%%%%%%%%%%%%%%
\subsection{A comment on topologic terms and parity breaking signatures}
%%%%%%%%%%%%%%%%%%%%
As we realized in  \cref{sec:12form}, the system involving a general coupling between the $1-$form and the $2-$form can be expressed as a model of a massive vector field with kinetic couplings as in \cref{Smass}. Aside of this, we also saw in \cref{sec:3formsec} that the $3-$form, when gauge invariance is respected, evolves homogeneously only as a function of time and do not contribute to the sourcing of statistical anisotropies. This implies that the possible cosmological signatures of the model involving all the possible coupling between the $p-$forms are equivalent to the signatures of a massive vector field model.  There is a great body of literature about about cosmological signatures of vector field models, an incomplete list of references on the subject include \cite{Yokoyama:2008xw, Dimopoulos:2009vu, Dimopoulos:2009am, Watanabe:2010bu, Bartolo:2012sd, Biagetti:2013qqa, Shiraishi:2013vja, Abolhasani:2013zya, Lyth:2013kah, Rodriguez:2013cj, Lyth:2013vha, Shiraishi:2013oqa, Chen:2014eua, Almeida:2014ava, Fujita:2017lfu} for parity conserving vector field models; \cite{Sorbo:2011rz, Dimopoulos:2012av, Anber:2012du, Bartolo:2014hwa, Caprini:2014mja, Bartolo:2015dga, Namba:2015gja, Shiraishi:2016yun, Caprini:2017vnn, Almeida:2017lrq, Almeida:2018pir} for parity violating vector field models and  \cite{Ohashi:2013qba, Ohashi:2013mka,  Obata:2018ilf} for $2-$forms models.  \\

As evidenced by the previous list of references, the issue of  the statistical anisotropies and parity breaking signatures in vector fields and $2-$form field models with kinetic couplings have been explored in great detail in recent literature, and there is nothing really new more to say about this subject here. Nevertheless, one aspect that we would like to mention here is the fact that parity violating signatures, in the presence of $p-$forms,  appears always as topological terms. In the four-dimensional case that we considered here, the only topological term which plays a decisive role in the dynamics of the system is  $F_{(1)}\wedge F_{(1)}$. This term is responsible for the breaking of parity in the transverse polarizations of the vector field
% while the $B_{(2)}\wedge F_{(1)}$, despite it looks like a parity breaking  term, it doesn't breaks parity but instead is responsible for the presence of a longitudinal polarization in a vector field model.  The parity breaking of the $F_{(1)}\wedge F_{(1)}$ term has been studied in detail in 
\cite{Sorbo:2011rz, Dimopoulos:2012av, Anber:2012du, Bartolo:2014hwa, Caprini:2014mja, Bartolo:2015dga, Namba:2015gja, Shiraishi:2016yun, Caprini:2017vnn, Almeida:2017lrq, Almeida:2018pir}. Some striking features of this model is the production of chiral gravitational waves \cite{Sorbo:2011rz, Anber:2012du} and the presence of odd terms in the multipolar expansion of the inflationary correlators such as the bispectrum of primordial curvature perturbations \cite{Bartolo:2015dga}, among others.

If the features mentioned before, turns out to be measurable by near term future missions, this could be seen as signatures of either topological defects, similar to topological defects in a continuous medium \cite{Julia:1979ur} during the inflationary expansion, or as global topological characteristics of the background spacetime during the inflationary era.

%%%%%%%%%%%%%%%%%%%%%%%%%%%%%%%%%%%%%%%%%%%%%%%%%%%%%%%%%%%
\section{Conclusions}\label{sec:Conclusions}
%%%%%%%%%%%%%%%%%%%%%%%%%%%%%%%%%%%%%%%%%%%%%%%%%%%%%%%%%%%

%\ale{Check this paragraph!}
In this paper we  provided a general construction of a Lagrangian based on coupled $p-$forms in $D$ dimensions. Assuming as building blocks the $p-$forms, their field strength and their  duals, we wrote down the general expression of the Lagrangian including general mixtures of the $p-$forms. Under the assumption of a Lagrangian based solely on first derivatives of the $p-$forms, and providing as well gauge invariance, the construction, of course,  becomes much more simpler, but not trivial, in comparison with the case of generalized $p-$forms Galileons, since no extra couplings with curvature are invoked. \\

We specialized to the case of a four dimensional space-time, and besides the standard $1-$form, its dual and the $2-$form, we allow the Lagrangian to have a contribution from the $3-$form. In addition to the general result which implies that the field strength of the $3-$form acts as a cosmological constant, we provide an expression for the energy momentum-tensor, probing effectively that this component %fluid 
has an equation of state $w_{(3)}=-1$.  We see how the $3-$form is absorbed in an effective potential for the scalar field and how it contributes to the dynamics of the system. %However, from the equations of motion, the coupling between the $3-$form and a general function of the scalar field, gives a non-trivial dynamics, which in some way can be seen as an extension of the model proposed in \cite{Kaloper:2008qs,Kaloper:2008fb}. 
In this sense, the $3-$form-scalar field system offers an interesting approach to the inflationary problem, and also to the current accelerated expansion of the universe.  We discuss as well the importance of topological terms, which are the seeds for parity violating signatures in the statistical correlators.   Working in four dimensions, we show that the system composed by interacting  $p-$forms is equivalent, under a particular parametrization of the coupling scalar functions, to a massive, parity violating vector field Lagrangian, giving thus an alternative mechanism to mass generation consistent with  gauge symmetry.
%Working with the $1-$ and $2-$forms, we had shown that the Lagrangian of the interacting system composed by the $1-$, $2-$ and $3-$forms, is equivalent, under a particular parametrization of the coupling scalar functions, to a massive, parity violating vector field Lagrangian, giving thus an alternative mechanism to mass generation consistent with  gauge symmetry. \\
%As a result, we realize that the system composed by the interacting $p-$forms in four dimensions is equivalent to a massive and parity breaking vector field.

Next, we provided some applications to cosmological backgrounds. We studied a very minimalistic system composed by a scalar field and a $3-$form. We isolated the analysis of the $3-$form due to its relevance in the context of an accelerated expansion driven by dark energy or a cosmological constant. We considered the late time evolution of this system in the presence of a perfect fluid with matter and radiation components, examined the dynamics and its critical points and their stability. We also analized, at the background level, the signatures of an anisotropic source such as a $1-$form field  in the late time evolution of the universe.  We plan to include the information of all the coupled $p-$form interacting system in a forthcoming publication.
%an analysis to cosmological backgrounds, focusing our attention in the $3$-form-scalar field coupling. Assuming a FLRW, we found expressions for the Friedmann equations and the particular solution of the e.o.m. of the three form. Further, we study the background dynamics, first of all, determining the fixed points and its stability. Since in the case of zero $3-$form, the system is only composed by the kinetic and potential energies of the scalar field (or equivalent to a  Quintessence field scenario), we concentrate in the additional fixed points from the extra degree of freedom, taking into account that the whole stability now depends  on the coupling $\gamma$ of the $3$-form. We found that within these new points, a scaling solution appears whit potential  applications to solve the concordance problem, seeing that in this case the dark energy density parameter scales proportionally to the matter one. Furthermore, we provide some examples for the solutions of the full dynamical system, exalting their potential use for cosmological applications. We plan to include the full information of the $3-$form system coupled to a scalar field, which includes the $\theta$ term of its dual, in a forthcoming publication. \\

Finally, we made some general statements about the signatures of $p-$forms in the correlation functions. As expenses of a homogeneous evolution in time of the  $3-$form, no statistical anisotropies are generated by this degree. In fact, all possible signatures of parity violation rely in the topological terms $F\wedge F$ of the $1-$form. Despite their appearance, the terms $\tilde{F}_{(3)}$ and  $B\wedge F$, don't introduce further parity violating signatures. Due to the vast literature concerning statistical anisotropies with single $1$-forms, no more studies in this paper were made in this direction. Nevertheless, the statistical signatures in the system of coupled $1-$ and $2-$forms was not studied in detail in the literature and we expect to come back to study these subject elsewhere. \\

% Contact with approaches related with massive higher-spin fields.

%%%%%%%%%%%%%%%%%%%%%%%%%%%%%%%%%%%%%%%%%%%%%%%%%%%%%%%%%%%%%%%%%%%%%%%%%%%%%%%%%%%%
\section*{Acknowledgments}

It is a pleasure to thank Hern\'an Ocampo Dur\'an for discussions and comments.  We are also in debt with Shinji Tsujikawa and Ryotaro Kase for valuable comments about the draft and with Lorenzo Sorbo for important discussions about the $B\wedge F$ model and for driving our attention to references \cite{Birmingham:1991ty,Quevedo:1996uu}. This work was supported by COLCIENCIAS grants numbers 123365843539 RC FP44842-081-2014 and 110671250405 RC FP44842-103-2016 and by COLCIENCIAS -- DAAD grand number 110278258747. JPBA acknowledge support from  Universidad Antonio Nari\~no grant number 2017239 and thanks Universidad del Valle for its warm hospitality during several stages of this project.

%%%%%%%%%%%%%%%%%%%%%%%%%%%%%%%%%%%%%%%%%%%%%%%%%%%%%%%%%%%%%%%%%%%%%%%%%%%%%%%%%%%%
\newpage
\bibliographystyle{utcaps}
\bibliography{bibli}

\providecommand{\href}[2]{#2}\begingroup\raggedright\begin{thebibliography}{100}

\bibitem{Guth:1982}
A.~H. Guth and S.-Y. Pi, ``Fluctuations in the New Inflationary Universe,''
  \href{http://dx.doi.org/10.1103/PhysRevLett.49.1110}{{\em Phys. Rev. Lett.}
  {\bfseries 49} (Oct, 1982) 1110--1113}.

\bibitem{Starobinsky1982}
A.~Starobinsky, ``Dynamics of phase transition in the new inflationary universe
  scenario and generation of perturbations,''
  \href{http://dx.doi.org/https://doi.org/10.1016/0370-2693(82)90541-X}{{\em
  Physics Letters B} {\bfseries 117} no.~3, (1982) 175 -- 178}.
  \url{http://www.sciencedirect.com/science/article/pii/037026938290541X}.

\bibitem{Hawking1982}
S.~Hawking, ``The development of irregularities in a single bubble inflationary
  universe,''
  \href{http://dx.doi.org/https://doi.org/10.1016/0370-2693(82)90373-2}{{\em
  Physics Letters B} {\bfseries 115} no.~4, (1982) 295 -- 297}.
  \url{http://www.sciencedirect.com/science/article/pii/0370269382903732}.

\bibitem{Aghanim:2018eyx}
{\bfseries Planck} Collaboration, N.~Aghanim {\em et~al.}, ``{Planck 2018
  results. VI. Cosmological parameters},''
\href{http://arxiv.org/abs/1807.06209}{{\ttfamily arXiv:1807.06209
  [astro-ph.CO]}}.
%%CITATION = ARXIV:1807.06209;%%.

\bibitem{Akrami:2019izv}
{\bfseries Planck} Collaboration, Y.~Akrami {\em et~al.}, ``{Planck 2018
  results. IX. Constraints on primordial non-Gaussianity},''
\href{http://arxiv.org/abs/1905.05697}{{\ttfamily arXiv:1905.05697
  [astro-ph.CO]}}.
%%CITATION = ARXIV:1905.05697;%%.

\bibitem{Akrami:2019bkn}
{\bfseries Planck} Collaboration, Y.~Akrami {\em et~al.}, ``{Planck 2018
  results. VII. Isotropy and Statistics of the CMB},''
\href{http://arxiv.org/abs/1906.02552}{{\ttfamily arXiv:1906.02552
  [astro-ph.CO]}}.
%%CITATION = ARXIV:1906.02552;%%.

\bibitem{Schwarz:2015cma}
D.~J. Schwarz, C.~J. Copi, D.~Huterer, and G.~D. Starkman, ``{CMB Anomalies
  after Planck},'' \href{http://dx.doi.org/10.1088/0264-9381/33/18/184001}{{\em
  Class. Quant. Grav.} {\bfseries 33} no.~18, (2016) 184001},
\href{http://arxiv.org/abs/1510.07929}{{\ttfamily arXiv:1510.07929
  [astro-ph.CO]}}.
%%CITATION = ARXIV:1510.07929;%%.

\bibitem{Perivolaropoulos:2014lua}
L.~Perivolaropoulos, ``{Large Scale Cosmological Anomalies and Inhomogeneous
  Dark Energy},'' \href{http://dx.doi.org/10.3390/galaxies2010022}{{\em
  Galaxies} {\bfseries 2} (2014) 22--61},
\href{http://arxiv.org/abs/1401.5044}{{\ttfamily arXiv:1401.5044
  [astro-ph.CO]}}.
%%CITATION = ARXIV:1401.5044;%%.

\bibitem{Dimastrogiovanni:2010sm}
E.~Dimastrogiovanni, N.~Bartolo, S.~Matarrese, and A.~Riotto,
  ``{Non-Gaussianity and Statistical Anisotropy from Vector Field Populated
  Inflationary Models},'' \href{http://dx.doi.org/10.1155/2010/752670}{{\em
  Adv. Astron.} {\bfseries 2010} (2010) 752670},
\href{http://arxiv.org/abs/1001.4049}{{\ttfamily arXiv:1001.4049
  [astro-ph.CO]}}.
%%CITATION = ARXIV:1001.4049;%%.

\bibitem{Soda:2012zm}
J.~Soda, ``{Statistical Anisotropy from Anisotropic Inflation},''
  \href{http://dx.doi.org/10.1088/0264-9381/29/8/083001}{{\em Class. Quant.
  Grav.} {\bfseries 29} (2012) 083001},
\href{http://arxiv.org/abs/1201.6434}{{\ttfamily arXiv:1201.6434 [hep-th]}}.
%%CITATION = ARXIV:1201.6434;%%.

\bibitem{Maleknejad:2012fw}
A.~Maleknejad, M.~M. Sheikh-Jabbari, and J.~Soda, ``{Gauge Fields and
  Inflation},'' \href{http://dx.doi.org/10.1016/j.physrep.2013.03.003}{{\em
  Phys. Rept.} {\bfseries 528} (2013) 161--261},
\href{http://arxiv.org/abs/1212.2921}{{\ttfamily arXiv:1212.2921 [hep-th]}}.
%%CITATION = ARXIV:1212.2921;%%.

\bibitem{Mulryne:2012ax}
D.~J. Mulryne, J.~Noller, and N.~J. Nunes, ``{Three-form inflation and
  non-Gaussianity},''
  \href{http://dx.doi.org/10.1088/1475-7516/2012/12/016}{{\em JCAP} {\bfseries
  1212} (2012) 016},
\href{http://arxiv.org/abs/1209.2156}{{\ttfamily arXiv:1209.2156
  [astro-ph.CO]}}.
%%CITATION = ARXIV:1209.2156;%%.

\bibitem{Ohashi:2013qba}
J.~Ohashi, J.~Soda, and S.~Tsujikawa, ``{Observational signatures of
  anisotropic inflationary models},''
  \href{http://dx.doi.org/10.1088/1475-7516/2013/12/009}{{\em JCAP} {\bfseries
  1312} (2013) 009},
\href{http://arxiv.org/abs/1308.4488}{{\ttfamily arXiv:1308.4488
  [astro-ph.CO]}}.
%%CITATION = ARXIV:1308.4488;%%.

\bibitem{Ohashi:2013mka}
J.~Ohashi, J.~Soda, and S.~Tsujikawa, ``{Anisotropic Non-Gaussianity from a
  Two-Form Field},'' \href{http://dx.doi.org/10.1103/PhysRevD.87.083520}{{\em
  Phys. Rev.} {\bfseries D87} no.~8, (2013) 083520},
\href{http://arxiv.org/abs/1303.7340}{{\ttfamily arXiv:1303.7340
  [astro-ph.CO]}}.
%%CITATION = ARXIV:1303.7340;%%.

\bibitem{Kumar:2016tdn}
K.~Sravan~Kumar, D.~J. Mulryne, N.~J. Nunes, J.~Marto, and P.~Vargas~Moniz,
  ``{Non-Gaussianity in multiple three-form field inflation},''
  \href{http://dx.doi.org/10.1103/PhysRevD.94.103504}{{\em Phys. Rev.}
  {\bfseries D94} no.~10, (2016) 103504},
\href{http://arxiv.org/abs/1606.07114}{{\ttfamily arXiv:1606.07114
  [astro-ph.CO]}}.
%%CITATION = ARXIV:1606.07114;%%.

\bibitem{Farakos:2017jme}
F.~Farakos, S.~Lanza, L.~Martucci, and D.~Sorokin, ``{Three-forms in
  Supergravity and Flux Compactifications},''
  \href{http://dx.doi.org/10.1140/epjc/s10052-017-5185-y}{{\em Eur. Phys. J.}
  {\bfseries C77} no.~9, (2017) 602},
\href{http://arxiv.org/abs/1706.09422}{{\ttfamily arXiv:1706.09422 [hep-th]}}.
%%CITATION = ARXIV:1706.09422;%%.

\bibitem{Obata:2018ilf}
I.~Obata and T.~Fujita, ``{Footprint of Two-Form Field: Statistical Anisotropy
  in Primordial Gravitational Waves},''
  \href{http://dx.doi.org/10.1103/PhysRevD.99.023513}{{\em Phys. Rev.}
  {\bfseries D99} no.~2, (2019) 023513},
\href{http://arxiv.org/abs/1808.00548}{{\ttfamily arXiv:1808.00548
  [astro-ph.CO]}}.
%%CITATION = ARXIV:1808.00548;%%.

\bibitem{Almeida:2019xzt}
J.~P. Beltr\'an~Almeida, A.~Guarnizo, R.~Kase, S.~Tsujikawa, and C.~A.
  Valenzuela-Toledo, ``{Anisotropic inflation with coupled $p-$forms},''
  \href{http://dx.doi.org/10.1088/1475-7516/2019/03/025}{{\em JCAP} {\bfseries
  1903} (2019) 025},
\href{http://arxiv.org/abs/1901.06097}{{\ttfamily arXiv:1901.06097 [gr-qc]}}.
%%CITATION = ARXIV:1901.06097;%%.

\bibitem{Arkani-Hamed:2015bza}
N.~Arkani-Hamed and J.~Maldacena, ``{Cosmological Collider Physics},''
\href{http://arxiv.org/abs/1503.08043}{{\ttfamily arXiv:1503.08043 [hep-th]}}.
%%CITATION = ARXIV:1503.08043;%%.

\bibitem{Kehagias:2017cym}
A.~Kehagias and A.~Riotto, ``{On the Inflationary Perturbations of Massive
  Higher-Spin Fields},''
  \href{http://dx.doi.org/10.1088/1475-7516/2017/07/046}{{\em JCAP} {\bfseries
  1707} no.~07, (2017) 046},
\href{http://arxiv.org/abs/1705.05834}{{\ttfamily arXiv:1705.05834 [hep-th]}}.
%%CITATION = ARXIV:1705.05834;%%.

\bibitem{Bartolo:2017sbu}
N.~Bartolo, A.~Kehagias, M.~Liguori, A.~Riotto, M.~Shiraishi, and V.~Tansella,
  ``{Detecting higher spin fields through statistical anisotropy in the CMB and
  galaxy power spectra},''
  \href{http://dx.doi.org/10.1103/PhysRevD.97.023503}{{\em Phys. Rev.}
  {\bfseries D97} no.~2, (2018) 023503},
\href{http://arxiv.org/abs/1709.05695}{{\ttfamily arXiv:1709.05695
  [astro-ph.CO]}}.
%%CITATION = ARXIV:1709.05695;%%.

\bibitem{Franciolini:2017ktv}
G.~Franciolini, A.~Kehagias, and A.~Riotto, ``{Imprints of Spinning Particles
  on Primordial Cosmological Perturbations},''
  \href{http://dx.doi.org/10.1088/1475-7516/2018/02/023}{{\em JCAP} {\bfseries
  1802} no.~02, (2018) 023},
\href{http://arxiv.org/abs/1712.06626}{{\ttfamily arXiv:1712.06626 [hep-th]}}.
%%CITATION = ARXIV:1712.06626;%%.

\bibitem{Baumann:2017jvh}
D.~Baumann, G.~Goon, H.~Lee, and G.~L. Pimentel, ``{Partially Massless Fields
  During Inflation},'' \href{http://dx.doi.org/10.1007/JHEP04(2018)140}{{\em
  JHEP} {\bfseries 04} (2018) 140},
\href{http://arxiv.org/abs/1712.06624}{{\ttfamily arXiv:1712.06624 [hep-th]}}.
%%CITATION = ARXIV:1712.06624;%%.

\bibitem{Franciolini:2018eno}
G.~Franciolini, A.~Kehagias, A.~Riotto, and M.~Shiraishi, ``{Detecting higher
  spin fields through statistical anisotropy in the CMB bispectrum},''
  \href{http://dx.doi.org/10.1103/PhysRevD.98.043533}{{\em Phys. Rev.}
  {\bfseries D98} no.~4, (2018) 043533},
\href{http://arxiv.org/abs/1803.03814}{{\ttfamily arXiv:1803.03814
  [astro-ph.CO]}}.
%%CITATION = ARXIV:1803.03814;%%.

\bibitem{Bordin:2018pca}
L.~Bordin, P.~Creminelli, A.~Khmelnitsky, and L.~Senatore, ``{Light Particles
  with Spin in Inflation},''
  \href{http://dx.doi.org/10.1088/1475-7516/2018/10/013}{{\em JCAP} {\bfseries
  1810} no.~10, (2018) 013},
\href{http://arxiv.org/abs/1806.10587}{{\ttfamily arXiv:1806.10587 [hep-th]}}.
%%CITATION = ARXIV:1806.10587;%%.

\bibitem{Anninos:2019nib}
D.~Anninos, V.~De~Luca, G.~Franciolini, A.~Kehagias, and A.~Riotto,
  ``{Cosmological Shapes of Higher-Spin Gravity},''
\href{http://arxiv.org/abs/1902.01251}{{\ttfamily arXiv:1902.01251 [hep-th]}}.
%%CITATION = ARXIV:1902.01251;%%.

\bibitem{Cai:2014vua}
Y.-F. Cai, F.~Chen, E.~G.~M. Ferreira, and J.~Quintin, ``{New model of axion
  monodromy inflation and its cosmological implications},''
  \href{http://dx.doi.org/10.1088/1475-7516/2016/06/027}{{\em JCAP} {\bfseries
  1606} no.~06, (2016) 027},
\href{http://arxiv.org/abs/1412.4298}{{\ttfamily arXiv:1412.4298 [hep-th]}}.
%%CITATION = ARXIV:1412.4298;%%.

\bibitem{Cai:2015xla}
Y.-F. Cai, E.~G.~M. Ferreira, B.~Hu, and J.~Quintin, ``{Searching for features
  of a string-inspired inflationary model with cosmological observations},''
  \href{http://dx.doi.org/10.1103/PhysRevD.92.121303}{{\em Phys. Rev.}
  {\bfseries D92} no.~12, (2015) 121303},
\href{http://arxiv.org/abs/1507.05619}{{\ttfamily arXiv:1507.05619
  [astro-ph.CO]}}.
%%CITATION = ARXIV:1507.05619;%%.

\bibitem{Ibanez:2015fcv}
L.~E. Ibanez, M.~Montero, A.~Uranga, and I.~Valenzuela, ``{Relaxion Monodromy
  and the Weak Gravity Conjecture},''
  \href{http://dx.doi.org/10.1007/JHEP04(2016)020}{{\em JHEP} {\bfseries 04}
  (2016) 020},
\href{http://arxiv.org/abs/1512.00025}{{\ttfamily arXiv:1512.00025 [hep-th]}}.
%%CITATION = ARXIV:1512.00025;%%.

\bibitem{Valenzuela:2016yny}
I.~Valenzuela, ``{Backreaction Issues in Axion Monodromy and Minkowski
  4-forms},'' \href{http://dx.doi.org/10.1007/JHEP06(2017)098}{{\em JHEP}
  {\bfseries 06} (2017) 098},
\href{http://arxiv.org/abs/1611.00394}{{\ttfamily arXiv:1611.00394 [hep-th]}}.
%%CITATION = ARXIV:1611.00394;%%.

\bibitem{Wald:1983ky}
R.~M. Wald, ``{Asymptotic behavior of homogeneous cosmological models in the
  presence of a positive cosmological constant},''
\href{http://dx.doi.org/10.1103/PhysRevD.28.2118}{{\em Phys. Rev.} {\bfseries
  D28} (1983) 2118--2120}.
%%CITATION = PHRVA,D28,2118;%%.

\bibitem{Yokoyama:2008xw}
S.~Yokoyama and J.~Soda, ``{Primordial statistical anisotropy generated at the
  end of inflation},''
  \href{http://dx.doi.org/10.1088/1475-7516/2008/08/005}{{\em JCAP} {\bfseries
  0808} (2008) 005},
\href{http://arxiv.org/abs/0805.4265}{{\ttfamily arXiv:0805.4265 [astro-ph]}}.
%%CITATION = ARXIV:0805.4265;%%.

\bibitem{Watanabe:2009ct}
M.-a. Watanabe, S.~Kanno, and J.~Soda, ``{Inflationary Universe with
  Anisotropic Hair},''
  \href{http://dx.doi.org/10.1103/PhysRevLett.102.191302}{{\em Phys. Rev.
  Lett.} {\bfseries 102} (2009) 191302},
\href{http://arxiv.org/abs/0902.2833}{{\ttfamily arXiv:0902.2833 [hep-th]}}.
%%CITATION = ARXIV:0902.2833;%%.

\bibitem{Dimopoulos:2009vu}
K.~Dimopoulos, M.~Karciauskas, and J.~M. Wagstaff, ``{Vector Curvaton without
  Instabilities},''
  \href{http://dx.doi.org/10.1016/j.physletb.2009.12.024}{{\em Phys. Lett.}
  {\bfseries B683} (2010) 298--301},
\href{http://arxiv.org/abs/0909.0475}{{\ttfamily arXiv:0909.0475 [hep-ph]}}.
%%CITATION = ARXIV:0909.0475;%%.

\bibitem{Dimopoulos:2009am}
K.~Dimopoulos, M.~Karciauskas, and J.~M. Wagstaff, ``{Vector Curvaton with
  varying Kinetic Function},''
  \href{http://dx.doi.org/10.1103/PhysRevD.81.023522}{{\em Phys. Rev.}
  {\bfseries D81} (2010) 023522},
\href{http://arxiv.org/abs/0907.1838}{{\ttfamily arXiv:0907.1838 [hep-ph]}}.
%%CITATION = ARXIV:0907.1838;%%.

\bibitem{Watanabe:2010bu}
M.-a. Watanabe, S.~Kanno, and J.~Soda, ``{Imprints of Anisotropic Inflation on
  the Cosmic Microwave Background},''
  \href{http://dx.doi.org/10.1111/j.1745-3933.2011.01010.x}{{\em Mon. Not. Roy.
  Astron. Soc.} {\bfseries 412} (2011) L83--L87},
\href{http://arxiv.org/abs/1011.3604}{{\ttfamily arXiv:1011.3604
  [astro-ph.CO]}}.
%%CITATION = ARXIV:1011.3604;%%.

\bibitem{Bartolo:2012sd}
N.~Bartolo, S.~Matarrese, M.~Peloso, and A.~Ricciardone, ``{Anisotropic power
  spectrum and bispectrum in the $f(\phi)F^2$ mechanism},''
  \href{http://dx.doi.org/10.1103/PhysRevD.87.023504}{{\em Phys. Rev.}
  {\bfseries D87} no.~2, (2013) 023504},
\href{http://arxiv.org/abs/1210.3257}{{\ttfamily arXiv:1210.3257
  [astro-ph.CO]}}.
%%CITATION = ARXIV:1210.3257;%%.

\bibitem{Biagetti:2013qqa}
M.~Biagetti, A.~Kehagias, E.~Morgante, H.~Perrier, and A.~Riotto, ``{Symmetries
  of Vector Perturbations during the de Sitter Epoch},''
  \href{http://dx.doi.org/10.1088/1475-7516/2013/07/030}{{\em JCAP} {\bfseries
  1307} (2013) 030},
\href{http://arxiv.org/abs/1304.7785}{{\ttfamily arXiv:1304.7785
  [astro-ph.CO]}}.
%%CITATION = ARXIV:1304.7785;%%.

\bibitem{Shiraishi:2013vja}
M.~Shiraishi, E.~Komatsu, M.~Peloso, and N.~Barnaby, ``{Signatures of
  anisotropic sources in the squeezed-limit bispectrum of the cosmic microwave
  background},'' \href{http://dx.doi.org/10.1088/1475-7516/2013/05/002}{{\em
  JCAP} {\bfseries 1305} (2013) 002},
\href{http://arxiv.org/abs/1302.3056}{{\ttfamily arXiv:1302.3056
  [astro-ph.CO]}}.
%%CITATION = ARXIV:1302.3056;%%.

\bibitem{Abolhasani:2013zya}
A.~A. Abolhasani, R.~Emami, J.~T. Firouzjaee, and H.~Firouzjahi, ``{$\delta N$
  formalism in anisotropic inflation and large anisotropic bispectrum and
  trispectrum},'' \href{http://dx.doi.org/10.1088/1475-7516/2013/08/016}{{\em
  JCAP} {\bfseries 1308} (2013) 016},
\href{http://arxiv.org/abs/1302.6986}{{\ttfamily arXiv:1302.6986
  [astro-ph.CO]}}.
%%CITATION = ARXIV:1302.6986;%%.

\bibitem{Lyth:2013kah}
D.~H. Lyth and M.~Karciauskas, ``{The statistically anisotropic curvature
  perturbation generated by $f(\phi)^2 F^2$},''
  \href{http://dx.doi.org/10.1088/1475-7516/2013/05/011}{{\em JCAP} {\bfseries
  1305} (2013) 011},
\href{http://arxiv.org/abs/1302.7304}{{\ttfamily arXiv:1302.7304
  [astro-ph.CO]}}.
%%CITATION = ARXIV:1302.7304;%%.

\bibitem{Rodriguez:2013cj}
Y.~Rodriguez, J.~P. Beltr\'an~Almeida, and C.~A. Valenzuela-Toledo, ``{The
  different varieties of the Suyama-Yamaguchi consistency relation and its
  violation as a signal of statistical inhomogeneity},''
  \href{http://dx.doi.org/10.1088/1475-7516/2013/04/039}{{\em JCAP} {\bfseries
  1304} (2013) 039},
\href{http://arxiv.org/abs/1301.5843}{{\ttfamily arXiv:1301.5843
  [astro-ph.CO]}}.
%%CITATION = ARXIV:1301.5843;%%.

\bibitem{Lyth:2013vha}
D.~H. Lyth, ``{The CMB modulation from inflation},''
  \href{http://dx.doi.org/10.1088/1475-7516/2013/08/007}{{\em JCAP} {\bfseries
  1308} (2013) 007},
\href{http://arxiv.org/abs/1304.1270}{{\ttfamily arXiv:1304.1270
  [astro-ph.CO]}}.
%%CITATION = ARXIV:1304.1270;%%.

\bibitem{Shiraishi:2013oqa}
M.~Shiraishi, E.~Komatsu, and M.~Peloso, ``{Signatures of anisotropic sources
  in the trispectrum of the cosmic microwave background},''
  \href{http://dx.doi.org/10.1088/1475-7516/2014/04/027}{{\em JCAP} {\bfseries
  1404} (2014) 027},
\href{http://arxiv.org/abs/1312.5221}{{\ttfamily arXiv:1312.5221
  [astro-ph.CO]}}.
%%CITATION = ARXIV:1312.5221;%%.

\bibitem{Chen:2014eua}
X.~Chen, R.~Emami, H.~Firouzjahi, and Y.~Wang, ``{The TT, TB, EB and BB
  correlations in anisotropic inflation},''
  \href{http://dx.doi.org/10.1088/1475-7516/2014/08/027}{{\em JCAP} {\bfseries
  1408} (2014) 027},
\href{http://arxiv.org/abs/1404.4083}{{\ttfamily arXiv:1404.4083
  [astro-ph.CO]}}.
%%CITATION = ARXIV:1404.4083;%%.

\bibitem{Almeida:2014ava}
J.~P. Beltr\'an~Almeida, Y.~Rodriguez, and C.~A. Valenzuela-Toledo, ``{Scale
  and shape dependent non-Gaussianity in the presence of inflationary vector
  fields},'' \href{http://dx.doi.org/10.1103/PhysRevD.90.103511}{{\em Phys.
  Rev.} {\bfseries D90} (2014) 103511},
\href{http://arxiv.org/abs/1405.7374}{{\ttfamily arXiv:1405.7374
  [astro-ph.CO]}}.
%%CITATION = ARXIV:1405.7374;%%.

\bibitem{Fujita:2017lfu}
T.~Fujita and I.~Obata, ``{Does anisotropic inflation produce a small
  statistical anisotropy?},''
  \href{http://dx.doi.org/10.1088/1475-7516/2018/01/049}{{\em JCAP} {\bfseries
  1801} no.~01, (2018) 049},
\href{http://arxiv.org/abs/1711.11539}{{\ttfamily arXiv:1711.11539
  [astro-ph.CO]}}.
%%CITATION = ARXIV:1711.11539;%%.

\bibitem{Thorsrud:2012mu}
M.~Thorsrud, D.~F. Mota, and S.~Hervik, ``{Cosmology of a Scalar Field Coupled
  to Matter and an Isotropy-Violating Maxwell Field},''
  \href{http://dx.doi.org/10.1007/JHEP10(2012)066}{{\em JHEP} {\bfseries 10}
  (2012) 066},
\href{http://arxiv.org/abs/1205.6261}{{\ttfamily arXiv:1205.6261 [hep-th]}}.
%%CITATION = ARXIV:1205.6261;%%.

\bibitem{Sorbo:2011rz}
L.~Sorbo, ``{Parity violation in the Cosmic Microwave Background from a
  pseudoscalar inflaton},''
  \href{http://dx.doi.org/10.1088/1475-7516/2011/06/003}{{\em JCAP} {\bfseries
  1106} (2011) 003},
\href{http://arxiv.org/abs/1101.1525}{{\ttfamily arXiv:1101.1525
  [astro-ph.CO]}}.
%%CITATION = ARXIV:1101.1525;%%.

\bibitem{Dimopoulos:2012av}
K.~Dimopoulos and M.~Karciauskas, ``{Parity Violating Statistical
  Anisotropy},'' \href{http://dx.doi.org/10.1007/JHEP06(2012)040}{{\em JHEP}
  {\bfseries 06} (2012) 040},
\href{http://arxiv.org/abs/1203.0230}{{\ttfamily arXiv:1203.0230 [hep-ph]}}.
%%CITATION = ARXIV:1203.0230;%%.

\bibitem{Anber:2012du}
M.~M. Anber and L.~Sorbo, ``{Non-Gaussianities and chiral gravitational waves
  in natural steep inflation},''
  \href{http://dx.doi.org/10.1103/PhysRevD.85.123537}{{\em Phys. Rev.}
  {\bfseries D85} (2012) 123537},
\href{http://arxiv.org/abs/1203.5849}{{\ttfamily arXiv:1203.5849
  [astro-ph.CO]}}.
%%CITATION = ARXIV:1203.5849;%%.

\bibitem{Bartolo:2014hwa}
N.~Bartolo, S.~Matarrese, M.~Peloso, and M.~Shiraishi, ``{Parity-violating and
  anisotropic correlations in pseudoscalar inflation},''
  \href{http://dx.doi.org/10.1088/1475-7516/2015/01/027}{{\em JCAP} {\bfseries
  1501} no.~01, (2015) 027},
\href{http://arxiv.org/abs/1411.2521}{{\ttfamily arXiv:1411.2521
  [astro-ph.CO]}}.
%%CITATION = ARXIV:1411.2521;%%.

\bibitem{Caprini:2014mja}
C.~Caprini and L.~Sorbo, ``{Adding helicity to inflationary magnetogenesis},''
  \href{http://dx.doi.org/10.1088/1475-7516/2014/10/056}{{\em JCAP} {\bfseries
  1410} no.~10, (2014) 056},
\href{http://arxiv.org/abs/1407.2809}{{\ttfamily arXiv:1407.2809
  [astro-ph.CO]}}.
%%CITATION = ARXIV:1407.2809;%%.

\bibitem{Bartolo:2015dga}
N.~Bartolo, S.~Matarrese, M.~Peloso, and M.~Shiraishi, ``{Parity-violating CMB
  correlators with non-decaying statistical anisotropy},''
  \href{http://dx.doi.org/10.1088/1475-7516/2015/07/039}{{\em JCAP} {\bfseries
  1507} no.~07, (2015) 039},
\href{http://arxiv.org/abs/1505.02193}{{\ttfamily arXiv:1505.02193
  [astro-ph.CO]}}.
%%CITATION = ARXIV:1505.02193;%%.

\bibitem{Namba:2015gja}
R.~Namba, M.~Peloso, M.~Shiraishi, L.~Sorbo, and C.~Unal, ``{Scale-dependent
  gravitational waves from a rolling axion},''
  \href{http://dx.doi.org/10.1088/1475-7516/2016/01/041}{{\em JCAP} {\bfseries
  1601} no.~01, (2016) 041},
\href{http://arxiv.org/abs/1509.07521}{{\ttfamily arXiv:1509.07521
  [astro-ph.CO]}}.
%%CITATION = ARXIV:1509.07521;%%.

\bibitem{Shiraishi:2016yun}
M.~Shiraishi, C.~Hikage, R.~Namba, T.~Namikawa, and M.~Hazumi, ``{Testing
  statistics of the CMB B -mode polarization toward unambiguously establishing
  quantum fluctuation of the vacuum},''
  \href{http://dx.doi.org/10.1103/PhysRevD.94.043506}{{\em Phys. Rev.}
  {\bfseries D94} no.~4, (2016) 043506},
\href{http://arxiv.org/abs/1606.06082}{{\ttfamily arXiv:1606.06082
  [astro-ph.CO]}}.
%%CITATION = ARXIV:1606.06082;%%.

\bibitem{Caprini:2017vnn}
C.~Caprini, M.~C. Guzzetti, and L.~Sorbo, ``{Inflationary magnetogenesis with
  added helicity: constraints from non-gaussianities},''
  \href{http://dx.doi.org/10.1088/1361-6382/aac143}{{\em Class. Quant. Grav.}
  {\bfseries 35} no.~12, (2018) 124003},
\href{http://arxiv.org/abs/1707.09750}{{\ttfamily arXiv:1707.09750
  [astro-ph.CO]}}.
%%CITATION = ARXIV:1707.09750;%%.

\bibitem{Almeida:2017lrq}
J.~P.~B. Almeida, J.~Motoa-Manzano, and C.~A. Valenzuela-Toledo, ``{de Sitter
  symmetries and inflationary correlators in parity violating scalar-vector
  models},'' \href{http://dx.doi.org/10.1088/1475-7516/2017/11/015}{{\em JCAP}
  {\bfseries 1711} no.~11, (2017) 015},
\href{http://arxiv.org/abs/1706.05099}{{\ttfamily arXiv:1706.05099
  [astro-ph.CO]}}.
%%CITATION = ARXIV:1706.05099;%%.

\bibitem{Almeida:2018pir}
J.~P. Beltrán~Almeida and N.~Bernal, ``{Nonminimally coupled pseudoscalar
  inflaton},'' \href{http://dx.doi.org/10.1103/PhysRevD.98.083519}{{\em Phys.
  Rev.} {\bfseries D98} no.~8, (2018) 083519},
\href{http://arxiv.org/abs/1803.09743}{{\ttfamily arXiv:1803.09743
  [astro-ph.CO]}}.
%%CITATION = ARXIV:1803.09743;%%.

\bibitem{Almeida:2019hhx}
J.~P.~B. Almeida, J.~Motoa-Manzano, and C.~A. Valenzuela-Toledo, ``{Correlation
  functions of sourced gravitational waves in inflationary scalar vector
  models. A symmetry based approach},''
  \href{http://dx.doi.org/10.1007/JHEP09(2019)118}{{\em JHEP} {\bfseries 09}
  (2019) 118},
\href{http://arxiv.org/abs/1905.00900}{{\ttfamily arXiv:1905.00900 [gr-qc]}}.
%%CITATION = ARXIV:1905.00900;%%.

\bibitem{Ackerman:2007nb}
L.~Ackerman, S.~M. Carroll, and M.~B. Wise, ``{Imprints of a Primordial
  Preferred Direction on the Microwave Background},''
  \href{http://dx.doi.org/10.1103/PhysRevD.75.083502,
  10.1103/PhysRevD.80.069901}{{\em Phys. Rev.} {\bfseries D75} (2007) 083502},
  \href{http://arxiv.org/abs/astro-ph/0701357}{{\ttfamily
  arXiv:astro-ph/0701357 [astro-ph]}}.
[Erratum: Phys. Rev.D80,069901(2009)].
%%CITATION = ASTRO-PH/0701357;%%.

\bibitem{Flauger:2009ab}
R.~Flauger, L.~McAllister, E.~Pajer, A.~Westphal, and G.~Xu, ``{Oscillations in
  the CMB from Axion Monodromy Inflation},''
  \href{http://dx.doi.org/10.1088/1475-7516/2010/06/009}{{\em JCAP} {\bfseries
  1006} (2010) 009},
\href{http://arxiv.org/abs/0907.2916}{{\ttfamily arXiv:0907.2916 [hep-th]}}.
%%CITATION = ARXIV:0907.2916;%%.

\bibitem{Flauger:2014ana}
R.~Flauger, L.~McAllister, E.~Silverstein, and A.~Westphal, ``{Drifting
  Oscillations in Axion Monodromy},''
  \href{http://dx.doi.org/10.1088/1475-7516/2017/10/055}{{\em JCAP} {\bfseries
  1710} no.~10, (2017) 055},
\href{http://arxiv.org/abs/1412.1814}{{\ttfamily arXiv:1412.1814 [hep-th]}}.
%%CITATION = ARXIV:1412.1814;%%.

\bibitem{Schmidt:2015xka}
F.~Schmidt, N.~E. Chisari, and C.~Dvorkin, ``{Imprint of inflation on galaxy
  shape correlations},''
  \href{http://dx.doi.org/10.1088/1475-7516/2015/10/032}{{\em JCAP} {\bfseries
  1510} no.~10, (2015) 032},
\href{http://arxiv.org/abs/1506.02671}{{\ttfamily arXiv:1506.02671
  [astro-ph.CO]}}.
%%CITATION = ARXIV:1506.02671;%%.

\bibitem{MoradinezhadDizgah:2017szk}
A.~Moradinezhad~Dizgah and C.~Dvorkin, ``{Scale-Dependent Galaxy Bias from
  Massive Particles with Spin during Inflation},''
  \href{http://dx.doi.org/10.1088/1475-7516/2018/01/010}{{\em JCAP} {\bfseries
  1801} no.~01, (2018) 010},
\href{http://arxiv.org/abs/1708.06473}{{\ttfamily arXiv:1708.06473
  [astro-ph.CO]}}.
%%CITATION = ARXIV:1708.06473;%%.

\bibitem{MoradinezhadDizgah:2018pfo}
A.~Moradinezhad~Dizgah, G.~Franciolini, A.~Kehagias, and A.~Riotto,
  ``{Constraints on long-lived, higher-spin particles from galaxy
  bispectrum},'' \href{http://dx.doi.org/10.1103/PhysRevD.98.063520}{{\em Phys.
  Rev.} {\bfseries D98} no.~6, (2018) 063520},
\href{http://arxiv.org/abs/1805.10247}{{\ttfamily arXiv:1805.10247
  [astro-ph.CO]}}.
%%CITATION = ARXIV:1805.10247;%%.

\bibitem{MoradinezhadDizgah:2018ssw}
A.~Moradinezhad~Dizgah, H.~Lee, J.~B. Mu\~noz, and C.~Dvorkin, ``{Galaxy
  Bispectrum from Massive Spinning Particles},''
  \href{http://dx.doi.org/10.1088/1475-7516/2018/05/013}{{\em JCAP} {\bfseries
  1805} no.~05, (2018) 013},
\href{http://arxiv.org/abs/1801.07265}{{\ttfamily arXiv:1801.07265
  [astro-ph.CO]}}.
%%CITATION = ARXIV:1801.07265;%%.

\bibitem{Normann:2017aav}
B.~D. Normann, S.~Hervik, A.~Ricciardone, and M.~Thorsrud, ``{Bianchi
  cosmologies with $p$-form gauge fields},''
  \href{http://dx.doi.org/10.1088/1361-6382/aab3a7}{{\em Class. Quant. Grav.}
  {\bfseries 35} no.~9, (2018) 095004},
\href{http://arxiv.org/abs/1712.08752}{{\ttfamily arXiv:1712.08752 [gr-qc]}}.
%%CITATION = ARXIV:1712.08752;%%.

\bibitem{Brown:1987dd}
J.~D. Brown and C.~Teitelboim, ``{Dynamical Neutralization of the Cosmological
  Constant},''
\href{http://dx.doi.org/10.1016/0370-2693(87)91190-7}{{\em Phys. Lett.}
  {\bfseries B195} (1987) 177--182}.
%%CITATION = PHLTA,B195,177;%%.

\bibitem{Brown:1988kg}
J.~D. Brown and C.~Teitelboim, ``{Neutralization of the Cosmological Constant
  by Membrane Creation},''
\href{http://dx.doi.org/10.1016/0550-3213(88)90559-7}{{\em Nucl. Phys.}
  {\bfseries B297} (1988) 787--836}.
%%CITATION = NUPHA,B297,787;%%.

\bibitem{Duncan:1989ug}
M.~J. Duncan and L.~G. Jensen, ``{Four Forms and the Vanishing of the
  Cosmological Constant},''
\href{http://dx.doi.org/10.1016/0550-3213(90)90344-D}{{\em Nucl. Phys.}
  {\bfseries B336} (1990) 100--114}.
%%CITATION = NUPHA,B336,100;%%.

\bibitem{Duff:1989ah}
M.~J. Duff, ``{The Cosmological Constant Is Possibly Zero, but the Proof Is
  Probably Wrong},'' \href{http://dx.doi.org/10.1016/0370-2693(89)90284-0}{{\em
  Phys. Lett.} {\bfseries B226} (1989) 36}.
[Conf. Proc.C8903131,403(1989)].
%%CITATION = PHLTA,B226,36;%%.

\bibitem{Bousso:2000xa}
R.~Bousso and J.~Polchinski, ``{Quantization of four form fluxes and dynamical
  neutralization of the cosmological constant},''
  \href{http://dx.doi.org/10.1088/1126-6708/2000/06/006}{{\em JHEP} {\bfseries
  06} (2000) 006},
\href{http://arxiv.org/abs/hep-th/0004134}{{\ttfamily arXiv:hep-th/0004134
  [hep-th]}}.
%%CITATION = HEP-TH/0004134;%%.

\bibitem{Dvali:2005an}
G.~Dvali, ``{Three-form gauging of axion symmetries and gravity},''
\href{http://arxiv.org/abs/hep-th/0507215}{{\ttfamily arXiv:hep-th/0507215
  [hep-th]}}.
%%CITATION = HEP-TH/0507215;%%.

\bibitem{Kaloper:2008qs}
N.~Kaloper and L.~Sorbo, ``{Where in the String Landscape is Quintessence},''
  \href{http://dx.doi.org/10.1103/PhysRevD.79.043528}{{\em Phys. Rev.}
  {\bfseries D79} (2009) 043528},
\href{http://arxiv.org/abs/0810.5346}{{\ttfamily arXiv:0810.5346 [hep-th]}}.
%%CITATION = ARXIV:0810.5346;%%.

\bibitem{Kaloper:2008fb}
N.~Kaloper and L.~Sorbo, ``{A Natural Framework for Chaotic Inflation},''
  \href{http://dx.doi.org/10.1103/PhysRevLett.102.121301}{{\em Phys. Rev.
  Lett.} {\bfseries 102} (2009) 121301},
\href{http://arxiv.org/abs/0811.1989}{{\ttfamily arXiv:0811.1989 [hep-th]}}.
%%CITATION = ARXIV:0811.1989;%%.

\bibitem{Koivisto:2009sd}
T.~S. Koivisto, D.~F. Mota, and C.~Pitrou, ``{Inflation from N-Forms and its
  stability},'' \href{http://dx.doi.org/10.1088/1126-6708/2009/09/092}{{\em
  JHEP} {\bfseries 09} (2009) 092},
\href{http://arxiv.org/abs/0903.4158}{{\ttfamily arXiv:0903.4158
  [astro-ph.CO]}}.
%%CITATION = ARXIV:0903.4158;%%.

\bibitem{Koivisto:2009ew}
T.~S. Koivisto and N.~J. Nunes, ``{Three-form cosmology},''
  \href{http://dx.doi.org/10.1016/j.physletb.2010.01.051}{{\em Phys. Lett.}
  {\bfseries B685} (2010) 105--109},
\href{http://arxiv.org/abs/0907.3883}{{\ttfamily arXiv:0907.3883
  [astro-ph.CO]}}.
%%CITATION = ARXIV:0907.3883;%%.

\bibitem{Koivisto:2009fb}
T.~S. Koivisto and N.~J. Nunes, ``{Inflation and dark energy from
  three-forms},'' \href{http://dx.doi.org/10.1103/PhysRevD.80.103509}{{\em
  Phys. Rev.} {\bfseries D80} (2009) 103509},
\href{http://arxiv.org/abs/0908.0920}{{\ttfamily arXiv:0908.0920
  [astro-ph.CO]}}.
%%CITATION = ARXIV:0908.0920;%%.

\bibitem{Bielleman:2015ina}
S.~Bielleman, L.~E. Ibanez, and I.~Valenzuela, ``{Minkowski 3-forms, Flux
  String Vacua, Axion Stability and Naturalness},''
  \href{http://dx.doi.org/10.1007/JHEP12(2015)119}{{\em JHEP} {\bfseries 12}
  (2015) 119},
\href{http://arxiv.org/abs/1507.06793}{{\ttfamily arXiv:1507.06793 [hep-th]}}.
%%CITATION = ARXIV:1507.06793;%%.

\bibitem{Amendola2010}
L.~Amendola and S.~Tsujikawa, {\em {Dark Energy: Theory and Observations}}.
\newblock Cambridge University Press, 2010.

\bibitem{Clifton:2011jh}
T.~Clifton, P.~G. Ferreira, A.~Padilla, and C.~Skordis, ``{Modified Gravity and
  Cosmology},'' \href{http://dx.doi.org/10.1016/j.physrep.2012.01.001}{{\em
  Phys.Rept.} {\bfseries 513} (2012) 1--189},
\href{http://arxiv.org/abs/1106.2476}{{\ttfamily arXiv:1106.2476
  [astro-ph.CO]}}.
%\%CITATION = ARXIV:1106.2476;\%\%.

\bibitem{Koivisto:2008xf}
T.~Koivisto and D.~F. Mota, ``{Vector Field Models of Inflation and Dark
  Energy},'' \href{http://dx.doi.org/10.1088/1475-7516/2008/08/021}{{\em JCAP}
  {\bfseries 0808} (2008) 021},
\href{http://arxiv.org/abs/0805.4229}{{\ttfamily arXiv:0805.4229 [astro-ph]}}.
%%CITATION = ARXIV:0805.4229;%%.

\bibitem{ArmendarizPicon:2004pm}
C.~Armendariz-Picon, ``{Could dark energy be vector-like?},''
  \href{http://dx.doi.org/10.1088/1475-7516/2004/07/007}{{\em JCAP} {\bfseries
  0407} (2004) 007},
\href{http://arxiv.org/abs/astro-ph/0405267}{{\ttfamily arXiv:astro-ph/0405267
  [astro-ph]}}.
%%CITATION = ASTRO-PH/0405267;%%.

\bibitem{Boehmer:2007qa}
C.~G. Boehmer and T.~Harko, ``{Dark energy as a massive vector field},''
  \href{http://dx.doi.org/10.1140/epjc/s10052-007-0210-1}{{\em Eur. Phys. J.}
  {\bfseries C50} (2007) 423--429},
\href{http://arxiv.org/abs/gr-qc/0701029}{{\ttfamily arXiv:gr-qc/0701029
  [gr-qc]}}.
%%CITATION = GR-QC/0701029;%%.

\bibitem{Fairlie:1991qe}
D.~B. Fairlie, J.~Govaerts, and A.~Morozov, ``{Universal field equations with
  covariant solutions},''
  \href{http://dx.doi.org/10.1016/0550-3213(92)90455-K}{{\em Nucl. Phys.}
  {\bfseries B373} (1992) 214--232},
\href{http://arxiv.org/abs/hep-th/9110022}{{\ttfamily arXiv:hep-th/9110022
  [hep-th]}}.
%%CITATION = HEP-TH/9110022;%%.

\bibitem{Fairlie:1992nb}
D.~B. Fairlie and J.~Govaerts, ``{Euler hierarchies and universal equations},''
  \href{http://dx.doi.org/10.1063/1.529904}{{\em J. Math. Phys.} {\bfseries 33}
  (1992) 3543--3566},
\href{http://arxiv.org/abs/hep-th/9204074}{{\ttfamily arXiv:hep-th/9204074
  [hep-th]}}.
%%CITATION = HEP-TH/9204074;%%.

\bibitem{Nicolis:2008in}
A.~Nicolis, R.~Rattazzi, and E.~Trincherini, ``{The Galileon as a local
  modification of gravity},''
  \href{http://dx.doi.org/10.1103/PhysRevD.79.064036}{{\em Phys. Rev.}
  {\bfseries D79} (2009) 064036},
\href{http://arxiv.org/abs/0811.2197}{{\ttfamily arXiv:0811.2197 [hep-th]}}.
%%CITATION = ARXIV:0811.2197;%%.

\bibitem{Deffayet:2009wt}
C.~Deffayet, G.~Esposito-Farese, and A.~Vikman, ``{Covariant Galileon},''
  \href{http://dx.doi.org/10.1103/PhysRevD.79.084003}{{\em Phys. Rev.}
  {\bfseries D79} (2009) 084003},
\href{http://arxiv.org/abs/0901.1314}{{\ttfamily arXiv:0901.1314 [hep-th]}}.
%%CITATION = ARXIV:0901.1314;%%.

\bibitem{Deffayet:2009mn}
C.~Deffayet, S.~Deser, and G.~Esposito-Farese, ``{Generalized Galileons: All
  scalar models whose curved background extensions maintain second-order field
  equations and stress-tensors},''
  \href{http://dx.doi.org/10.1103/PhysRevD.80.064015}{{\em Phys. Rev.}
  {\bfseries D80} (2009) 064015},
\href{http://arxiv.org/abs/0906.1967}{{\ttfamily arXiv:0906.1967 [gr-qc]}}.
%%CITATION = ARXIV:0906.1967;%%.

\bibitem{Deffayet:2010zh}
C.~Deffayet, S.~Deser, and G.~Esposito-Farese, ``{Arbitrary $p$-form
  Galileons},'' \href{http://dx.doi.org/10.1103/PhysRevD.82.061501}{{\em Phys.
  Rev.} {\bfseries D82} (2010) 061501},
\href{http://arxiv.org/abs/1007.5278}{{\ttfamily arXiv:1007.5278 [gr-qc]}}.
%%CITATION = ARXIV:1007.5278;%%.

\bibitem{Deffayet:2017eqq}
C.~Deffayet, S.~Garcia-Saenz, S.~Mukohyama, and V.~Sivanesan, ``{Classifying
  Galileon $p$-form theories},''
  \href{http://dx.doi.org/10.1103/PhysRevD.96.045014}{{\em Phys. Rev.}
  {\bfseries D96} no.~4, (2017) 045014},
\href{http://arxiv.org/abs/1704.02980}{{\ttfamily arXiv:1704.02980 [hep-th]}}.
%%CITATION = ARXIV:1704.02980;%%.

\bibitem{Ito:2015sxj}
A.~Ito and J.~Soda, ``{Designing Anisotropic Inflation with Form Fields},''
  \href{http://dx.doi.org/10.1103/PhysRevD.92.123533}{{\em Phys. Rev.}
  {\bfseries D92} no.~12, (2015) 123533},
\href{http://arxiv.org/abs/1506.02450}{{\ttfamily arXiv:1506.02450 [hep-th]}}.
%%CITATION = ARXIV:1506.02450;%%.

\bibitem{Fleury:2014qfa}
P.~Fleury, J.~P.~B. Almeida, C.~Pitrou, and J.-P. Uzan, ``{On the stability and
  causality of scalar-vector theories},''
  \href{http://dx.doi.org/10.1088/1475-7516/2014/11/043}{{\em JCAP} {\bfseries
  1411} no.~11, (2014) 043},
\href{http://arxiv.org/abs/1406.6254}{{\ttfamily arXiv:1406.6254 [hep-th]}}.
%%CITATION = ARXIV:1406.6254;%%.

\bibitem{Holland:2017cza}
J.~Holland, S.~Kanno, and I.~Zavala, ``{Anisotropic Inflation with Derivative
  Couplings},'' \href{http://dx.doi.org/10.1103/PhysRevD.97.103534}{{\em Phys.
  Rev.} {\bfseries D97} no.~10, (2018) 103534},
\href{http://arxiv.org/abs/1711.07450}{{\ttfamily arXiv:1711.07450 [hep-th]}}.
%%CITATION = ARXIV:1711.07450;%%.

\bibitem{Chern:1974ft}
S.-S. Chern and J.~Simons, ``{Characteristic forms and geometric invariants},''
\href{http://dx.doi.org/10.2307/1971013}{{\em Annals Math.} {\bfseries 99}
  (1974) 48--69}.
%%CITATION = ANMAA,99,48;%%.

\bibitem{Blau:1989dh}
M.~Blau and G.~Thompson, ``{A New Class of Topological Field Theories and the
  Ray-singer Torsion},''
\href{http://dx.doi.org/10.1016/0370-2693(89)90526-1}{{\em Phys. Lett.}
  {\bfseries B228} (1989) 64--68}.
%%CITATION = PHLTA,B228,64;%%.

\bibitem{Horowitz:1989ng}
G.~T. Horowitz, ``{Exactly Soluble Diffeomorphism Invariant Theories},''
\href{http://dx.doi.org/10.1007/BF01218410}{{\em Commun. Math. Phys.}
  {\bfseries 125} (1989) 417}.
%%CITATION = CMPHA,125,417;%%.

\bibitem{Blau:1989bq}
M.~Blau and G.~Thompson, ``{Topological Gauge Theories of Antisymmetric Tensor
  Fields},''
\href{http://dx.doi.org/10.1016/0003-4916(91)90240-9}{{\em Annals Phys.}
  {\bfseries 205} (1991) 130--172}.
%%CITATION = APNYA,205,130;%%.

\bibitem{Birmingham:1991ty}
D.~Birmingham, M.~Blau, M.~Rakowski, and G.~Thompson, ``{Topological field
  theory},''
\href{http://dx.doi.org/10.1016/0370-1573(91)90117-5}{{\em Phys. Rept.}
  {\bfseries 209} (1991) 129--340}.
%%CITATION = PRPLC,209,129;%%.

\bibitem{Aydemir:2010tna}
U.~Aydemir, L.~Grisa, and L.~Sorbo, ``{Dynamical Four-Form Fields},''
  \href{http://dx.doi.org/10.1103/PhysRevD.83.063516}{{\em Phys. Rev.}
  {\bfseries D83} (2011) 063516},
\href{http://arxiv.org/abs/1009.5690}{{\ttfamily arXiv:1009.5690 [hep-th]}}.
%%CITATION = ARXIV:1009.5690;%%.

\bibitem{Maleknejad:2011jw}
A.~Maleknejad and M.~M. Sheikh-Jabbari, ``{Gauge-flation: Inflation From
  Non-Abelian Gauge Fields},''
  \href{http://dx.doi.org/10.1016/j.physletb.2013.05.001}{{\em Phys. Lett.}
  {\bfseries B723} (2013) 224--228},
\href{http://arxiv.org/abs/1102.1513}{{\ttfamily arXiv:1102.1513 [hep-ph]}}.
%%CITATION = ARXIV:1102.1513;%%.

\bibitem{Maleknejad:2011sq}
A.~Maleknejad and M.~M. Sheikh-Jabbari, ``{Non-Abelian Gauge Field
  Inflation},'' \href{http://dx.doi.org/10.1103/PhysRevD.84.043515}{{\em Phys.
  Rev.} {\bfseries D84} (2011) 043515},
\href{http://arxiv.org/abs/1102.1932}{{\ttfamily arXiv:1102.1932 [hep-ph]}}.
%%CITATION = ARXIV:1102.1932;%%.

\bibitem{Adshead:2012kp}
P.~Adshead and M.~Wyman, ``{Chromo-Natural Inflation: Natural inflation on a
  steep potential with classical non-Abelian gauge fields},''
  \href{http://dx.doi.org/10.1103/PhysRevLett.108.261302}{{\em Phys. Rev.
  Lett.} {\bfseries 108} (2012) 261302},
\href{http://arxiv.org/abs/1202.2366}{{\ttfamily arXiv:1202.2366 [hep-th]}}.
%%CITATION = ARXIV:1202.2366;%%.

\bibitem{Namba:2013kia}
R.~Namba, E.~Dimastrogiovanni, and M.~Peloso, ``{Gauge-flation confronted with
  Planck},'' \href{http://dx.doi.org/10.1088/1475-7516/2013/11/045}{{\em JCAP}
  {\bfseries 1311} (2013) 045},
\href{http://arxiv.org/abs/1308.1366}{{\ttfamily arXiv:1308.1366
  [astro-ph.CO]}}.
%%CITATION = ARXIV:1308.1366;%%.

\bibitem{Nieto:2016gnp}
C.~M. Nieto and Y.~Rodriguez, ``{Massive Gauge-flation},''
  \href{http://dx.doi.org/10.1142/S0217732316400058}{{\em Mod. Phys. Lett.}
  {\bfseries A31} no.~21, (2016) 1640005},
\href{http://arxiv.org/abs/1602.07197}{{\ttfamily arXiv:1602.07197 [gr-qc]}}.
%%CITATION = ARXIV:1602.07197;%%.

\bibitem{Adshead:2016omu}
P.~Adshead, E.~Martinec, E.~I. Sfakianakis, and M.~Wyman, ``{Higgsed
  Chromo-Natural Inflation},''
  \href{http://dx.doi.org/10.1007/JHEP12(2016)137}{{\em JHEP} {\bfseries 12}
  (2016) 137},
\href{http://arxiv.org/abs/1609.04025}{{\ttfamily arXiv:1609.04025 [hep-th]}}.
%%CITATION = ARXIV:1609.04025;%%.

\bibitem{Allys:2016kbq}
E.~Allys, P.~Peter, and Y.~Rodriguez, ``{Generalized SU(2) Proca Theory},''
  \href{http://dx.doi.org/10.1103/PhysRevD.94.084041}{{\em Phys. Rev.}
  {\bfseries D94} no.~8, (2016) 084041},
\href{http://arxiv.org/abs/1609.05870}{{\ttfamily arXiv:1609.05870 [hep-th]}}.
%%CITATION = ARXIV:1609.05870;%%.

\bibitem{Maleknejad:2014wsa}
A.~Maleknejad, ``{Chiral Gravity Waves and Leptogenesis in Inflationary Models
  with non-Abelian Gauge Fields},''
  \href{http://dx.doi.org/10.1103/PhysRevD.90.023542}{{\em Phys. Rev.}
  {\bfseries D90} no.~2, (2014) 023542},
\href{http://arxiv.org/abs/1401.7628}{{\ttfamily arXiv:1401.7628 [hep-th]}}.
%%CITATION = ARXIV:1401.7628;%%.

\bibitem{Maleknejad:2016dci}
A.~Maleknejad, ``{Gravitational leptogenesis in axion inflation with SU(2)
  gauge field},'' \href{http://dx.doi.org/10.1088/1475-7516/2016/12/027}{{\em
  JCAP} {\bfseries 1612} no.~12, (2016) 027},
\href{http://arxiv.org/abs/1604.06520}{{\ttfamily arXiv:1604.06520 [hep-ph]}}.
%%CITATION = ARXIV:1604.06520;%%.

\bibitem{Allen:1990gb}
T.~J. Allen, M.~J. Bowick, and A.~Lahiri, ``{Topological mass generation in
  (3+1)-dimensions},''
\href{http://dx.doi.org/10.1142/S0217732391000580}{{\em Mod. Phys. Lett.}
  {\bfseries A6} (1991) 559--572}.
%%CITATION = MPLAE,A6,559;%%.

\bibitem{Dvali:2005ws}
G.~Dvali, R.~Jackiw, and S.-Y. Pi, ``{Topological mass generation in four
  dimensions},'' \href{http://dx.doi.org/10.1103/PhysRevLett.96.081602}{{\em
  Phys. Rev. Lett.} {\bfseries 96} (2006) 081602},
\href{http://arxiv.org/abs/hep-th/0511175}{{\ttfamily arXiv:hep-th/0511175
  [hep-th]}}.
%%CITATION = HEP-TH/0511175;%%.

\bibitem{Tasinato:2014eka}
G.~Tasinato, ``{Cosmic Acceleration from Abelian Symmetry Breaking},''
  \href{http://dx.doi.org/10.1007/JHEP04(2014)067}{{\em JHEP} {\bfseries 04}
  (2014) 067},
\href{http://arxiv.org/abs/1402.6450}{{\ttfamily arXiv:1402.6450 [hep-th]}}.
%%CITATION = ARXIV:1402.6450;%%.

\bibitem{Heisenberg:2014rta}
L.~Heisenberg, ``{Generalization of the Proca Action},''
  \href{http://dx.doi.org/10.1088/1475-7516/2014/05/015}{{\em JCAP} {\bfseries
  1405} (2014) 015},
\href{http://arxiv.org/abs/1402.7026}{{\ttfamily arXiv:1402.7026 [hep-th]}}.
%%CITATION = ARXIV:1402.7026;%%.

\bibitem{Allys:2015sht}
E.~Allys, P.~Peter, and Y.~Rodriguez, ``{Generalized Proca action for an
  Abelian vector field},''
  \href{http://dx.doi.org/10.1088/1475-7516/2016/02/004}{{\em JCAP} {\bfseries
  1602} no.~02, (2016) 004},
\href{http://arxiv.org/abs/1511.03101}{{\ttfamily arXiv:1511.03101 [hep-th]}}.
%%CITATION = ARXIV:1511.03101;%%.

\bibitem{Jimenez:2016isa}
J.~Beltran~Jimenez and L.~Heisenberg, ``{Derivative self-interactions for a
  massive vector field},''
  \href{http://dx.doi.org/10.1016/j.physletb.2016.04.017}{{\em Phys. Lett.}
  {\bfseries B757} (2016) 405--411},
\href{http://arxiv.org/abs/1602.03410}{{\ttfamily arXiv:1602.03410 [hep-th]}}.
%%CITATION = ARXIV:1602.03410;%%.

\bibitem{Allys:2016jaq}
E.~Allys, J.~P. Beltr\'an~Almeida, P.~Peter, and Y.~Rodr\'iguez, ``{On the 4D
  generalized Proca action for an Abelian vector field},''
  \href{http://dx.doi.org/10.1088/1475-7516/2016/09/026}{{\em JCAP} {\bfseries
  1609} no.~09, (2016) 026},
\href{http://arxiv.org/abs/1605.08355}{{\ttfamily arXiv:1605.08355 [hep-th]}}.
%%CITATION = ARXIV:1605.08355;%%.

\bibitem{Julia:1979ur}
B.~Julia and G.~Toulouse, ``{The Many Defect Problem: Gauge Like Variables for
  Ordered Media Containing Defects},''
\href{http://dx.doi.org/10.1051/jphyslet:019790040016039500}{{\em J. Phys.
  Lett.} {\bfseries 40} (1979) 396}.
%%CITATION = JPSLB,40,396;%%.

\bibitem{Quevedo:1996uu}
F.~Quevedo and C.~A. Trugenberger, ``{Phases of antisymmetric tensor field
  theories},'' \href{http://dx.doi.org/10.1016/S0550-3213(97)00337-4}{{\em
  Nucl. Phys.} {\bfseries B501} (1997) 143--172},
\href{http://arxiv.org/abs/hep-th/9604196}{{\ttfamily arXiv:hep-th/9604196
  [hep-th]}}.
%%CITATION = HEP-TH/9604196;%%.

\bibitem{Koivisto:2012xm}
T.~S. Koivisto and N.~J. Nunes, ``{Coupled three-form dark energy},''
  \href{http://dx.doi.org/10.1103/PhysRevD.88.123512}{{\em Phys. Rev.}
  {\bfseries D88} no.~12, (2013) 123512},
\href{http://arxiv.org/abs/1212.2541}{{\ttfamily arXiv:1212.2541
  [astro-ph.CO]}}.
%%CITATION = ARXIV:1212.2541;%%.

\bibitem{Almeida:2019iqp}
J.~P. Beltrán~Almeida, A.~Guarnizo, R.~Kase, S.~Tsujikawa, and C.~A.
  Valenzuela-Toledo, ``{Anisotropic $2$-form dark energy},''
  \href{http://dx.doi.org/10.1016/j.physletb.2019.05.008}{{\em Phys. Lett.}
  {\bfseries B793} (2019) 396--404},
\href{http://arxiv.org/abs/1902.05846}{{\ttfamily arXiv:1902.05846 [hep-th]}}.
%%CITATION = ARXIV:1902.05846;%%.

\bibitem{Guarnizo:2019mwf}
A.~Guarnizo, J.~P.~B. Almeida, and C.~A. Valenzuela-Toledo, ``{$p$-form
  quintessence: exploring dark energy of $p-$forms coupled to a scalar
  field},'' in {\em {15th Marcel Grossmann Meeting on Recent Developments in
  Theoretical and Experimental General Relativity, Astrophysics, and
  Relativistic Field Theories (MG15) Rome, Italy, July 1-7, 2018}}.
\newblock 2019.
\newblock
\href{http://arxiv.org/abs/1910.10499}{{\ttfamily arXiv:1910.10499 [gr-qc]}}.
\newblock
%%CITATION = ARXIV:1910.10499;%%.

\bibitem{Bahamonde:2017ize}
S.~Bahamonde, C.~G. Böhmer, S.~Carloni, E.~J. Copeland, W.~Fang, and
  N.~Tamanini, ``{Dynamical systems applied to cosmology: dark energy and
  modified gravity},''
  \href{http://dx.doi.org/10.1016/j.physrep.2018.09.001}{{\em Phys. Rept.}
  {\bfseries 775-777} (2018) 1--122},
\href{http://arxiv.org/abs/1712.03107}{{\ttfamily arXiv:1712.03107 [gr-qc]}}.
%%CITATION = ARXIV:1712.03107;%%.

\bibitem{wainwright:1997}
J.~Wainwright and G.~F.~R. Ellis,
  \href{http://dx.doi.org/10.1017/CBO9780511524660}{{\em Dynamical Systems in
  Cosmology}}.
\newblock Cambridge University Press, 1997.

\bibitem{Bean:2001wt}
R.~Bean, S.~H. Hansen, and A.~Melchiorri, ``{Early universe constraints on a
  primordial scaling field},''
  \href{http://dx.doi.org/10.1103/PhysRevD.64.103508}{{\em Phys. Rev.}
  {\bfseries D64} (2001) 103508},
\href{http://arxiv.org/abs/astro-ph/0104162}{{\ttfamily arXiv:astro-ph/0104162
  [astro-ph]}}.
%%CITATION = ASTRO-PH/0104162;%%.

\bibitem{Ohashi:2009xw}
J.~Ohashi and S.~Tsujikawa, ``{Assisted dark energy},''
  \href{http://dx.doi.org/10.1103/PhysRevD.80.103513}{{\em Phys. Rev.}
  {\bfseries D80} (2009) 103513},
\href{http://arxiv.org/abs/0909.3924}{{\ttfamily arXiv:0909.3924 [gr-qc]}}.
%%CITATION = ARXIV:0909.3924;%%.

\bibitem{Ade:2015rim}
{\bfseries Planck} Collaboration, P.~A.~R. Ade {\em et~al.}, ``{Planck 2015
  results. XIV. Dark energy and modified gravity},''
  \href{http://dx.doi.org/10.1051/0004-6361/201525814}{{\em Astron. Astrophys.}
  {\bfseries 594} (2016) A14},
\href{http://arxiv.org/abs/1502.01590}{{\ttfamily arXiv:1502.01590
  [astro-ph.CO]}}.
%%CITATION = ARXIV:1502.01590;%%.

\bibitem{Amendola:2012ys}
{\bfseries Euclid Theory Working Group} Collaboration, L.~Amendola {\em
  et~al.}, ``{Cosmology and fundamental physics with the Euclid satellite},''
  \href{http://dx.doi.org/10.12942/lrr-2013-6}{{\em Living Rev. Rel.}
  {\bfseries 16} (2013) 6},
\href{http://arxiv.org/abs/1206.1225}{{\ttfamily arXiv:1206.1225
  [astro-ph.CO]}}.
%%CITATION = ARXIV:1206.1225;%%.

\bibitem{Riess:2016jrr}
A.~G. Riess {\em et~al.}, ``{A 2.4\% Determination of the Local Value of the
  Hubble Constant},'' \href{http://dx.doi.org/10.3847/0004-637X/826/1/56}{{\em
  Astrophys. J.} {\bfseries 826} no.~1, (2016) 56},
\href{http://arxiv.org/abs/1604.01424}{{\ttfamily arXiv:1604.01424
  [astro-ph.CO]}}.
%%CITATION = ARXIV:1604.01424;%%.

\bibitem{Mortsell:2018mfj}
E.~Mörtsell and S.~Dhawan, ``{Does the Hubble constant tension call for new
  physics?},'' \href{http://dx.doi.org/10.1088/1475-7516/2018/09/025}{{\em
  JCAP} {\bfseries 1809} no.~09, (2018) 025},
\href{http://arxiv.org/abs/1801.07260}{{\ttfamily arXiv:1801.07260
  [astro-ph.CO]}}.
%%CITATION = ARXIV:1801.07260;%%.

\bibitem{DiValentino:2017rcr}
E.~Di~Valentino, E.~V. Linder, and A.~Melchiorri, ``{Vacuum phase transition
  solves the $H_0$ tension},''
  \href{http://dx.doi.org/10.1103/PhysRevD.97.043528}{{\em Phys. Rev.}
  {\bfseries D97} no.~4, (2018) 043528},
\href{http://arxiv.org/abs/1710.02153}{{\ttfamily arXiv:1710.02153
  [astro-ph.CO]}}.
%%CITATION = ARXIV:1710.02153;%%.

\bibitem{DiValentino:2017iww}
E.~Di~Valentino, A.~Melchiorri, and O.~Mena, ``{Can interacting dark energy
  solve the $H_0$ tension?},''
  \href{http://dx.doi.org/10.1103/PhysRevD.96.043503}{{\em Phys. Rev.}
  {\bfseries D96} no.~4, (2017) 043503},
\href{http://arxiv.org/abs/1704.08342}{{\ttfamily arXiv:1704.08342
  [astro-ph.CO]}}.
%%CITATION = ARXIV:1704.08342;%%.

\bibitem{Guo:2018ans}
R.-Y. Guo, J.-F. Zhang, and X.~Zhang, ``{Can the $H_0$ tension be resolved in
  extensions to $\Lambda$CDM cosmology?},''
  \href{http://dx.doi.org/10.1088/1475-7516/2019/02/054}{{\em JCAP} {\bfseries
  1902} (2019) 054},
\href{http://arxiv.org/abs/1809.02340}{{\ttfamily arXiv:1809.02340
  [astro-ph.CO]}}.
%%CITATION = ARXIV:1809.02340;%%.

\bibitem{Pourtsidou:2016ico}
A.~Pourtsidou and T.~Tram, ``{Reconciling CMB and structure growth measurements
  with dark energy interactions},''
  \href{http://dx.doi.org/10.1103/PhysRevD.94.043518}{{\em Phys. Rev.}
  {\bfseries D94} no.~4, (2016) 043518},
\href{http://arxiv.org/abs/1604.04222}{{\ttfamily arXiv:1604.04222
  [astro-ph.CO]}}.
%%CITATION = ARXIV:1604.04222;%%.

\bibitem{Khosravi:2017hfi}
N.~Khosravi, S.~Baghram, N.~Afshordi, and N.~Altamirano, ``{$H_0$ tension as a
  hint for a transition in gravitational theory},''
  \href{http://dx.doi.org/10.1103/PhysRevD.99.103526}{{\em Phys. Rev.}
  {\bfseries D99} no.~10, (2019) 103526},
\href{http://arxiv.org/abs/1710.09366}{{\ttfamily arXiv:1710.09366
  [astro-ph.CO]}}.
%%CITATION = ARXIV:1710.09366;%%.

\end{thebibliography}\endgroup

%\appendix

\end{document}